\documentclass[preprint]{elsarticle}

\usepackage[ruled]{algorithm2e}

\usepackage{multirow}
\usepackage{caption}
\usepackage{subfigure}
\usepackage{graphics}
\usepackage{graphicx}
\usepackage{amsmath}
\usepackage{amssymb}
\usepackage{amsfonts}
\usepackage{epstopdf}
\usepackage{booktabs}
\usepackage{url}
\usepackage{wrapfig}

\begin{document}
%

\title{Ear-Phone: A Context-Aware Noise Mapping using Smart Phones.}

 \author[rvt]{Rajib Rana\corref{cor1}}
\ead{rajib.rana@csiro.au}

 \author[focal]{Chun Tung Chou}
 \ead{ctchou@cse.unsw.edu.au}

 \author[els]{Nirupama Bulusu}
 \ead{nbulusu@cs.pdx.edu}

 \author[focal]{Salil Kanhere}
 \ead{salilk@cse.unsw.edu.au}
\author[rvt]{Wen Hu}
 \ead{wen.hu@csiro.au}

\cortext[cor1]{Corresponding author. \\Tel.: +61 7 3327 4471; Fax: +61 7 3327 4455. \\Address: 1 Technology ct, Pullenvale, QLD-4069, Australia.}

\address[rvt]{Autonomous Systems Laboratory, CSIRO, Australia.}
 \address[focal]{University of New South Wales, Sydney, Australia.}
 \address[els]{Portland State University, USA.}

\newcommand{\laeqt}[1]{\mbox{LA$_{\rm eq,{#1}}$}}
\newcommand{\laeq}{\mbox{LA$_{\rm eq}$} }
\newcommand{\commercial}{\mbox{{RefSLM}}\hspace{.1cm}}
\newcommand{\mobile}{\mbox{MobSLM}\hspace{.1cm}}
\newcommand{\mobiles}{\mbox{{MobSLM}s} }
\newcommand{\vbart}{\mbox{$\bar{v}_A(T)$}}
\newcommand{\PWSN}{\mbox{\textsc{pwsn}} }
\newcommand{\WSN}{\mbox{\textsc{wsn}} }
\newcommand{\CS}{\mbox{\textsc{cs}} }


\begin{abstract}
A noise map facilitates the monitoring of environmental noise pollution in urban areas. 
It can raise citizen awareness of noise pollution levels, and aid in the development of
mitigation strategies to cope with the adverse effects. However, state-of-the-art 
techniques for rendering noise maps in urban areas are expensive and rarely updated (for months or even years), as they rely on population and traffic models rather than 
on real data. Smart phone based urban sensing can be leveraged to create an open and inexpensive platform 
for rendering up-to-date noise maps. In this paper, we present the design, implementation and performance evaluation of 
an \emph{end-to-end, context-aware,}  noise mapping system 
called Ear-Phone. 

Ear-Phone investigates the use of different interpolation and regularization methods to address the fundamental  problem of recovering the noise map 
from incomplete and random samples obtained by crowdsourcing data collection. Ear-Phone, 
implemented on Nokia N95, N97 and HP iPAQ, HTC One mobile devices, also addresses the challenge of 
collecting accurate noise pollution readings at a mobile device. A major challenge
of using smart phones as sensors is that even at the same location, 
the sensor reading may vary depending on the phone orientation and user 
context (for example, whether the user is carrying the phone in a bag or holding it in her palm).
To address this problem, Ear-Phone leverages context-aware sensing. We develop classifiers to accurately determine the phone
sensing context. Upon context discovery, Ear-Phone automatically
decides whether to sense or not. Ear-phone also implements in-situ calibration which performs simple calibration that can be carried out without any technical skills whatsoever required on the user's part. Extensive 
 simulations and outdoor experiments demonstrate 
that Ear-Phone is a feasible platform to assess noise pollution, incurring reasonable system resource 
consumption at mobile devices and providing high reconstruction accuracy of the noise map.
\end{abstract}

\begin{keyword}
Noise map reconstruction \sep participatory urban sensing \sep context-aware sensing 
\end{keyword}



%
%


\maketitle

\section{Introduction}
\label{sec: intro}
At present, a large number of people around the world are exposed to high levels of noise pollution, which 
can cause serious problems ranging from hearing impairment to negatively influencing productivity and social behavior~\cite{eu}. As an abatement strategy, a number of countries, such as the United Kingdom~\cite{defra} and Germany~\cite{germany}, have started monitoring noise 
pollution. These countries typically use a \emph{noise map} to assess noise pollution 
levels. A noise map is like a weather map for noise but it shows areas which are relatively louder or quieter~\cite{defra}. It is computed using simulations based 
on inputs such as traffic flow data, road or rail type, and vehicle type. Since the collection of such input data 
is very expensive, these maps can be updated only after a long period of time (e.g. $5$ years for UK~\cite{defra}). To alleviate this problem, a recent study~\cite{santinis_realwsn08} proposes the deployment of wireless sensor networks to monitor noise pollution. Wireless sensor networks can certainly eliminate the requirement of sending acoustic engineers to collect real noise measurements, but the deployment cost of a dedicated sensor network in a large urban space will also be prohibitively expensive.

In this paper, we instead propose an opportunistic sensing approach, where noise level data contributed by \emph{pedestrians'} mobile phones are used for monitoring environmental noise, especially roadside ambient noise\footnote{We focus on roads because typically
noise pollution is most severe on busy roads.} . The key idea in opportunistic sensing~\cite{metro_sense} )  is to ``crowdsource'' the collection of environmental data in urban spaces to people, who carry smart phones equipped with sensors and Global Positioning System (GPS) receivers. Due to the ubiquity of mobile phones, the proposed approach can offer a large spatial-temporal sensing coverage at a small cost. Therefore, a noise map based on participatory data collection can be updated within hours or days compared to months or years, making information provided by such a noise map significantly more current than that provided by traditional approaches.

It is non-trivial to build a noise pollution monitoring system based on mobile phones. 
Mobile phones are intended for 
communication, rather than for acoustic signal processing.\footnote{For example, devices such as the Nokia N95, N97 or HP iPAQ do not 
support floating-point arithmetic, which must be emulated with fixed point operations. } To be credible, noise pollution data 
collected on mobile phones should be comparable in accuracy to commercial sound level meters used to measure 
noise pollution. For practical deployability, a detailed analysis must be carried out to find out the life of calibration offset, furthermore, an \emph{in-situ calibration} technique is inevitably required, which can be easily carried out by non-technical users whenever needed.  

Since a people-centric noise monitoring system relies on volunteers contributing noise pollution measurements, these measurements can only come from the place and time where the volunteers are present. Note that volunteers may prioritize the use of the microphone on their mobile phones for conversation. Furthermore, they may choose to collect data only when the phone has sufficient energy. Consequently, samples collected from mobile phones are typically randomly distributed in space and time, and are incomplete. To develop a useful noise pollution monitoring application, the noise map need to be recovered from \emph{random} and \emph{incomplete} samples obtained via crowdsourcing. 

It is also unrealistic to expect that volunteers will always carry the phones in their palm, with the microphones correctly positioned for sampling ambient noise. Research conducted by 
Nokia \cite{nokia_phone} suggests that people tend to carry the mobile phone in their trouser pockets, bags, belt cases and in their palm. In this paper we use the term ``sensing context'' to refer to the phone carrying locations.
Because volunteers may contribute samples when the phone is in different sensing contexts, it is necessary to investigate if the sensing context has a significant impact on the noise level recorded by the phone and address it if so. 
In this paper, we address these challenges. Our key contributions are summarized as follows:

\begin{enumerate}
\item We present the design and implementation of an \emph{end-to-end} noise mapping system, which we name 
\emph{Ear-Phone}, to generate the noise map of an area 
using people-centric urban sensing. The data collection method of this new noise mapping system is expected to cost less than the traditional noise monitoring systems.

\item We study the impact of sensing context on the measured noise level. 
We find that when the phone is held in palm, the deviation from ground truth in the noise level is insignificant, but the deviation is
quite significant when the phone is placed in trousers' pocket or in a bag. To enable data collection only in appropriate sensing contexts, we develop an energy-efficient classification algorithm, which detects the sensing context with 84\% accuracy. 
\item We evaluate Ear-Phone with extensive simulations and real-world outdoor experiments. The results based on our datasets
show that (depending on the compressibility of the noise profiles) Ear-Phone can offer an accuracy of up to 3 dB\footnote{A difference of 3 dB is barely perceptible by human ear~\cite{US_DOT}.} while incurring affordable resource consumption.
\end{enumerate}

The rest of the paper is organized as follows. In the next section, Section~\ref{sec:review}, we contrast our work with the existing literature. We then
describe the Ear-Phone architecture in Section~\ref{sec:archi}, followed by the system design in Section~\ref{sec:design}. 
Then, we evaluate Ear-Phone with both outdoor experiments (Section~\ref{sec:implementation}) and
extensive simulations (Section~\ref{sec:simulation}). Finally, we conclude in Section~\ref{sec:conclude} after a discussion in Section~\ref{sec:discussion}.

\section{Related Work}
\label{sec:review}

We contrast Ear-Phone with the existing literature on the basis of the contributions claimed in the paper. To facilitate the comparison, we discuss the literature in four different areas:  mobile phones for noise mapping in general, noise reconstruction algorithms, context classification algorithms, and finally, microphone (mobile phone's) calibration. 

Work conducted by Santini et al.~\cite{santinis_inss2009_ontheuse}, is one of the very first work, where mobile phones were used for assessing environmental noise. The authors survey 
technical issues influencing the design and implementation of systems that use mobile phones to assess noise pollution, however, their focus is not on developing end-to-end system and reconstructing noise map from incomplete and random samples.
In~\cite{Schweizer:2012} the authors present a noise mapping system, however, the proposed work is  at very preliminary stage as it does not apply A-weighting on the equivalent noise level. Furthermore, we could not find any information about the accuracy of the phone based sound level meter. Noisetube~\cite{noisetube,Maisonneuve:2010} and da sense~\cite{Schweizer:2012} are recently developed systems to generate the noise map by aggregating  
measurements collected from public. However, we could not find any details on the data 
aggregation methods they use in these papers; therefore, we cannot contrast EarPhone with their work.

Recent research in plenacoustic functions \cite{plenacoustic} studies the sampling requirement of an acoustic field. 
While the work in \cite{plenacoustic} deals with a continuous signal, our work considers a discrete signal over time 
and space. Specifically, we consider the equivalent noise level over a physical area and time duration.
Work presented in~\cite{cs_audio} studies the compressibility of acoustic signals in both spatial and temporal dimensions. 
This work is based on a single acoustic source in a laboratory setting, where the authors reconstruct the pressure waveform. This is different from our focus on studying the compressibility of temporal-spatial 
field of noise levels in an outdoor environment, which are influenced by multiple acoustic sources. Reconstruction using $\ell_1$ norm minimization has been used in various application domains, e.g. in~\cite{citeulike:7562715}, methods involving $\ell_1$-norm minimization have been proposed for estimating unknown ``spatial'' fields, such as facies distributions or porosity maps.  Our work is different as we consider to reconstruct the temporal-spatial noise map.
Recently, $\ell_1$-minimization has also been used in some other wireless sensor network applications. For example, \cite{Misra:2012} uses $\ell_1$-minimization for ranging based localization. The authors use $\ell_1$-minimization for classification, whereas we use $\ell_1$-minimization for reconstruction. $\ell_1$-minimization has also been used in~\cite{Wu:2012}. This paper is centred around proposing  a sparsifying basis function where the moisture signal has sparse representation. However, we show that environmental noise is sufficiently sparse in standard DCT basis and focus on reconstruction. 

Phone sensors have been used to detect the context of the person~\cite{cencme}, however, the contexts used in the paper are different and defined in terms of determining the user activity such as running, dancing, etc.
In contrast, our work defines context as the locations people tend to carry mobile phone e.g. hand, bag, or pocket. In~\cite{miluzzo:pocket}, similar to our work, authors present some preliminary results on classification of phone sensing contexts. The authors use audio modalities to detect in or out of pocket context, whereas we use the accelerometer and proximity sensor for context classification. Recently, authors in~\cite{wiese2013phoneprioception} used special hardware capacitive sensing grid to improve the accuracy of context detection. However, we wanted to use the existing sensors on the phone for automatic context detection. In addition, we have taken an analytical approach to determine the features for different sensing modalities, therefore, we have achieved slightly higher accuracy compared to the proposed method. For example, fusing the accelerometer and the proximity sensor data we achieve 84\% accuracy, whereas the authors achieve 83\% accuracy for similar sensor combination. 


Hondt et al. in~\cite{d2011participatory} have put numerous efforts on calibration by calibrating each phone independently in an anechoic chamber.  As we discuss later in the paper, the calibration offset of each phone is different. Our goal is to study this calibration offset and introduce in-situ calibration which can be operated without requiring special skills on the user's part.

\section{Ear-Phone Architecture}
\label{sec:archi}
In this section, we provide an overview of Ear-Phone. A detailed description 
of the system components is presented in Section \ref{sec:design}. 
\begin{figure}[thp]
\centering
\includegraphics[width=1\columnwidth]{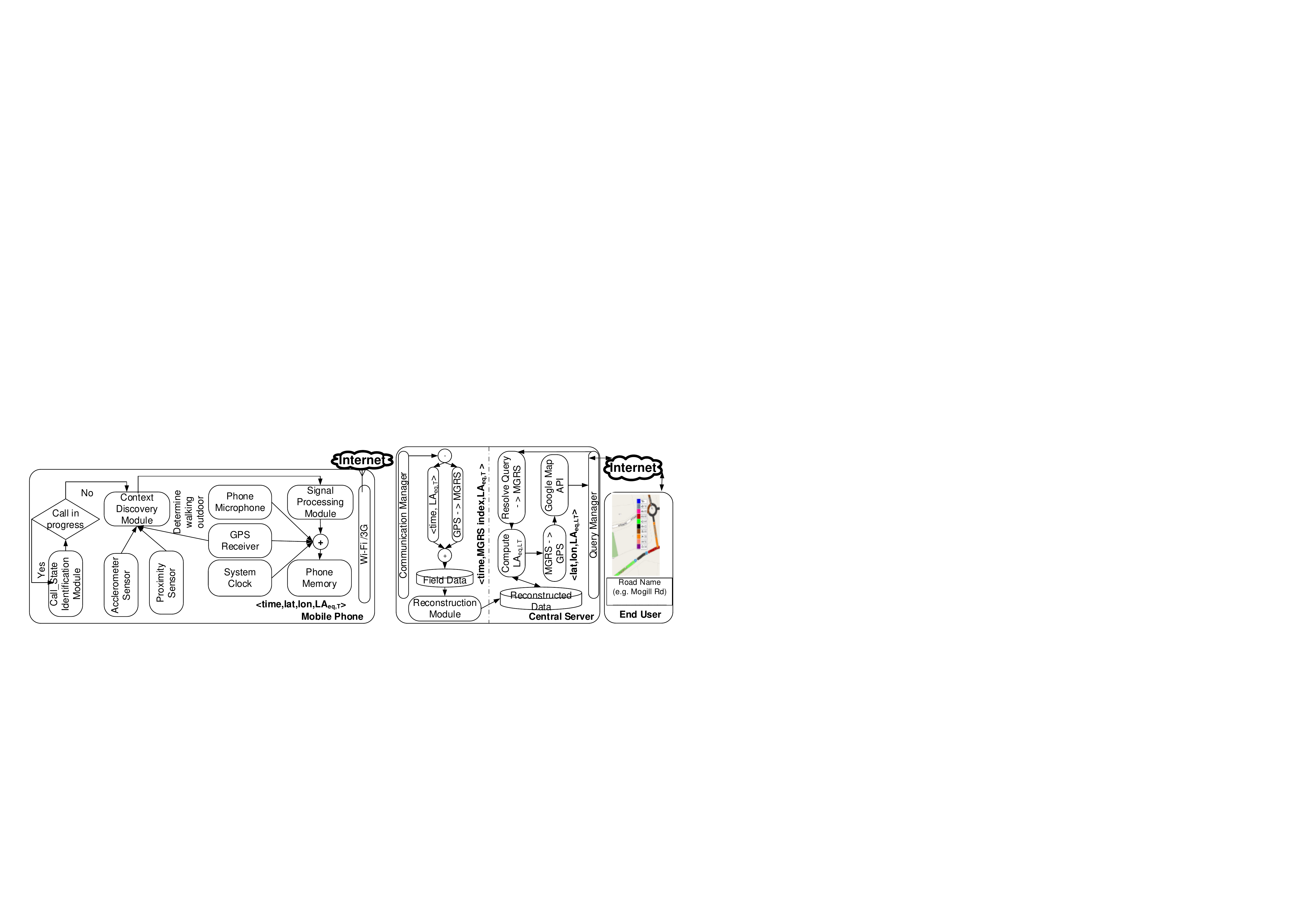}

\caption{Ear-Phone Architecture.}
\label{fig:archi}
\end{figure}

The overall Ear-Phone architecture, depicted in Fig.~\ref{fig:archi} 
consists of a mobile phone component and a central server component. 
Noise levels are assessed on the mobile phones and then transmitted to the central server. 
The central server reconstructs the noise map from the available noise measurements. Note that 
reconstruction is required because the urban sensing framework cannot guarantee that noise measurements 
are available at all times and locations. 

Let us begin with a user who is carrying a mobile phone. 
We call the Ear-Phone application on the mobile phone a \mobile\hspace{-0.1cm}, where SLM stands for ``sound level meter'' which is the instrument used by acoustic engineers to measure environmental noise level. When the mobile phone is not used for conversation, the \mobile is turned on. This is implemented through a call detection module (see Section~\ref{sec:design}).
When turned on, the context discovery module first discovers the sensing context (such as hand, pocket or bag) and decides 
whether to trigger the speed detection module. Note that we record noise level only for certain sensing contexts 
because only certain contexts provide accurate noise measurements (see Section \ref{sec:measurement_context} for details). The context discovery module does not run continuously, instead it runs at a regular interval. In Section~\ref{sec:ContextClassificationPerformance}  we determine an optimal execution interval for the context classifier.   

Upon activation, the context discovery module first uses the GPS signal to determine whether the user is outdoor. 
If the user is outdoor, the context discovery module invokes the signal processing module to start recording noise levels. If there is a change in the context or speed and location, the recording decision is recalculated. 

When triggered, the signal processing module starts computing a loudness characteristic known as 
the equivalent noise level (\laeqt{T}) over a time interval $T$ from the raw acoustic samples collected 
by the microphone over the corresponding time 
interval. The computed noise level is further tagged with the GPS coordinates (Latitude (\emph{lat}) and Longitude (\emph{lon}) ) and system time before being stored in the phone memory.
The stored records \emph{$\langle$ time-stamp, lat, lon, \laeqt{T}}$\rangle$ are uploaded to the central server when the mobile phone detects an open WiFi access point. 3G services on mobile phones can also be used to upload data.

The communication manager at the central server waits for user transmissions. When it receives user data, 
it converts the GPS coordinates of a record to a Military Grid Reference System (MGRS, see 
Section \ref{subsec:mgrs} for the detailed description) grid index 
and stores the information
\emph{$\langle$ time, grid index, \laeqt{T}} $\rangle$ in a data repository. Reconstruction is performed at (predefined) periodic intervals\footnote{Note that in this paper we primarily focus on the accuracy of the noise map obtained from participatory sensing. Determination of a suitable update interval is 
left for future work.}. When triggered, the reconstruction module is invoked to reconstruct the missing data. The reconstructed data is then stored in the data repository.

A query from an end user (e.g. what is the noise level on Oxford Street at 5pm on 28 October 2009?) is processed by a query manager at the central server. The location information (e.g. Oxford Street) of the query is first resolved into grid indices and the reconstructed data associated with those grid indices is fetched from the data repository. Then, the grid indices 
are converted back to GPS coordinates and the corresponding noise levels are overlaid on a web-based map before being displayed to the end user.

\section{System Components}
\label{sec:design}
In this section, we describe the major components of Ear-Phone in detail. 
\subsection{Call Detection Module}
This module detects whether there is an active call in progress. We use Python S60 to query the Symbian phone system via a module called {\bf telephone}. 
This module returns the state of the phone using the function
{\bf stateInformation}. The Ear-Phone module is invoked when the phone is in idle state, which happens when the following expression is TRUE

\begin{eqnarray}
\mbox{\textit{stateInformation == telephone.EStatusIdle}}.\nonumber
\end{eqnarray}
If the phone is in the idle state, context classifier (see Section \ref{subsec:class_1}) is invoked by the call detection module, otherwise it checks the phone status at a regular interval until the phone becomes idle. 
              
\subsection{Signal Processing Module}
The aim of the signal processing module is to 
measure the environmental noise as an A-weighted equivalent continuous sound level or \emph{\laeqt{T}}. 
The International standard ``IEC 61672:2003'' defines 5 different frequency weightings (called A, B, C, D and Z) for measuring sound pressure level. Among these, A-weighting is now commonly used for the measurement of environmental noise and industrial noise, as well as when assessing potential hearing damage and other noise health effects at all sound levels~\cite{a_weight}.

Measured in A-weighted decibels, \emph{\laeqt{T}} captures the A-weighted sound pressure level of a constant noise source over the time interval $T$ that has the same acoustic energy as the actual varying sound pressure level over the
same interval. The A-weighted equivalent sound level \emph{\laeqt{T}} in time interval $T$ is given by
\begin{eqnarray}
\laeqt{T} & = & 10 \log_{10} ( \underbrace{ \frac{1}{T} \int_{0}^T (v_A(t))^2 dt}_{\bar{v}_A(T)} )+\Delta
\label{eq:laeqv}
\end{eqnarray}
where $v_A(t)$ is the result of passing the induced voltage $v(t)$ in the microphone through an A-weighting filter and $\Delta$ is a constant offset determined by calibrating the microphone against a standard sound level meter. Since the sound level meter supports several sampling rates and observation intervals, average signal power (weighted average) is used to compute \emph{\laeqt{T}}.

To compute $v_A(t)$, we design a tenth-order digital filter (coefficients of the filter are given in Table~\ref{tab:filter_coeff} in the Appendix.) whose frequency response matches with that of A weighting over the range 0--8 kHz. This range is 
chosen because the acoustic standard, IEC651 Type 2 SLM~\cite{iec}, requires measurement of environmental noise between 0 and 8 kHz. For details on the algorithm to compute \emph{\laeqt{T}}, please refer to \cite{ear_phone_ipsn}. In order to safeguard user privacy, after the signal processing module computes the equivalent noise level, the audio streams are immediately deleted from the phone.

\subsection{Speech Detection module}
\label{sec:Speech_Detection_module}
\begin{figure}[ht]
\centering
\subfigure[Traffic noise with speech]{
\includegraphics[width=0.45\linewidth]{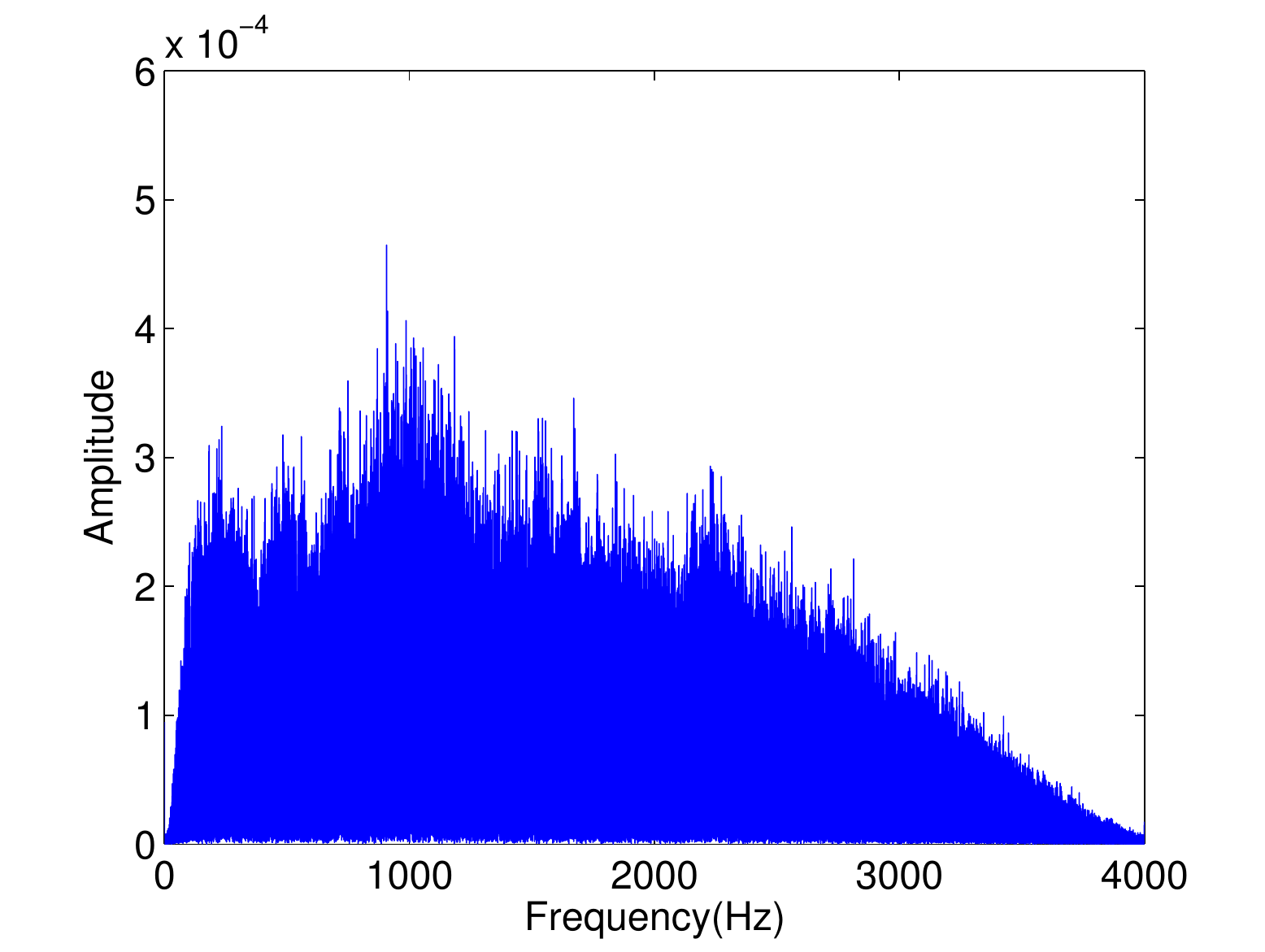}
\label{fig:energyVoice}
}
\subfigure[Traffic noice]{
\includegraphics[width=0.45\linewidth]{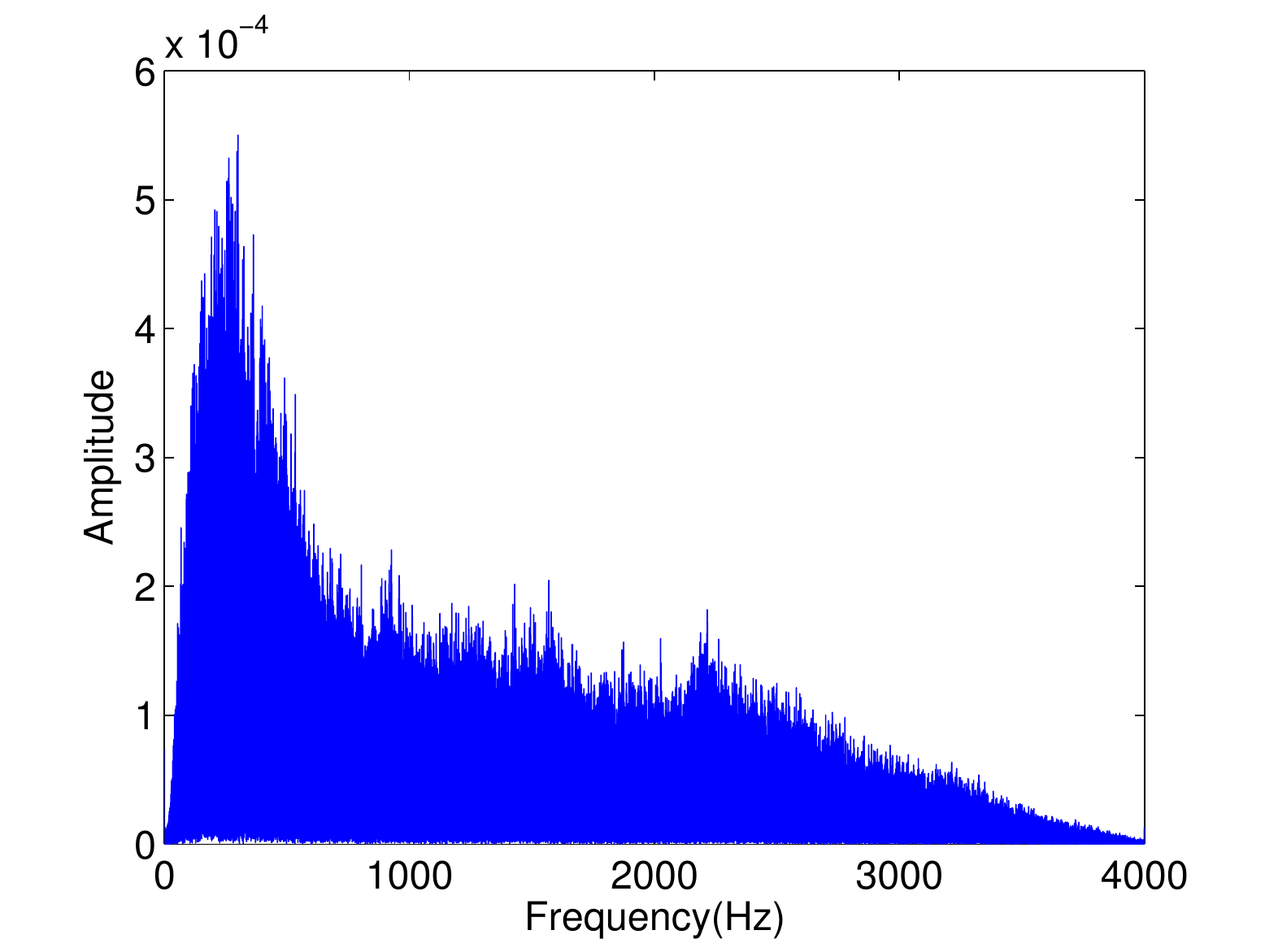}
\label{fig:energyRoadsideNoise}
}
\subfigure[Median Amplitudes. M stands for Male and F stands for female.]{
\includegraphics[width=0.75\linewidth]{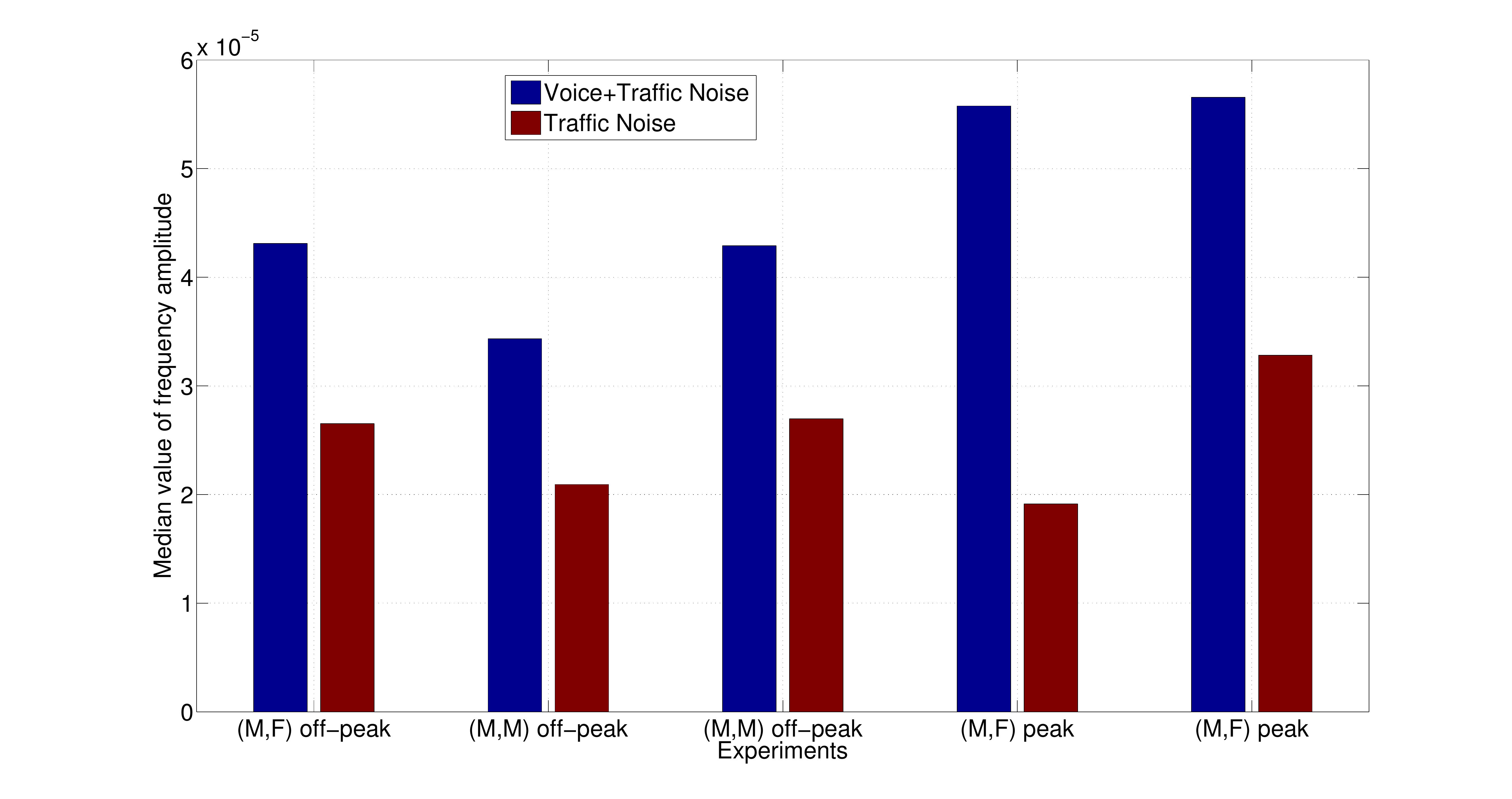}
\label{fig:medianSound}
}
\caption{Frequency spectrum of voice$+$noise and noise alone.}
\label{fig:signalEnergy}
\end{figure}
Since Ear-Phone will be used while the mobile phone users go about their own business, the microphone on the mobile phone unit may pick up ambient noise, which Ear-Phone wants to measure, but also conversations that are taking place around the phone. However, conversations should not be considered as an ambient noise, especially along the roadside. It will only increase the noise level locally but will not contribute significantly to the ambient noise. Ear-Phone includes a speech detection module to detect whether the sound it picks up is speech, and if so, the samples are discarded. 

The speech detection module makes use of spectral characteristics to classify speech from noise. Figure \ref{fig:energyVoice} and \ref{fig:energyRoadsideNoise} show the frequency spectrum of, respectively, traffic noise with simultaneous conversation and traffic noise alone. These plots are generated from Ear-Phone measurements using two phones: one phone was with a couple engaged in conversation along roadside, the other phone was sitting on a bus stop bench, 5 meters away from the first phone. It can be seen that the combined voice and traffic spectrogram is quite different from traffic noise spectrum. We tested three different features: mean, median and variance of the amplitude of the spectrogram within 0-4kHz and found that median gave the best results to detect whether a conversation is present. Figure \ref{fig:medianSound} shows the median amplitude in 0-4kHz for both voice with traffic noise and traffic noise alone for 5 different experiments. 

\subsection{Context Discovery Module}
The usability of the sound data collected from the mobile phone depends on the phone sensing context, that is where the phone is being carried. An extensive survey~\cite{nokia_phone} conducted by Nokia which queried people from over 11 different countries reports that the trouser's pocket, bag and hand are three common places to carry mobile phone. This study also identifies belt-case as a popular choice for carrying mobile phone. However, due to the increased size (width and height) of the latest smart phones, it is impractical to carry the phone in belt-case. In fact, belt-cases are not commercially available for many latest smart phones.
%
%
We therefore consider hand, pocket and bag as three possible sensing contexts. 

Later in Section~\ref{sec:Measurement Accuracy}, we report that the best quality data can be recorded when the phone is held in the palm. The equivalent noise level recorded in trousers' pocket and bag has a large deviation from the ground truth and is therefore not useful. We therefore cluster the sensing contexts into two groups - (i) hand (ii) pocket or bag. We seek to develop a context classifier that can distinguish these two groups.

\subsubsection{Feature Selection}
\begin{figure}
\centering
\includegraphics[width = 0.8\linewidth]{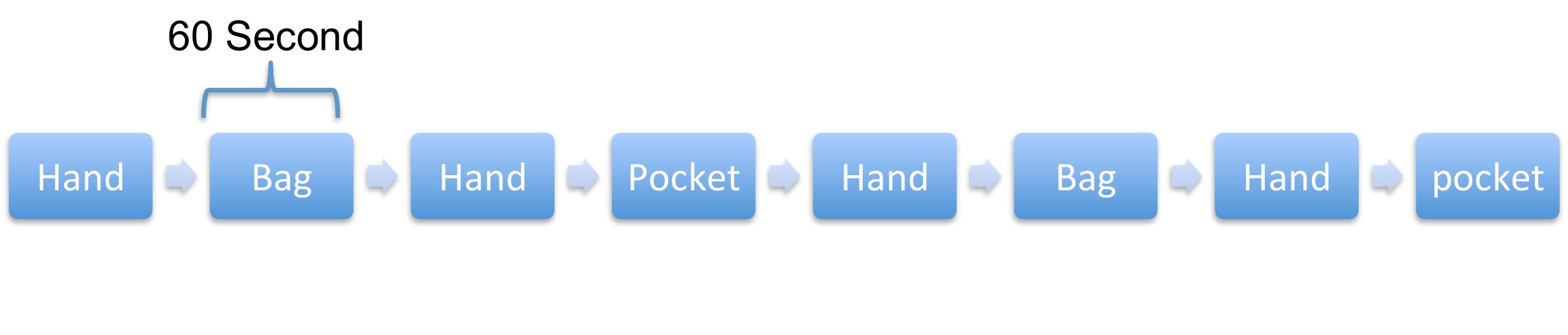}
\caption{Sequence of context switch.}
\label{fig:contextSwitchFlow}
\end{figure}

We investigate the feasibility of using the data from 3-axis accelerometer, the proximity sensor and the light sensor as features for context classification. We recruit 10 subjects to collect readings from these sensors while keeping the phone in different sensing contexts. Each subject was given one Android phone - ``HTC One" for data collection. While walking along the street each subject kept the phone in  various sensing contexts in the order reported in Fig.~\ref{fig:contextSwitchFlow} to carry the phone. 
In each sensing context the phone stayed for 60 seconds. The subjects were given stop watch to control the time. 

\begin{figure}
\centering
\includegraphics[width=1\linewidth]{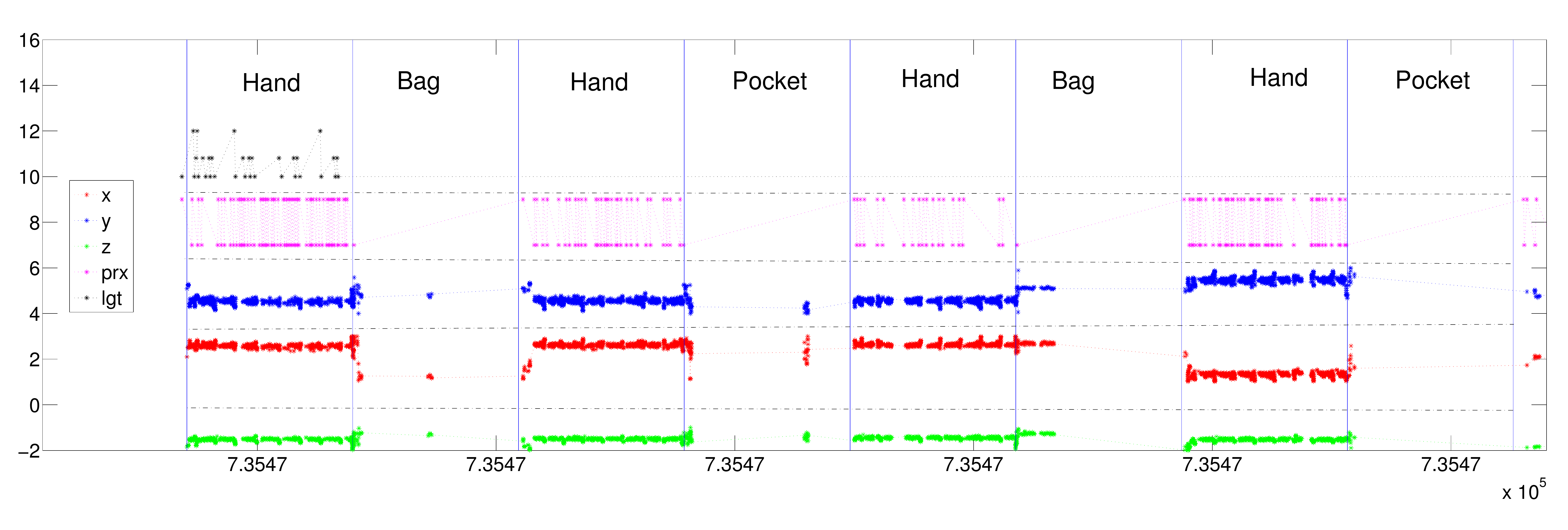}
\caption{Plot of various sensor data normalised in non-overlapping bands ($y$ axis values). $x$,$y$ and $z$ represent the three axis of the accelrometer. ``prx'' and ``lgt'' represent the proximity and light sensor respectively. }
\label{fig:contextSwitch}
\end{figure}

We normalize various sensor readings in non-overlapping bands to represent them in one figure (Fig~\ref{fig:contextSwitch}) without much cluttering. This plot is representative of all 10 subjects. Based on the response of various sensors, we decide to exclude the light sensor as it mostly remains unchanged, except some initial variation at the start of the experiment. We nominate the accelerometer and the proximity sensor readings for context classification.

{\bf Accelerometer Feature }In order to use accelerometer data for context classification, we first investigate the distribution of various axis readings in various contexts. For various sensing contexts, we plot the histogram of various axes and then find the best fit for the histogram. For choose to fit Normal distribution because of its generality. We observe that the z-axis data fits normal distribution for various contexts. We also observe that the mean  for various contexts is different, therefore, mean of ``z-axis'' is our primary accelerometer feature. However, we also study other features, such as, various percentiles and median for classification.

\begin{figure}[tbp]
\centering
\subfigure[X-axis]{
\includegraphics[width=0.45\columnwidth]{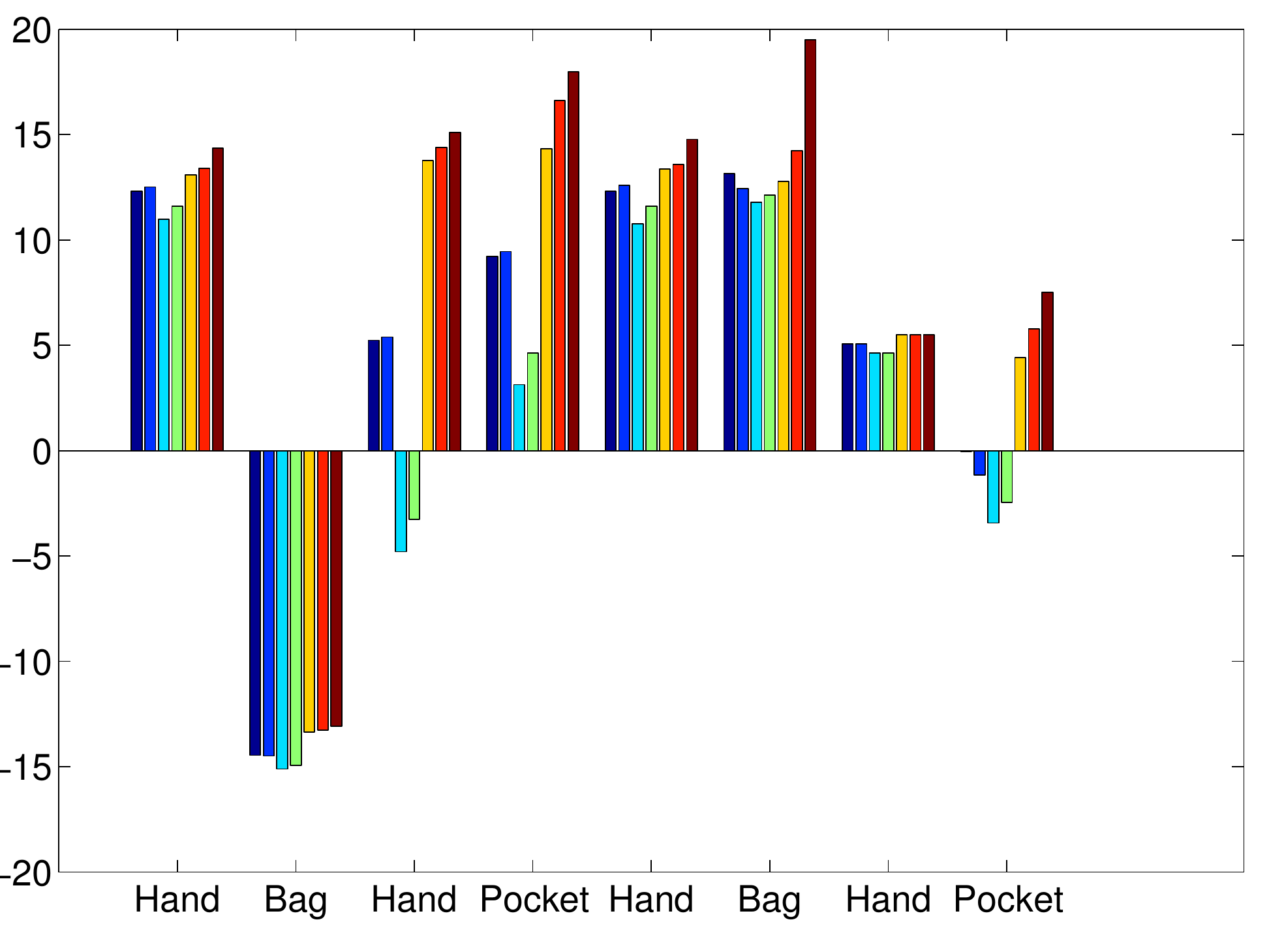}
\label{fig:xAxisStat}
}
\subfigure[Y-axis]{
\includegraphics[width=0.45\columnwidth]{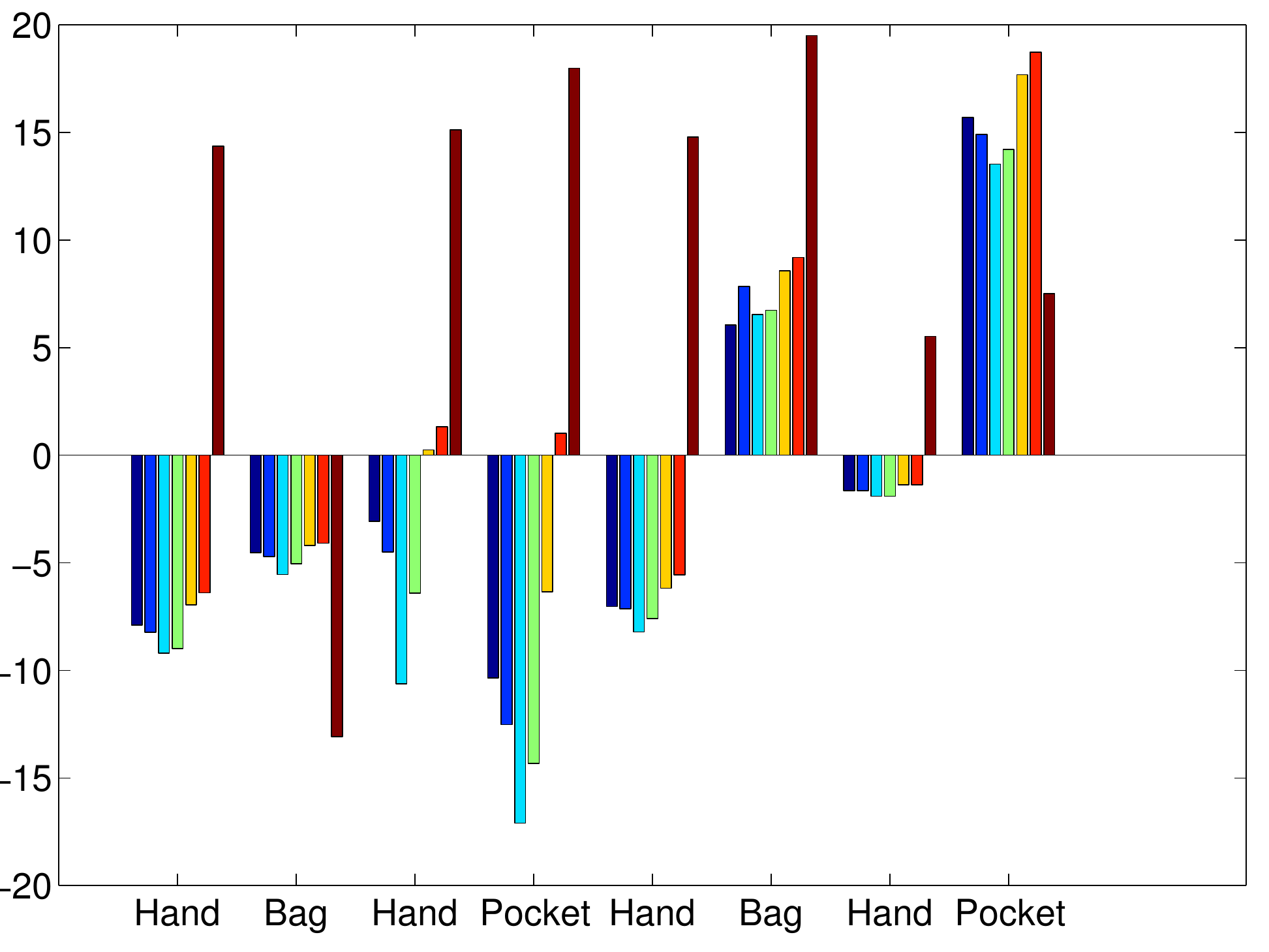}
\label{fig:yAxisStat}
}
\subfigure[Z-axis]{
\includegraphics[width=0.7\linewidth]{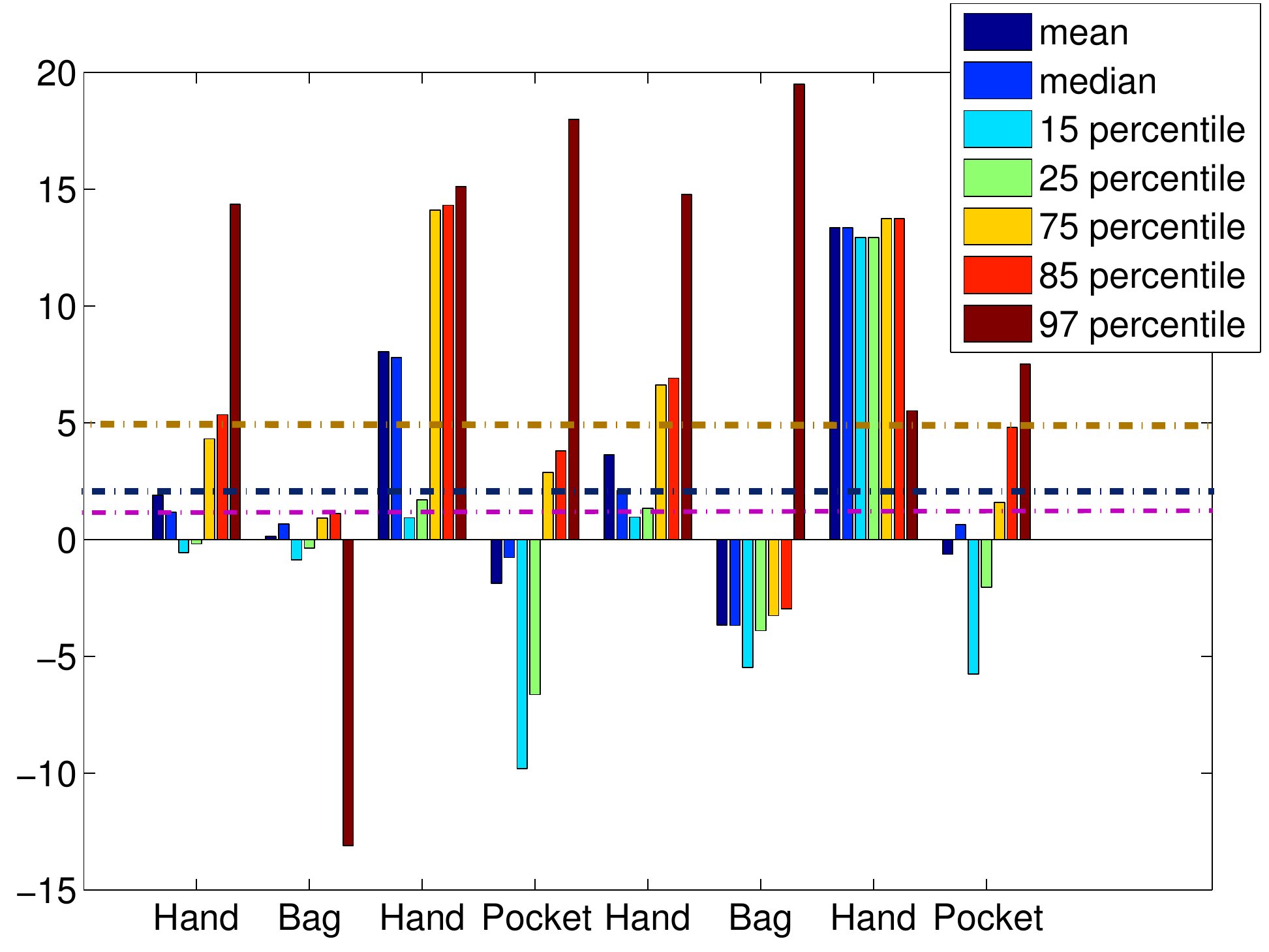}
\label{fig:zAxisStat}
}
\caption{Acceleromter feature for context classification. 
}
\label{fig:contextStat}
\end{figure}

In Fig.~\ref{fig:contextStat}, for various axes and for various sensing contexts we plot the values of various accelerometer features. The idea is to identify feature(s) that shows clear distinction amongst groups. We find that features from $x$ and $y$ axis data do not reveal any clear distinctions. However, features from $z$ axis data show some clear distinctions. For example, we draw a blue dotted line which suggests a threshold for hand sensing context. Precisely, ``mean'' value of z-axis readings above this threshold indicate the hand sensing context and mean value below this threshold indicate bag or pocket sensing context. Similarly, the orange and magenta dotted lines suggest a threshold for hand sensing context while using 75 percentile and median, respectively. 

We further plot a scatter including all the 10 participants data for various accelerometer features in Fig~\ref{fig:xAxisStat} to Fig.~\ref{fig:zAxisStat}. For ``mean", the data points from hand  sensing context and pocket or bag sensing contexts are well separable, however, for median and 75 percentile, the data points from the three sensing contexts are overlapped. 

\begin{figure}
\centering
\includegraphics[width = 1\linewidth]{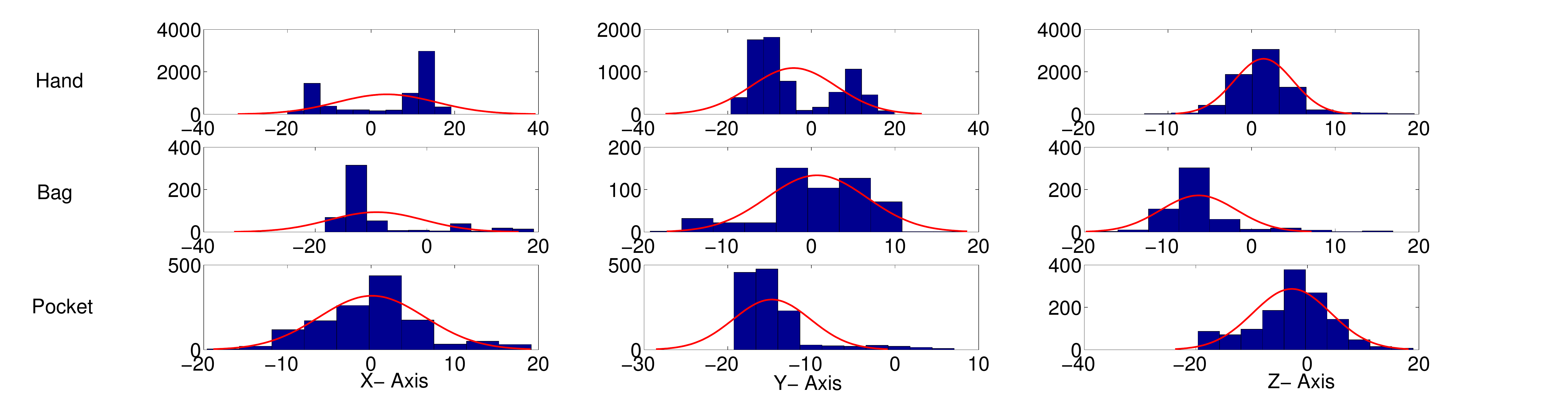}
\caption{Distribution of various axis data in various sensing contexts.}
\label{fig:normalFitted}
\end{figure}

%


\begin{figure}[ht]
\centering
\subfigure[mean]{
\includegraphics[width=0.45\columnwidth]{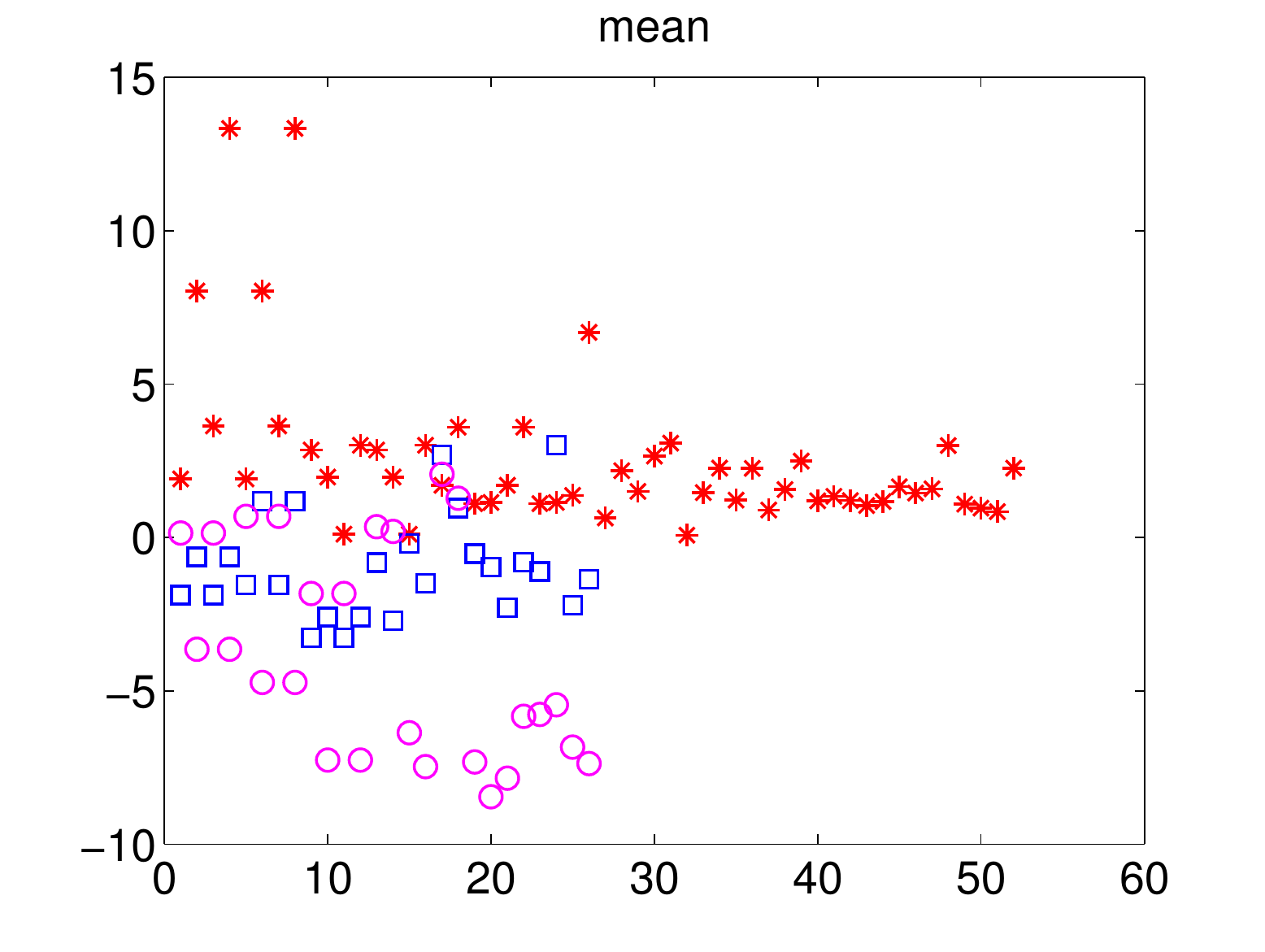}
\label{fig:xAxisStat}
}
\subfigure[median]{
\includegraphics[width=0.45\columnwidth]{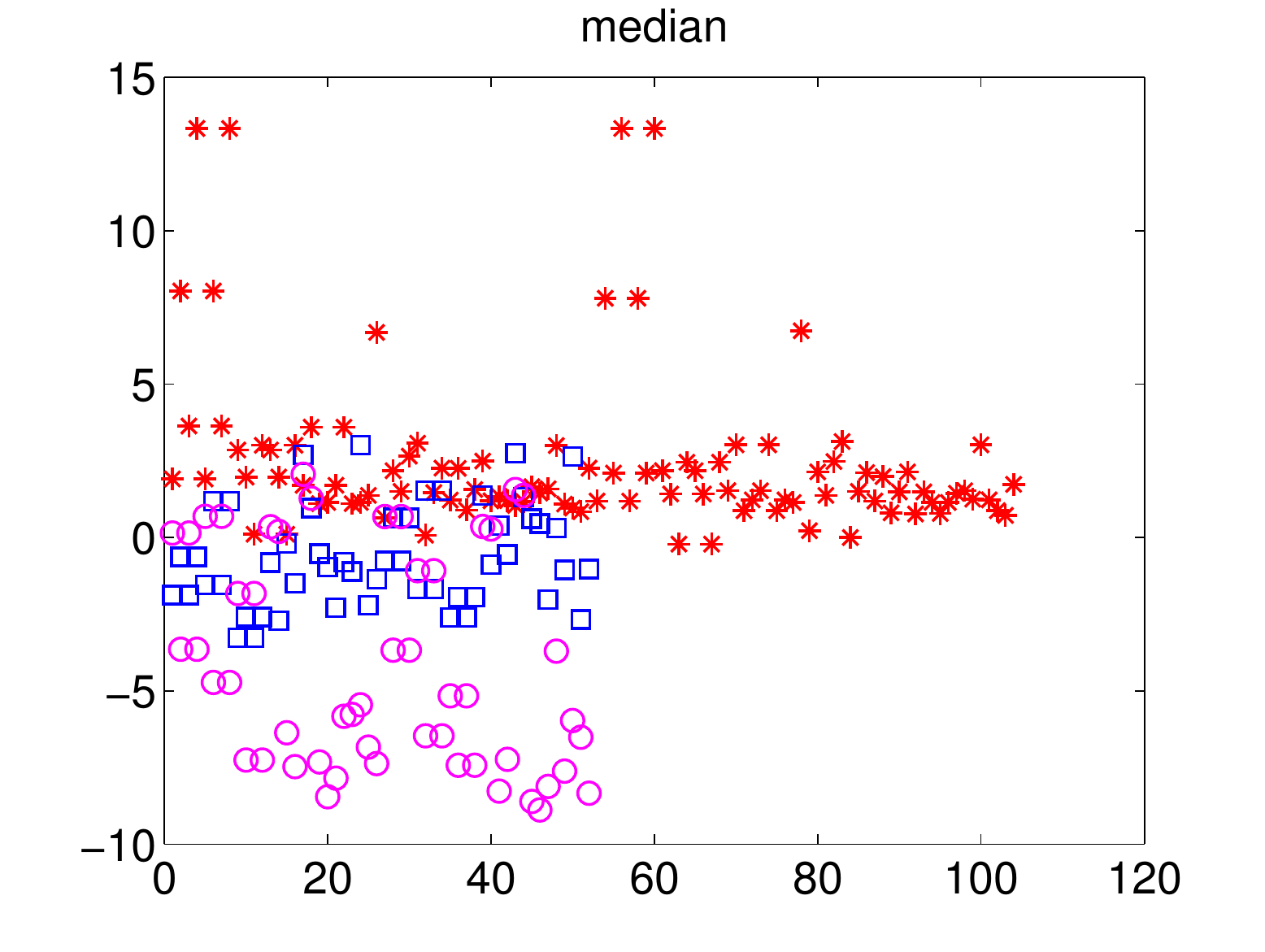}
\label{fig:yAxisStat}
}
\subfigure[75 percentile]{
\includegraphics[width=0.45\linewidth]{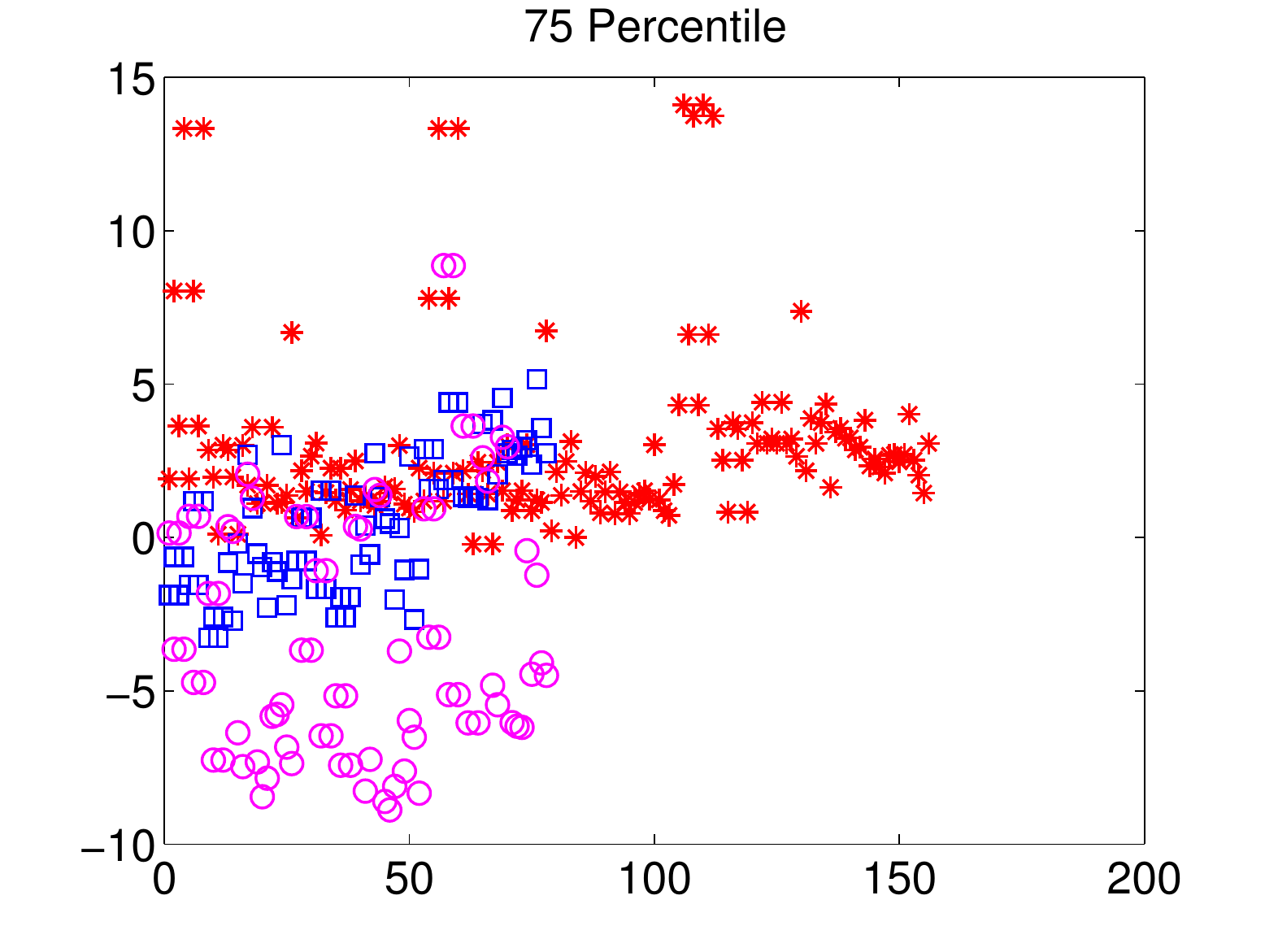}
\label{fig:zAxisStat}
}
\subfigure[Proximity]{
\includegraphics[width=0.45\linewidth]{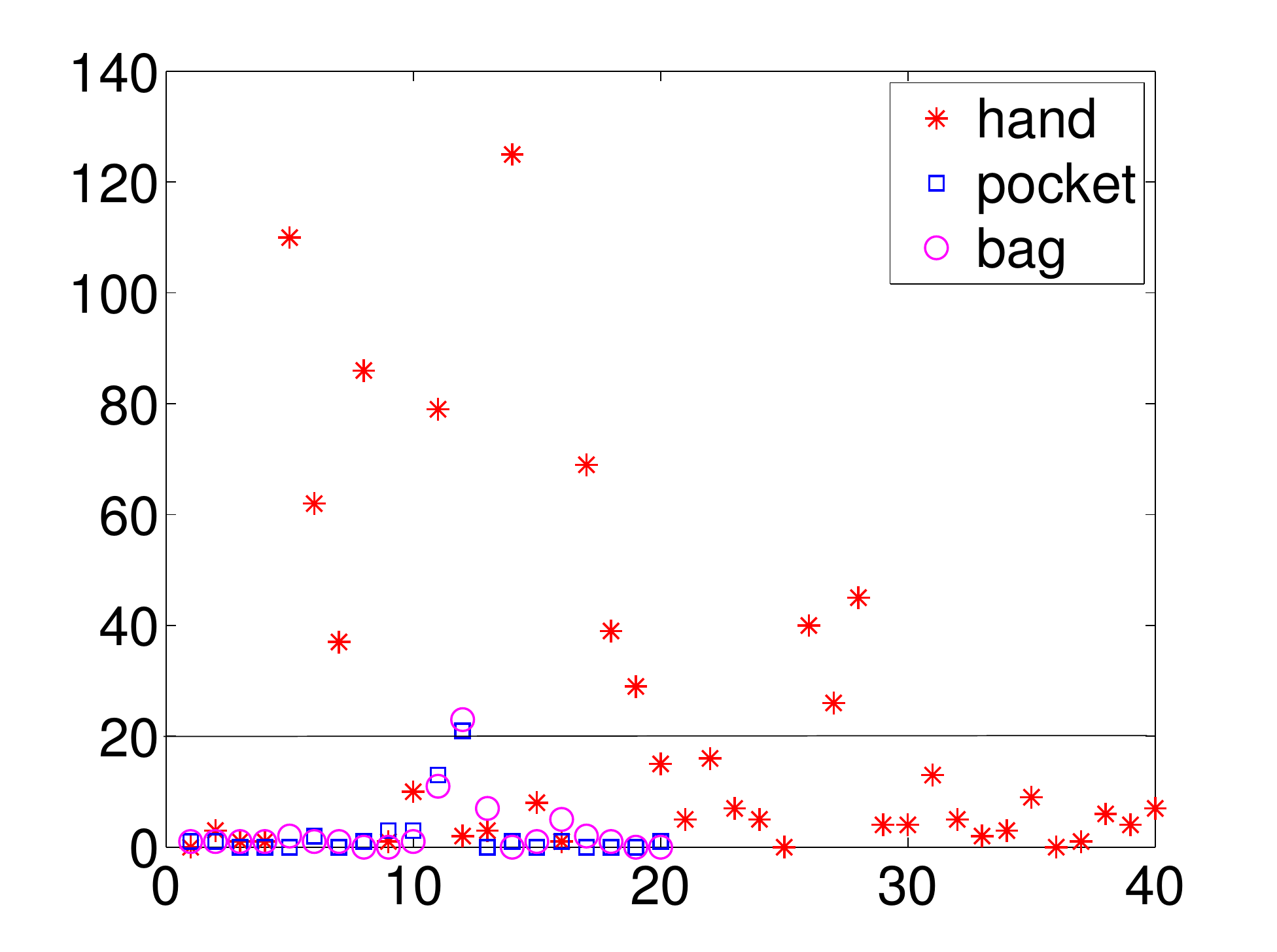}
\label{fig:ProximityClassifier}
}
\caption{Context switching feature for accelerometer and proximity sensor across all 10 subjects.}
\label{fig:contextAllsubjects}
\end{figure}	

{\bf Proximity Sensor Feature } In order to extract a feature from the proximity sensor data, we refer back to Fig.~\ref{fig:contextSwitch}. The number of times the proximity sensor is being triggered is a distinctive feature among hand and pocket or bag. We plot the number of times the proximity sensor is triggered in various sensing contexts for all ten subjects in Fig.~\ref{fig:ProximityClassifier}. We observe a good separation between hand sensing context and pocket or bag sensing context. 

In Section~\ref{sec:ContextClassificationPerformance}, we report the classification performance of proximity is poor while we use it to classify various sensing contexts. However, further investigation reveals that applying a threshold on the number of firings we can improve the classification performance. In addition, we also observe that the proximity sensor is particularly suitable to determine the hand sensing context, not the bag and pocket sensing contexts. We therefore, apply a threshold on the proximity sensor data and apply it to only determine hand sensing context.

{\bf Sensor Fusion}
We also investigate the feasibility of fusing the classification results form accelerometer sensor and the proximity sensor. In the fusion mechanism, we override the classification decision taken from the accelerometer sensor data with the classification decision taken by the proximity sensor data, if the context is classified (by proximity sensor) as hand sensing context. However, for rest of the sensing contexts the proximity sensor data is not accounted for classification.

\subsubsection{Context Classification}
We use k-nearest neighbour (kNN) algorithm for context classification. We choose kNN as it is light weight compared to support vector classification, yet produce good classification accuracy. We use leave-one-out cross fold validation for calculating the classification accuracy.

\subsubsection{ Automatic Context Switch Detection}
\label{subsec:class_1}
The context classifier runs in the background on the phone. We find that it is wasteful to run the classifier every second. We conduct experiments to determine an interval ($\Delta$) for executing the context classifier. We describe these experiments in Section~\ref{subsec:classification_performance}. The context switch detection process is presented in Algorithm~\ref{alg:contextSwitch}. 

\begin{algorithm}[H]
 \SetAlgoLined
 \KwData{Acceleromter and proximity sensor feed}
 \KwResult{Senisng context }

 \While{The phone is identified in outdoor}{
 - acquire $\Delta$ seconds of accelerometer z-axis and proximity sensor reading and compute mean and number of prolixity sensor trggers\;
- use kNN to classify the sensing context using accelerometer data and use threshold on the proximity sensor data to determine hand sensing context\;
  \eIf{proximity sensor determines ``in hand''}{
   context = ``in hand''\;
      }{
  context = context  determined by the accelerometer\;
  }
 }
 \caption{Automatic Context Switch Detection}
\label{alg:contextSwitch}
\end{algorithm}

\subsection{GPS, MGRS conversions}
\label{subsec:mgrs}
The reasons for approximating GPS with square areas are two fold. First, computing the \emph{\laeqt{T}} for every possible GPS coordinate is impractical because there are infinite GPS coordinates. Secondly, the acoustic standards for monitoring noise pollution recommend measuring the pollution in square areas (Section 5.3.1(a) in \cite{iec3}) assuming that the noise level is constant over that area. We followed the Australian acoustic standard to determine an appropriate size of the measuring square. We assume that the volunteers walk along the pavement (or sidewalk) and measure ambient noise on the street level, which is the aggregate of the noise generated by multiple moving vehicles. The Australian acoustic standard restricts 
the noise level difference between two adjacent grids to be no more than $5$ dB (Section 5.3.2 in \cite{iec3}). Therefore, we conducted a number of experiments where we put a \mobile at a static position and put another \mobile at varying distances from the first \mobile and recorded the difference in \emph{\laeqt{1s}} readings for each distance. For grid sizes of 10 m $\times$ 10 m, 20 m $\times$ 20 m, 30 m $\times$ 30 m, 40 m $\times$ 40 m and 50 m $\times$ 50 m, the corresponding
noise level differences between adjacent grids were found to be $2.26 \pm .06$, $3.82  \pm .05$, $3.86 \pm .03$, $4.11 \pm .02$ and $4.97 \pm .03$ dB, respectively. We can therefore use square grids which are less than or equal to 50 meters in each dimension. In order to minimize the error, we chose to use a grid size of $10$ m $\times$ 10 m.

To approximate GPS with grids, we use the Military Grid Reference System (MGRS)~\cite{mgrs}. MGRS can divide the earth surface into squares of $100$ m $\times$ $100$ m, $10$ m $\times$ $10$ m or $1$ m $\times$ $1$ m. As we used a grid size of 10 m$\times$10 m,  we can simply plug in the MGRS grid reference system without any further modification, might be required if we would have used other reference systems, such as, UTM
(Universal Transverse Mercator), UPS (Universal Polar Stereographic)~\cite{utmAndUps} etc. Equivalent noise level \laeqt{1s}
can be recorded at any point (A) of a MGRS grid. We assume that at any point
in the grid, the value of \laeqt{1s} would be same as of point A.

\begin{figure} 
\centering
\includegraphics[width=.7\columnwidth]{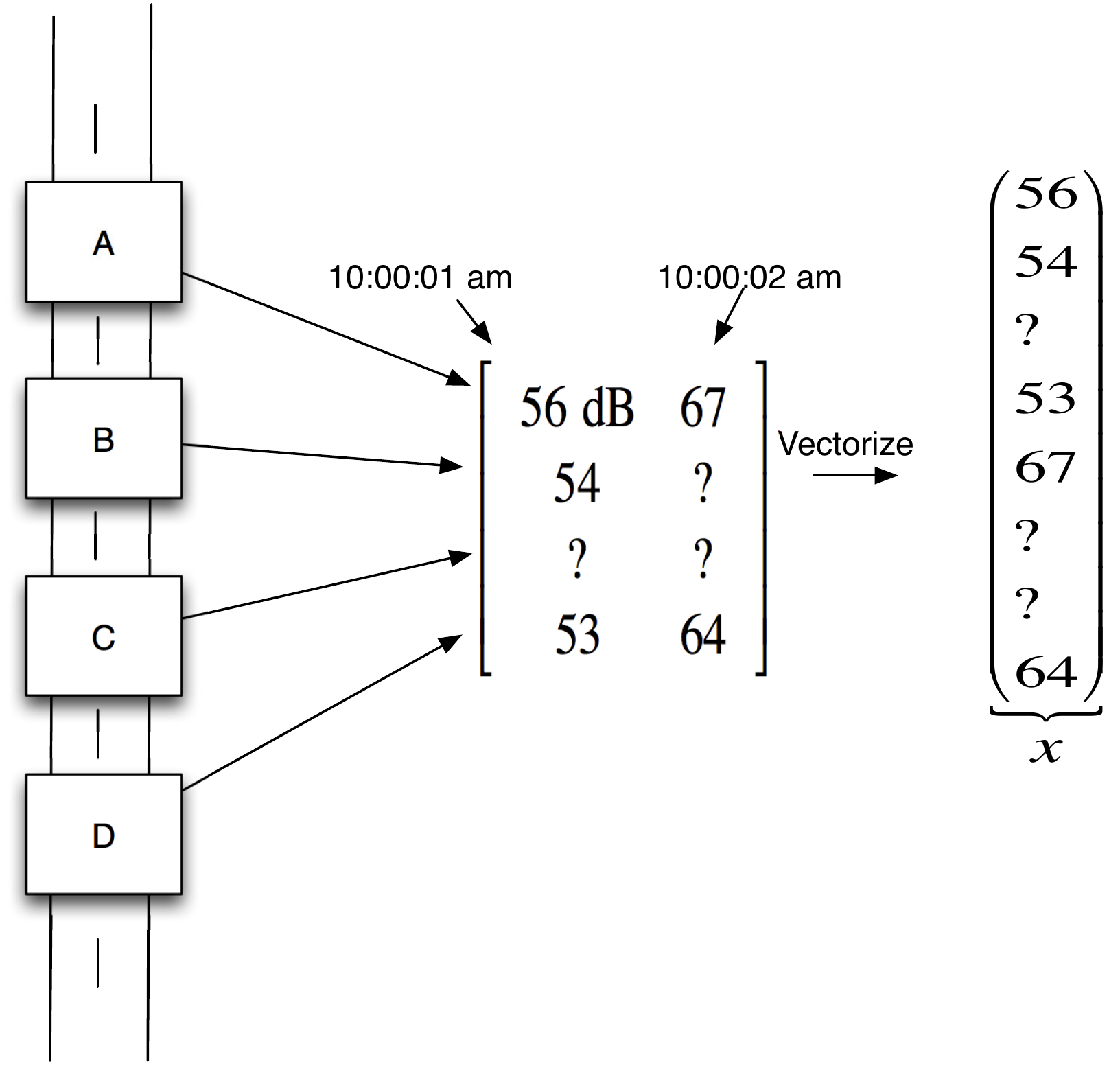}
\caption{Reconstruction Problem.}
\label{fig:raw-data_explained_3}
\end{figure}
\subsection{Signal Reconstruction Module}
\label{subsec:recontruction}
In the participatory noise mapping system, the availability of samples is not strictly guaranteed. The missing samples need to be reconstructed to recover the complete noise map. We will use Fig.~\ref{fig:raw-data_explained_3} as an illustrative example to help with the explanation. In this example, we assume the road segment consists of four MGRS grids of resolution 10m $\times$ 10m, which are labelled by A, B, C and D. We consider a two second  (10:00:01 and 10:00:02)  noise from spanning over these four grids. If there is a \mobile returning noise measurement from each MGRS grid every second, then there is no missing data. However, this is not always possible in participatory noise mapping. In our example, the question marks indicate missing data points. The reconstruction problem refers to recovering the complete noise map from these incomplete samples.

For computation purpose, we use a vector to represent the output of the entire noise map. For our example, we can equivalently represent the noise map by a vector with 8 elements where each element is the noise level in a MGRS grid. The middle and rightmost drawings in Fig.~\ref{fig:raw-data_explained_3} explains how we derive the vector from the noise map. 


In order to recover the missing \laeqt{1s} samples we investigate the feasibility of four different reconstruction algorithms which include basic interpolation techniques such as, linear interpolation and nearest neighbor interpolation and advanced interpolation such as, Gaussian process interpolation (widely known as Kriging) and regularization technique namely $\ell_1$-norm minimization. We study algorithms with different capability so that we can select the appropriate one for our problem. Below we provide a brief description of these algorithms and discuss how we apply them to solve the noise map reconstruction problem.

\subsubsection{Linear Interpolation}
We assume that noise levels from $n$ MGRS grids are available. Let us denote the available data by $(g_i,t_i,x_i)$ where $x_i$ is the noise level in grid $g_i$ at time $t_i$ with $i = 1,...,n$. Since noise level is specific to time and location, we assume that $x_i$ is a multivariate linear function of $g_i$ and $t_i$ as follows:
\begin{eqnarray}
x_i = a g_i + b t_i + c, i = 1,..,n \label{eqn:li}
\end{eqnarray}
and use least-squares to determine the coefficients $a$, $b$ and $c$. Missing measurements are then estimated using \eqref{eqn:li}. For example, if noise measurement in grid $g_{n + 1}$ at time $t_{n + 1}$ is not available, we estimate that by evaluating $\hat{a} g_{n + 1} + \hat{b} t_{n + 1} + \hat{c}$ where $\hat{a}$, $\hat{b}$ and $\hat{c}$ are the least-squares estimates of $a$, $b$ and $c$.


\subsubsection{Nearest Neighbor interpolation}
Let us assume that the noise samples are available at the 
grid time pair $\{(g_1,t_1),(g_2,t_2),…,(g_n,t_n)\}$ and the noise level at the grid time pair $(g_{n+1},t_{n+1})$ is unknown. The Nearest Neighbour Interpolation (NNI) algorithm assigns the noise level of grid time pair $(g_i,t_i)$ to $(g_{n+1},t_{n+1})$, if 
\begin{eqnarray}
i\ast & = & \arg \min_{i = 1,…,n} \rm{dist}((g_i,t_i),(g_{n+1},t_{n+1})), \nonumber
\end{eqnarray} 
where  $ \rm{dist}((g_i,t_i),(g_{n+1},t_{n+1})) = \sqrt{a \Delta g^2 + b \Delta t^2}$.   Here $\Delta g$ is the distance between the centres of grids $g_i$ and $g_{n+1}$ divided by spatial resolution, and $\Delta t = t_i - t_{n+1}$. Similar to linear interpolation, we use least square to estimate the coefficients $a$ and $b$. 

\subsubsection{Gaussian process Interpolation}
Gaussian process Interpolation (GPI)~\cite{Williams97predictionwith,Rasmussen06gaussianprocesses} performs little bit differently from NNI. Let us consider that the noise level at grid time pair $(g_i,t_i)$ is denoted by $x_i$. In order to recover an unknown noise level $x_{i+1}$, GPI performs the following steps.

Let us define e a number of notation first. Let $\bar{x}$ be the column vector $[x_1,x_2,....,x_n]^T$, $\mu_n$ be the mean $\bar{x}$, $\Sigma_{n}$ be an $n$-by-$n$ matrix whose $(i,j)$-element (where $1 \leq i,j \leq n$) is the covariance between $x_i$ and $x_j$, $\mu_{n+1}$ be the mean of $x_{n+1}$ without the knowledge of $x$, $k$ be a $n \times 1$ column vector whose $i$-th element is the covariance between $x_i$ and $x_{n+1}$, and $\sigma_{n+1}$ be the auto-covariance of $x_{n+1}$ without the knowledge of $\bar{x}$. The aim of the Gaussian process interpolation is to use the knowledge of the available data $\bar{x}$ to estimate $x_{n+1}$. Without the knowledge of $\bar{x}$, $x_{n+1}$ is Gaussian distributed with mean $\mu_{n+1}$ and covariance $\sigma_{n+1}$. With the knowledge of $\bar{x}$, $x_{n+1}$ is Gaussian distributed with mean $\mu_{n+1 | x}$ and covariance $\sigma_{{n+1} | x}$, where
\begin{eqnarray}
\mu_{n+1 | x} & = & \mu_{n+1} + k^T \Sigma_n (\bar{x} - \mu_n) \\
\sigma_{{n+1} | x} & = & \sigma_{n+1} - k^T \Sigma_n k 
\label{eq:gis}
\end{eqnarray}
One can readily see that Gaussian process interpolation uses the knowledge of $\bar{x}$ to reduce the variance of the estimate of $x_{n+1}$.

\subsubsection{$\ell_1$-norm minimization}
The $\ell_1$-norm minimization method is useful when the underlying data is sparse or compressible. In order to explain the concept of sparsity and compressibility, we need to introduce the notion of transform basis. As earlier, let us represent the noise level data by a vector $x$ with $N$ elements. A transform basis $\Psi$ can be represented by an orthonormal $N \times N$ matrix whose columns form the basis vectors. Given the transform basis $\Psi$, let $v$ be  an equivalent representation of $x$ in this basis, where $x = \Psi v$. The vector $x$ is said to be sparse (resp. compressible), if its equivalent representation $v$ in $\Psi$ has a small number non-zero elements. In other words, let $S$ be the number of non-zero elements in $v$, then the vector $x$ is sparse (compressible) if $\Psi$ if $S \ll N$. 

Basis selection is important for successful $\ell_1$ recovery, since a good basis offers better compressibility and thus better recovery. In Section~\ref{sec:basis_selection}, we use noise maps collected from real-life experiments to show that they have compressible representations in the Discrete Cosine Transform (DCT) basis. Hence we use DCT as a transform. 

For the $\ell_1$-norm minimization method, we also represent the noise map as a vector $x$. We assume that $M$ ($< N$) elements of the vector $x$ are known. We denote the indices of these known elements by $i_1$, $i_2$, ..., $i_M$. Let $\bar{x}$ be the $M\times 1$-vector that contains the $i_1$-th, $i_2$-th, ..., $i_M$-th elements of $x$. Let also $\bar{\Psi}$ be the $M\times N$ matrix consisting of the $i_1$-th, $i_2$-th, ..., $i_M$-th rows of $\Psi$. Let $\tilde{v}$ denote an unknown vector with $N$ elements. If the vector $\tilde{v}$ is chosen such that $\bar{x} = \bar{\Psi} \tilde{v}$, then we say that this $\tilde{v}$ is consistent with the available data in $x$. This is because, for any such $\tilde{v}$, the $i_1$-th, $i_2$-th, ..., $i_M$-th elements of $\Psi \tilde{v}$ are equal to the $i_1$-th, $i_2$-th, ..., $i_M$-th elements of $x$. Since $M < N$, there are many choices of $\tilde{v}$ that are consistent with the data in $x$ and our goal is to choose the sparest $\tilde{v}$. Since the $\ell_0$ norm of a vector returns the number of non-zero element in that vector, we can express our problem as an optimization problem:
\begin{eqnarray}
\arg \min_{\tilde{v}} ||\tilde{v}||_0, \mbox{s.t. } \bar{x} = \bar{\Psi} \tilde{v}\label{eqn:l0norm}
\end{eqnarray}
Unfortunately, the problem \eqref{eqn:l0norm} is NP hard. A computationally viable solution which approximates the solution given by $\ell_0$-norm, uses the $\ell_1$-norm minimization~\cite{donoho} and is expressed as: 
\begin{eqnarray}
\arg \min_{\tilde{v}} ||\tilde{v}||_1, \mbox{s.t. } \bar{x} = \bar{\Psi} \tilde{v}\label{eqn:l1norm}
\end{eqnarray}
Whether the recovery of a signal $x$ from $\bar{x}$ succeeds depends on the choice of $M$. Typically, the signal is reconstructed asymptotically when $M > O(S.\log^4{N/M})$. 

\section{Implementation and Evaluation}
\label{sec:implementation}
In this section, we first describe the implementation details of various Ear-Phone component.  Then, we evaluate the system performance in terms of measurement accuracy( noise-level), classification performance, resource (CPU, RAM and energy) usage and noise-map generation. 

\subsection{System Implementation}
We have implemented the mobile phone system components on five hardware platforms - Nokia N95, Nokia N96,  Nokia N97, HTC One and Google Nexus 4S. 
We chose Java for the Nokia phones and android for the others. Note that Android supports most of the Java libraries, however, there are a number of differences between these two programming languages. Such as, Java code compiles to Java bytecode, while Android code compiles in to Davilk opcode etc. We implement the context classification module on the Anrdoid phones. Various mobile components are implemented as separate application threads (e.g. signal processing thread) in Java. 

The server component consists of a 
MySQL database and PHP server-side scripting. We used the MySQL database to store both the collected noise level data
and the reconstructed noise level data. We used a PHP script to implement the server-side modules such
as the communication manager, the GPS to MGRS converter and the query manager. The signal reconstruction module was implemented using Matlab.
\subsection{Calibration and Measurement Accuracy}
\label{sec:Measurement Accuracy}

Recall from Section~\ref{sec:design} (Eq.(\ref{eq:laeqv})) that calibration offset $\Delta$ needs to be estimated to calculate the correct \emph{\laeqt{T}}. Comparing the sound level responses of different phone microphones with that of a commercial  sound level meter, Center-322 SLM~\cite{center}, we observe that even for the same phone model, the calibration offset varies from phone to phone. For example, as shown in Fig.~\ref{fig:offsetChange}, the calibration offset for phone 2, 3 and 4 are different (just compare day 1), although they are all Nokia N97. We therefore, introduce and implement an {\bf in-situ or on-phone calibration} system. In-situ calibration is used to determine a calibration offset specific to each phone.


We construct a calibration tone using freely available Audacity~\cite{audacity} software. The calibration tone is made up of a series of 1 kHz sinusodial tones\footnote{We use a pure tone for calibration, since standard calibrattion process uses one single frequency, namely 1 kHz.} (spanning the complete amplitude scale:  0.20 to  1) and silence. Note that we change amplitude and add silence to obtain a robust offset. The total length of the calibration tone is five minutes, consisting of five segments of one minute duration. The first segment is comprised of 30 seconds of Sine tone with amplitude 0.20 followed by 30 seconds of silence. The amplitude of sine tone increases by 0.2 units in consecutive segments. 

\begin{figure}[ht]
\centering
\resizebox{!}{4 cm}{
\includegraphics[width=0.6\columnwidth]{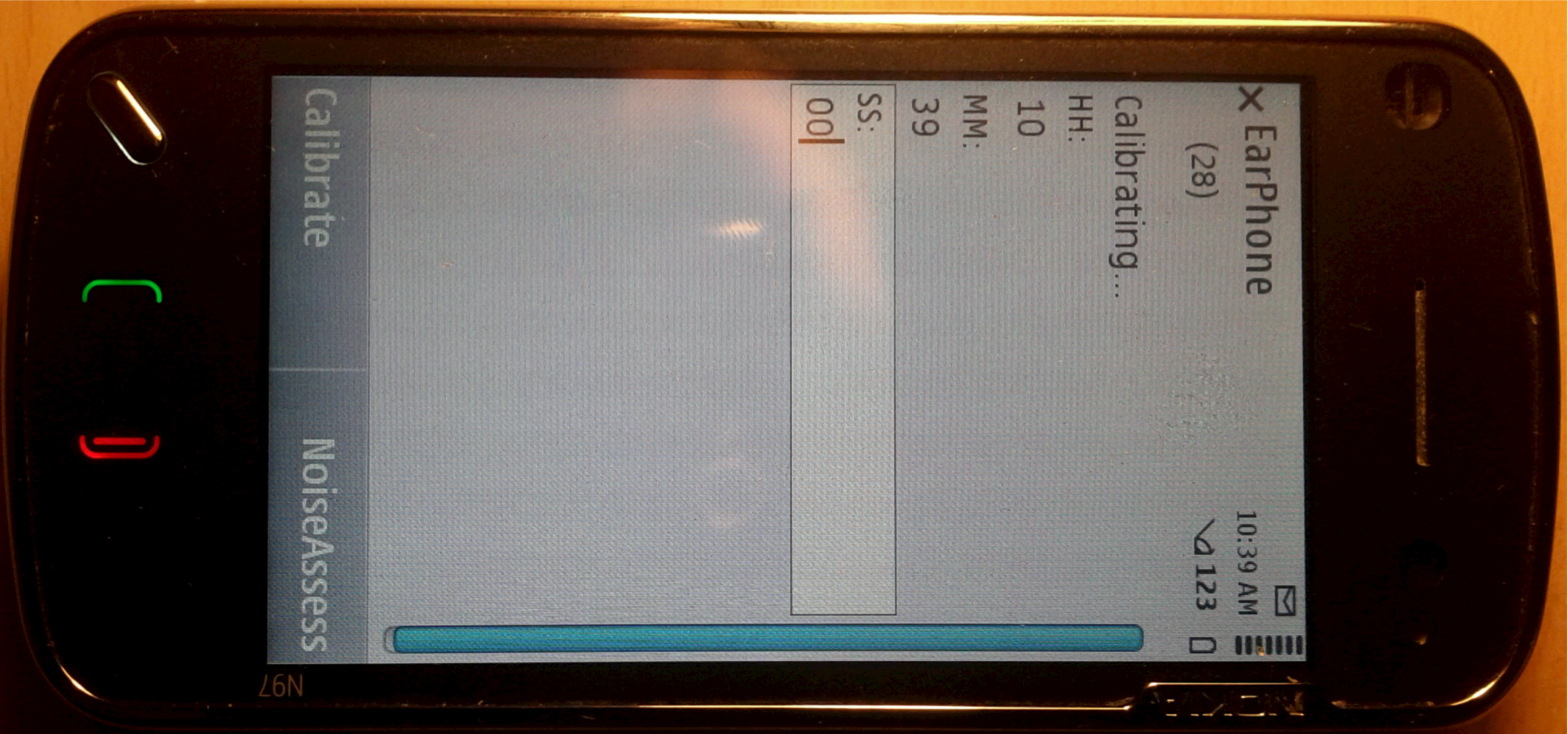}
\label{fig:calibratingPhoto}
}
\caption{Changes of calibration offset.}

\label{fig:context}
\end{figure}

A user invokes the calibration process by pressing the calibrate button on the Ear-Phone interface (see Fig.~\ref{fig:calibratingPhoto}). We suggest to leave the phone in a quite room for best performance of the calibration. A convenient time can be chosen on the interface when the user wants the calibration process to start. When calibration starts, the phone plays the calibration tone and at the same time makes a recording. At the end of the calibration tone, mean absolute difference between the recorded and reference tone sound levels is calculated and used as the offset.

\begin{figure}[ht]
\centering
\resizebox{!}{5 cm}{
\subfigure[]{
\includegraphics[width=0.33\columnwidth]{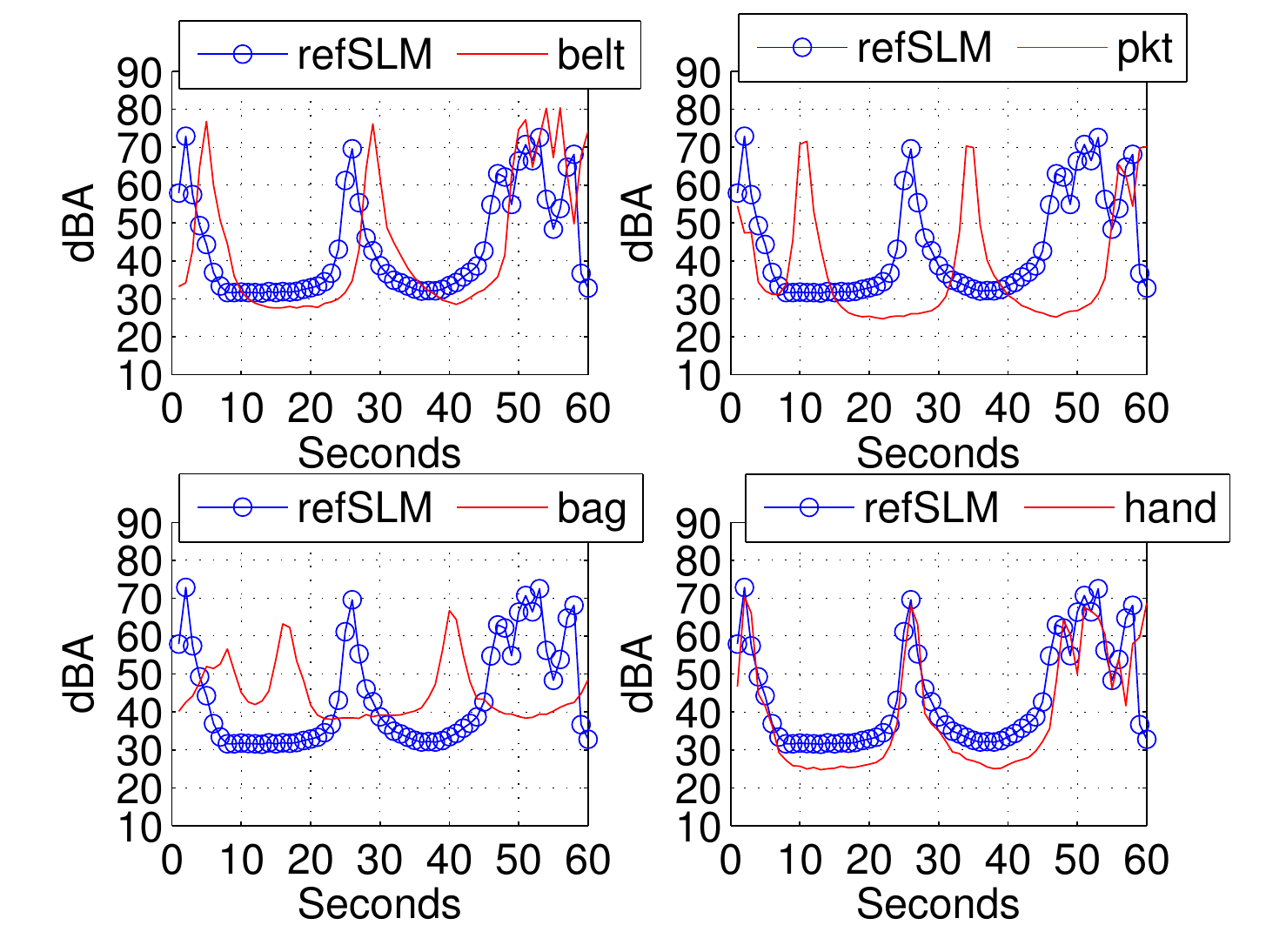}
\label{fig:context_exp_1721}
}
\subfigure[]{
\includegraphics[width=0.33\columnwidth]{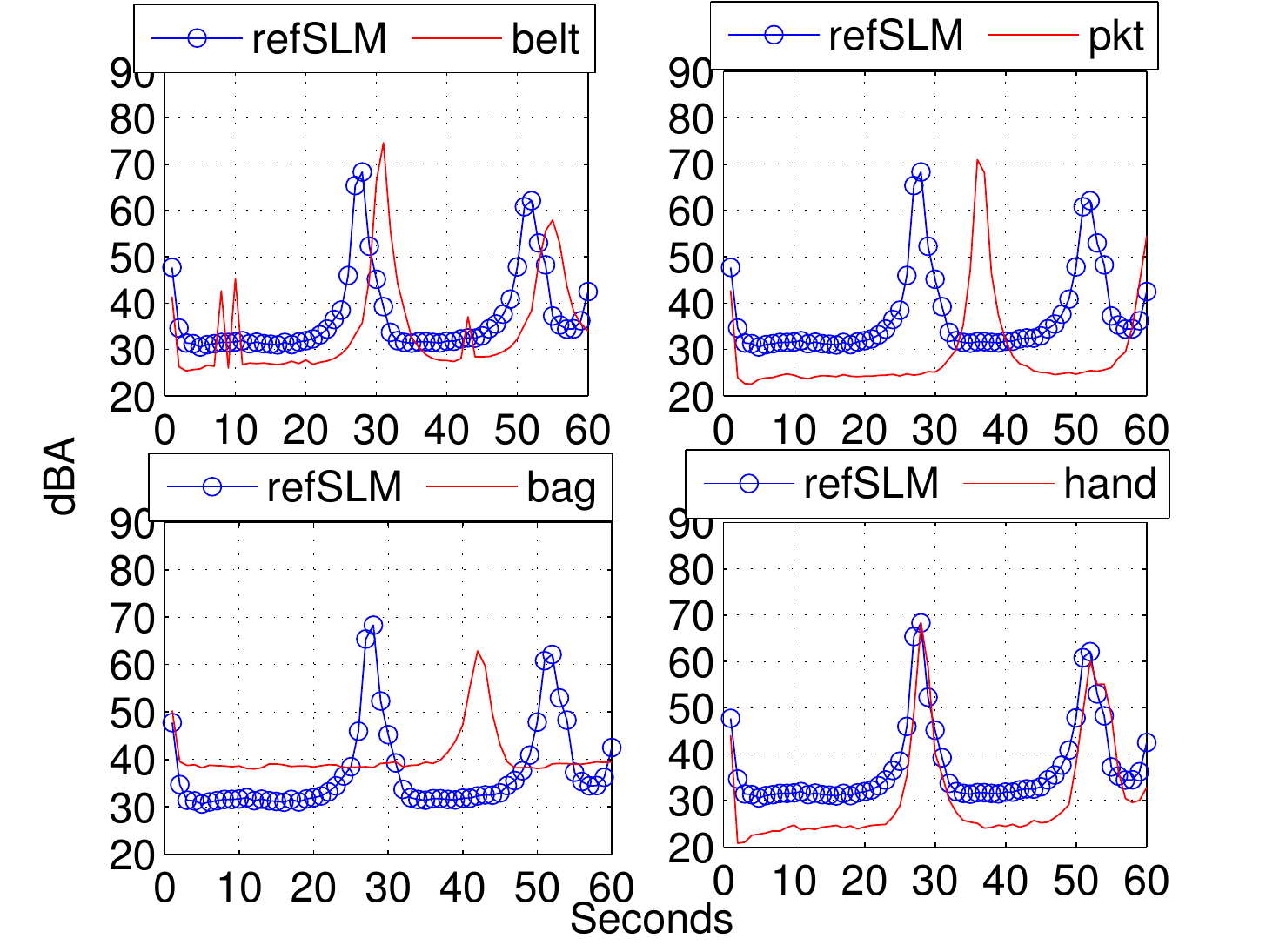}
\label{fig:context_exp_2218}
}
}
\caption{Measurement accuracy of mobile phone in different sensing contexts during (a) peak (b) off-peak hour.}
\label{fig:context}
\end{figure}	
\subsubsection{Measurement Accuracy in Different Sensing Contexts}
\label{sec:measurement_context}
In the above calibration experiments, we compute the calibration offset $\Delta$ and use it to calculate the \emph{\laeqt{T}}. We then conduct experiments to investigate how the positioning of the phone affects the measurement accuracy. 
In these experiments, we placed four \mobile in three different locations: palm, trouser's pocket and bag, and recorded equivalent noise levels while walking along the road. The experiments were conducted outdoor on the street during both peak and off-peak traffic. It was done three times during peak and three times during off-peak and each time the duration of experiment was 10 minutes. To compare the noise measurements with ground truth, the volunteer also carried a \commercial\hspace{.01cm} in his other hand.

%

%
\begin{table}
\centering
\small
\caption{Measurement accuracy (mean) in various sensing contexts. Experiment time is 10 minutes. For both pocket we present two set of results: for ``lead removed'', we remove the time lead before computing the mean-square error while ``with lead'' is simply the mean-square error. All results are reported in dBA. }
\resizebox{!}{1cm}{
\begin{tabular}{|c|c|c|c|c|}\hline
&  \mobile hand& \multicolumn{2}{|c|}{ \mobile pocket}& \mobile bag\\\hline
&&lead removed &with lead&\\ \hline
Peak&2.5&3.12&4.01&4.91\\\hline
Off-peak&2.96&3.11&3.89&4.68 \\\hline
\end{tabular}
}
\label{tab:measurement_accuracy}
\end{table}

Table~\ref{tab:measurement_accuracy} reports the mean absolute difference between ground truth (\commercial reading) and the \mobile measurements while the phone is placed in the three sensing contexts. We also plot the first 60 seconds of data in Fig.~\ref{fig:context_exp_1721} and ~\ref{fig:context_exp_2218} which corresponds to peak hour and off-peak hour results, respectively. There are a few interesting observations from this plot: while the mobile phone is held in hand, the response of the \commercial and \mobile are quite aligned and the mean difference is about $2.5$ dBA. While the mobile phone is kept inside pocket, the responses of \commercial and \mobile are aligned, however the pocket measurements lead the \commercial by a time interval of about 10s. 
Finally, when the mobile phone is carried inside a bag the responses are not similar to that of \commercial\hspace{-0.1cm}; there are noises and the overall magnitude of the signal is lowered. 

Based on these observations we decided to keep the {\laeqt{1s} samples from hand and discard the samples from pocket and bag. In future, we plan to devise a mechanism to automatically remove the time lead so that we can use the \emph{\laeqt{1s}} recorded from pocket.

\subsection{Performance of the Context Classification module}
\label{sec:ContextClassificationPerformance}
We run simulations on the data collected from the 10 participants to determine the context classification accuracy. We first determine context classification accuracy individually for accelerometer and proximity sensor. Then we determine context classification accuracy while fusing these two sensor data. 

For accelerometer we use time window as a control parameter. We use the data in a time window to calculate the features, such as mean, median etc.  In Table~\ref{tab:classificationAccuracyAcc}, we report the classification accuracy of the accelerometer sensor for two time windows ($\Delta$): 60 second and 30 second. We obtain the best accuracy when the feature is mean and the length of the  time window is 60 second.

For proximity sensor, we perform similar experiments to compute classification accuracy. Within the time window we calculate the number of triggers and use it as a feature for classification amongst hand enclosure and pocket or bag enclosure. The 60 second interval provides better accuracy, however, the overall accuracy is quite poor. We apply ``threshold'' on the number of firings of the proximity sensor and use it only for hand sensing context. Note that in the other sensing contexts the sensor firings are not significant enough or does not have pattern to use as a classifier. 

We use thresholds $5$, $10$, $20$ and $30$. A threshold $5$ means if the number of triggers is more than or equal to 5, we consider it as a hand sensing context. We fuse the classification decision with that of accelerometer and the accuracy for various selection of interval ($\Delta$), accelerometer feature and threshold is reported in Table~\ref{tab:fusion}. We obtain the best classification accuracy as 84\%. It is obtained when interval is $60$ second and threshold is 10. The accuracy is significantly higher compared to using only accelerometer sensor which is reported $76.86\%$ in Table~\ref{tab:classificationAccuracyAcc}. 

\begin{table}

\begin{minipage}{0.3\linewidth}
\caption{Classification Accuracy:Accelerometer z-axis.}
\begin{tabular}{|c|c|c|} \hline
&60 S&30 S\\ \hline
mean&75&73.57\\ \hline
median&76.86&75.23\\ \hline
75 Percentile &67.91&71.66\\ \hline
\end{tabular}
\label{tab:classificationAccuracyAcc}
\end{minipage}
\hspace{2.5cm}
\begin{minipage}{0.3\linewidth}
\caption{Classification Accuracy:Proximity Sensor.}
\centering
\begin{tabular}{|c|c|c|}\hline
&Mean Accuracy&Std \\\hline
30 S&51.04&1.44\\ \hline
60 S&52.08&2.88\\ \hline
\end{tabular}
\label{tab:classificationAccuracyPrx}
\end{minipage}
\end{table}

\begin{table}
\caption{Classification Accuracy: Sensor fusion of the proximity sensor and z-axis of the accelerometer.}
\centering
\begin{tabular}{|c|c|c|c|c|c|c|c|c|c|}\hline
&\multicolumn{4}{c|}{60 Second}&\multicolumn{4}{c|}{30 Second}\\\hline
&5&10 &20&30&5&10&20&30\\\hline
mean&83&{\bf 84}&77&79&74.13&75.43&73.91&73.91\\ \hline
median&80.16&78.33&75.83&76.33&73.13&72.3&73.27&72\\ \hline
75 Percentile &71.9&65.20&65.90&68.50&63.33&67.04&67.15&67.31 \\ \hline
\end{tabular}
\label{tab:fusion}
\end{table}

\subsection{Performance of Speech Detection Module}
We conducted experiments to determine the accuracy of the threshold based speech detection module discussed in Section~\ref{sec:Speech_Detection_module}. Four couples: two including two males and the other two including one male and one female, were used in the experiments. The training set was constructed from similar experiments described  in Section 4.3: conversation of two of the couples along roadside was contrasted to traffic noise alone recording to determine a threshold, where the recording took place in both peak and off-peak hours. 

In the test phase, the other two couples (those, who were not used in the training) were asked to engage in conversation in five 15 minutes sessions at 5 different times of a day, which include peak and off-peak periods. Two phones with the Ear-Phone client were used for each couple. One phone was held by the participant to measure conversation with traffic noise, while the other phone was placed 5 meters away to measure traffic noise only. Classifier was executed every minute\footnote{We tested various intervals, for example, 30 seconds, one minute, 90 seconds and observed one minute interval provides best accuracy. } on each of the phone, therefore, in total there were 150 events of conversation with traffic noise and 150 events of traffic noise alone. The confusion matrix is reported in Table~\ref{table:speechDetection}. In summary, both the precision and recall of the classifier is 94.33\%.
\begin{table}[t]
\centering
\caption{Clasisification accuracy of speech detection module.}
\resizebox{!}{0.75cm}{
\begin{tabular}{|c|c|c|c|} \hline
&&\multicolumn{2}{|c|}{Classified Context}\\\hline
&&traffic noise with human voice&traffic noise\\\hline
\multirow{2}{*}{Actual Context}&traffic noise with human voice & 139&11\\\cline{2-4}
&traffic noise & 6 & 144\\\hline
\end{tabular}
}
\label{table:speechDetection}
\end{table}

\subsection{Resource Usage}
\begin{table}[t]
\centering
\caption{CPU and RAM usage.}
\small
\begin{tabular}{|l|c|c|}
\hline
& CPU Load (\%)& RAM (MB)\\\hline
Signal processing &35.99$\pm$29.72&90.21 \\
+GPS threads &&\\ \hline
GPS thread &1.606$\pm$1.34&80.199\\ \hline
Ambient light sensor&1.29$\pm$0.59&79.66\\ \hline
Signal processing thread&43.68$\pm$35.22&86.55\\ \hline
\end{tabular}
\label{tab:resource_usage}
\end{table}

\subsubsection{Memory and CPU Benchmarks}
We carried out benchmark experiments to quantify the RAM and CPU usage of Ear-Phone running on the N97 using the Nokia Energy Profiler tool~\cite{energyProfiler}. 
We first measured the CPU and RAM usage for individual components: ambient light sensor, GPS thread and signal processing thread. Since the GPS thread and signal processing thread can run simultaneously, we then measured the combined CPU and RAM usage of these two components. The results are summarized in Table~\ref{tab:resource_usage}. 
Encouragingly, overall CPU utilization is quite low since the combination of GPS thread and signal processing thread (which will be running most of the time) use approximately 36\% of CPU. However, the memory usage is high, since it takes up about 90\% of RAM when the GPS and signal processing thread run simultaneously.

\subsubsection{Power Draw}

We use the Nokia Energy Profiler, which is a standard software tool to measure energy usage of Nokia hardware, to calculate the power draw on the Nokia N97 phone. We assume a seven minute cycle, where the first two and a half minutes are used to get a GPS lock (or fix) and read the co-ordinates and speed, the following half a minute is used for context discovery and the remaining four minutes is used for the signal processing module. Due to negligible power draw, we did not consider the phone status discovery module in this cycle. As expected, the GPS module is the dominant power consumer, it draws approximately 0.5  Watt-Hour. The context classifier draws 0.312 Watt-Hour and finally the signal processing module draws 0.237 Watt-Hour.
 
Note that once the GPS gets the lock, the consecutive steps would be just reading co-ordinates and speed from GPS receiver, however, we take a conservative approach and consider that the GPS gets a lock every time. In the urban areas due to high-rise buildings, the GPS may often lose the fix. We note that most modern smart phones interpolate locations between GPS readings from WiFi signals to conserve energy. A similar approach would greatly reduce the energy footprint of Ear-Phone

Nokia N97 uses a Lithium Polymer (abbreviated Li-poly) battery of 5.6 Watt-Hour. If the above cycle is repeated continuously, the battery will last for approximately for 5.33 hours. Note that we did not optimize the mobile phone module for energy consumption. A number of energy optimisation techniques have been suggested in~\cite{cencme}, which we will consider to adopt in our future study.

\begin{figure*}[tbp]
\centering
\subfigure[]{
\includegraphics[height=.4\columnwidth, width=.6\linewidth]{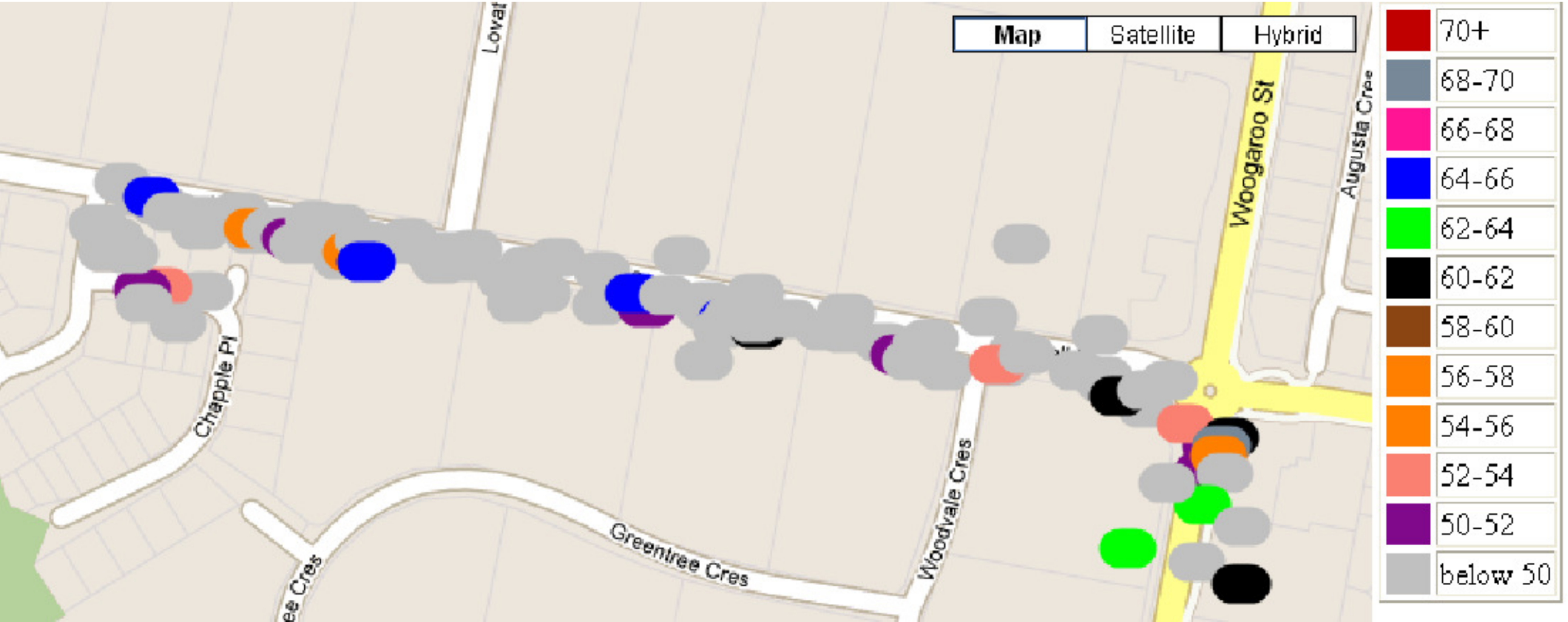}
\label{fig:1peak}
}
\subfigure[]{
\includegraphics[height=.2\columnwidth,width=.45\linewidth]{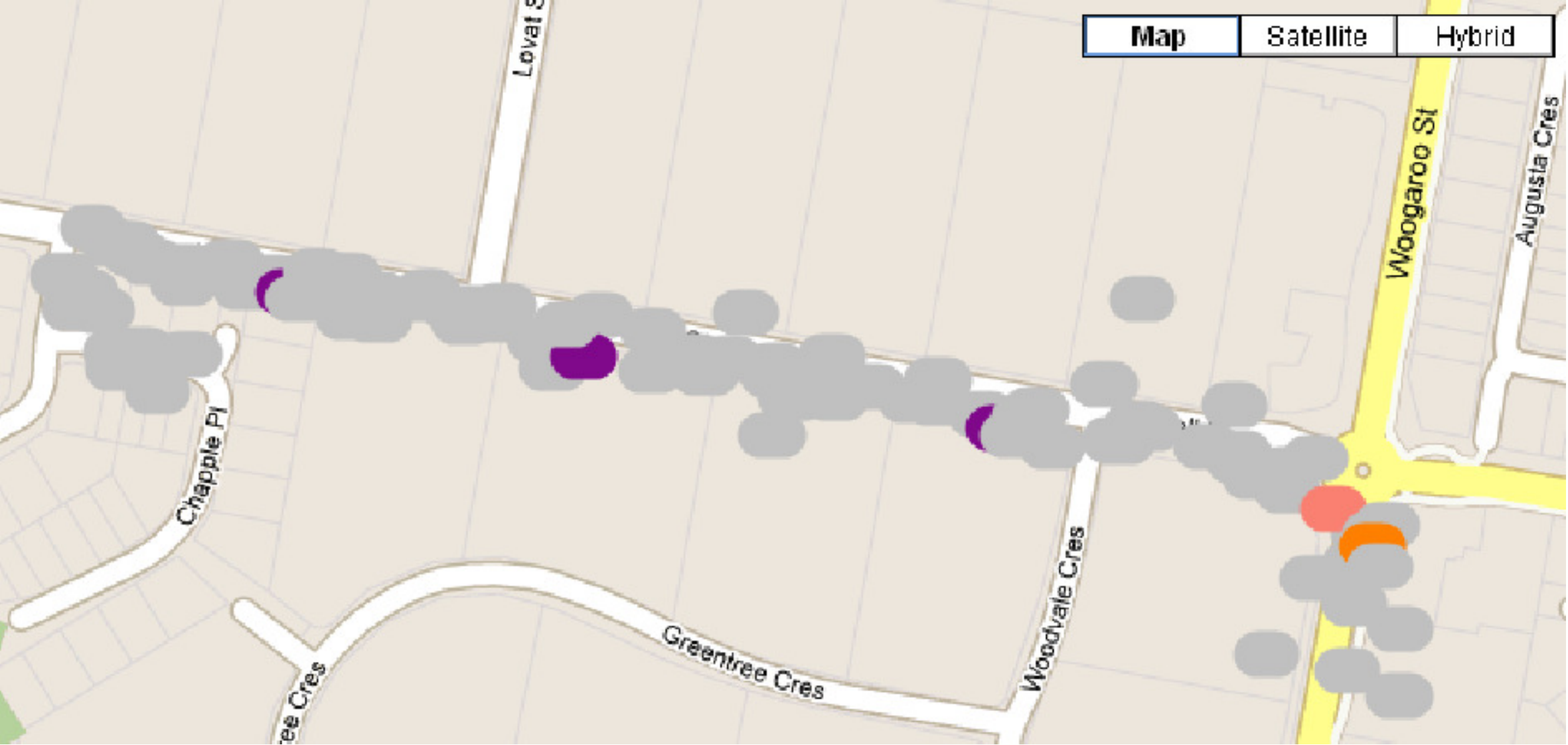}
\label{fig:3peak}
}
\subfigure[]{
\includegraphics[height=.2\columnwidth,width=.45\linewidth]{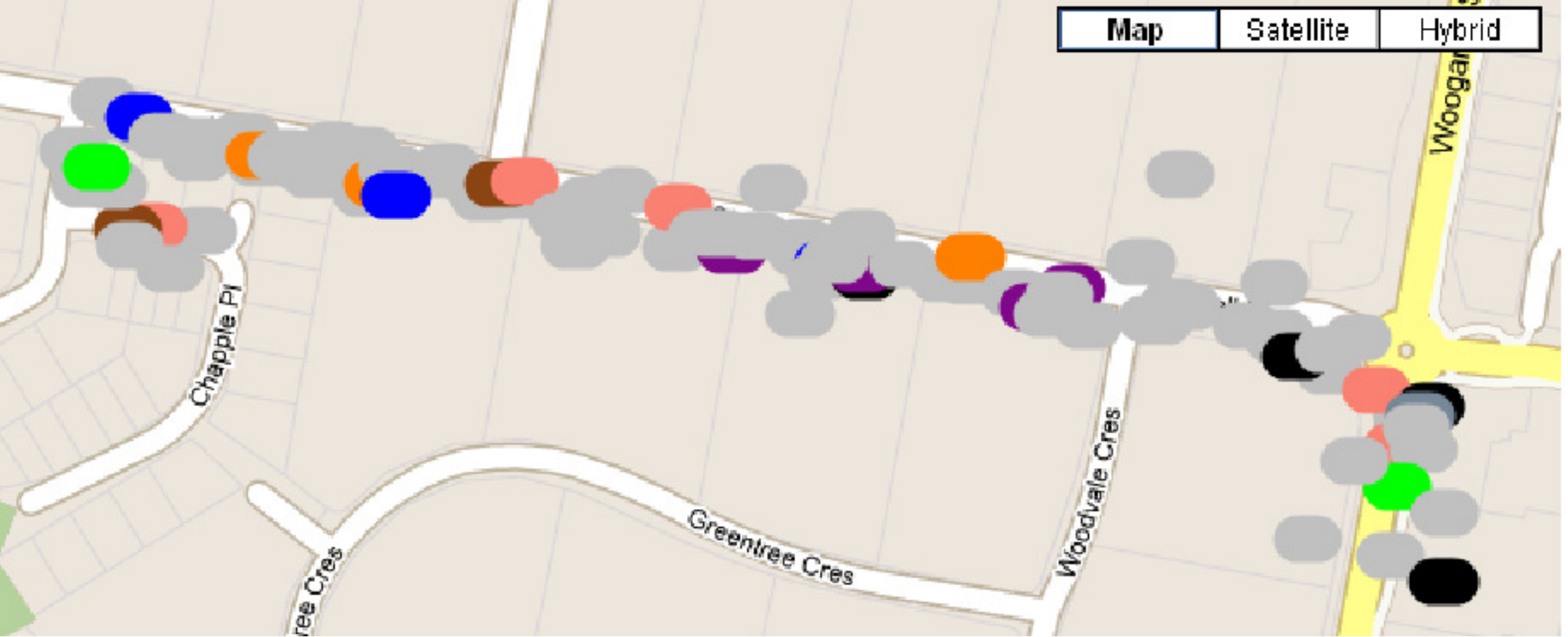}
\label{fig:5peak}
}
\subfigure[]{
\includegraphics[height=.2\columnwidth,width=.45\linewidth]{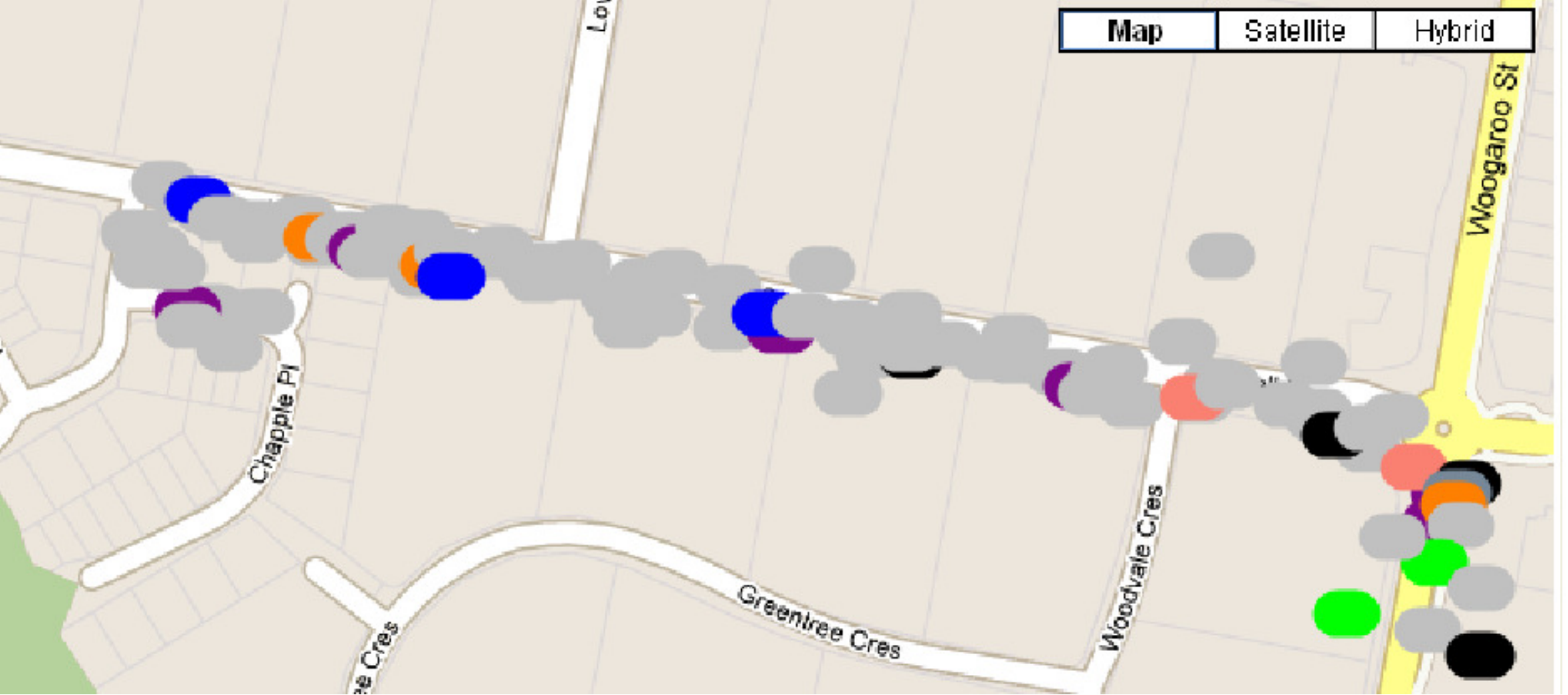}
\label{fig:7peak}
}
\caption{Noise map reconstruction during off-peak traffic hour (20:00 -21:00) at the Roxell Street intersection. (a) Ground Truth (b) Reconstruction using 90\% missing samples, (c) Reconstruction using 50\% missing samples, and (d) Reconstruction (very close to the ground truth) using 30\% missing samples. }
\label{fig:offpeak}
\end{figure*}

\begin{figure*}[tbp]
\centering
\subfigure[]{
\includegraphics[height=.4\columnwidth,width=.6\linewidth]{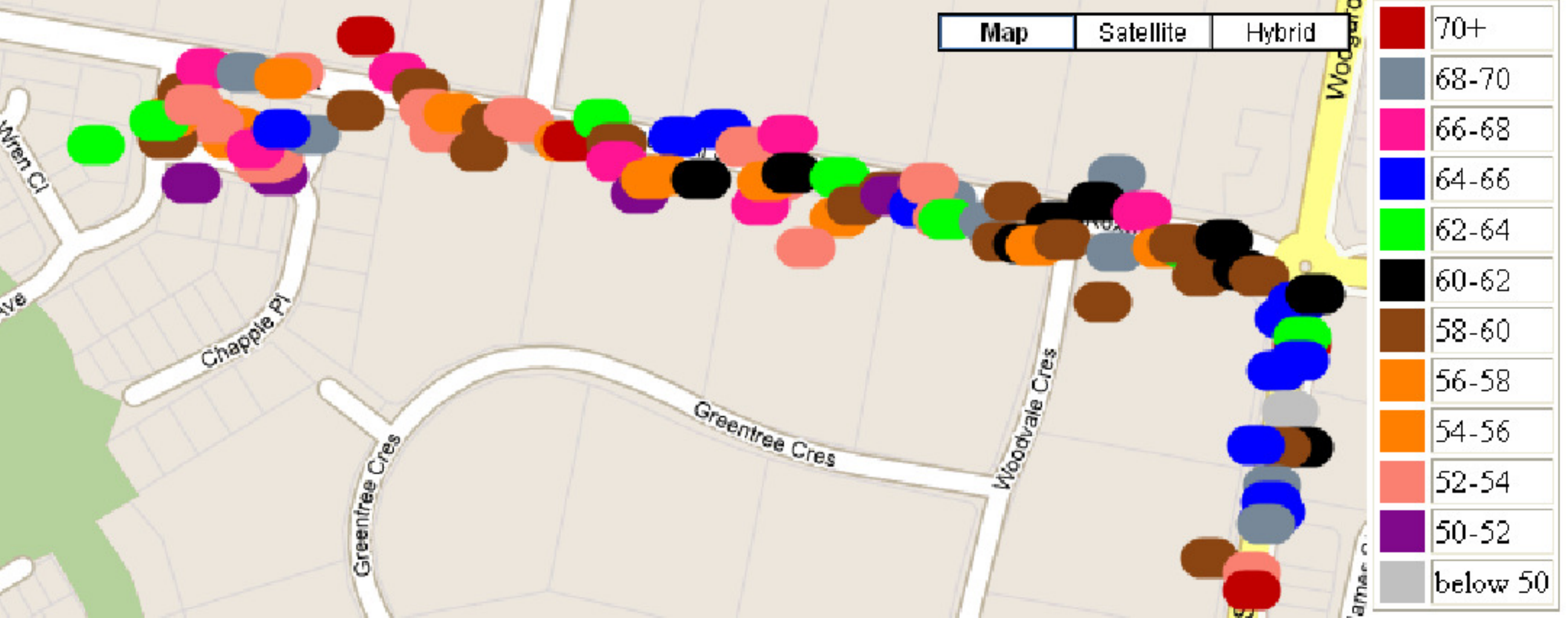}
\label{fig:1offpeak}
}
\subfigure[]{
\includegraphics[height=.2\columnwidth,width=.45\linewidth]{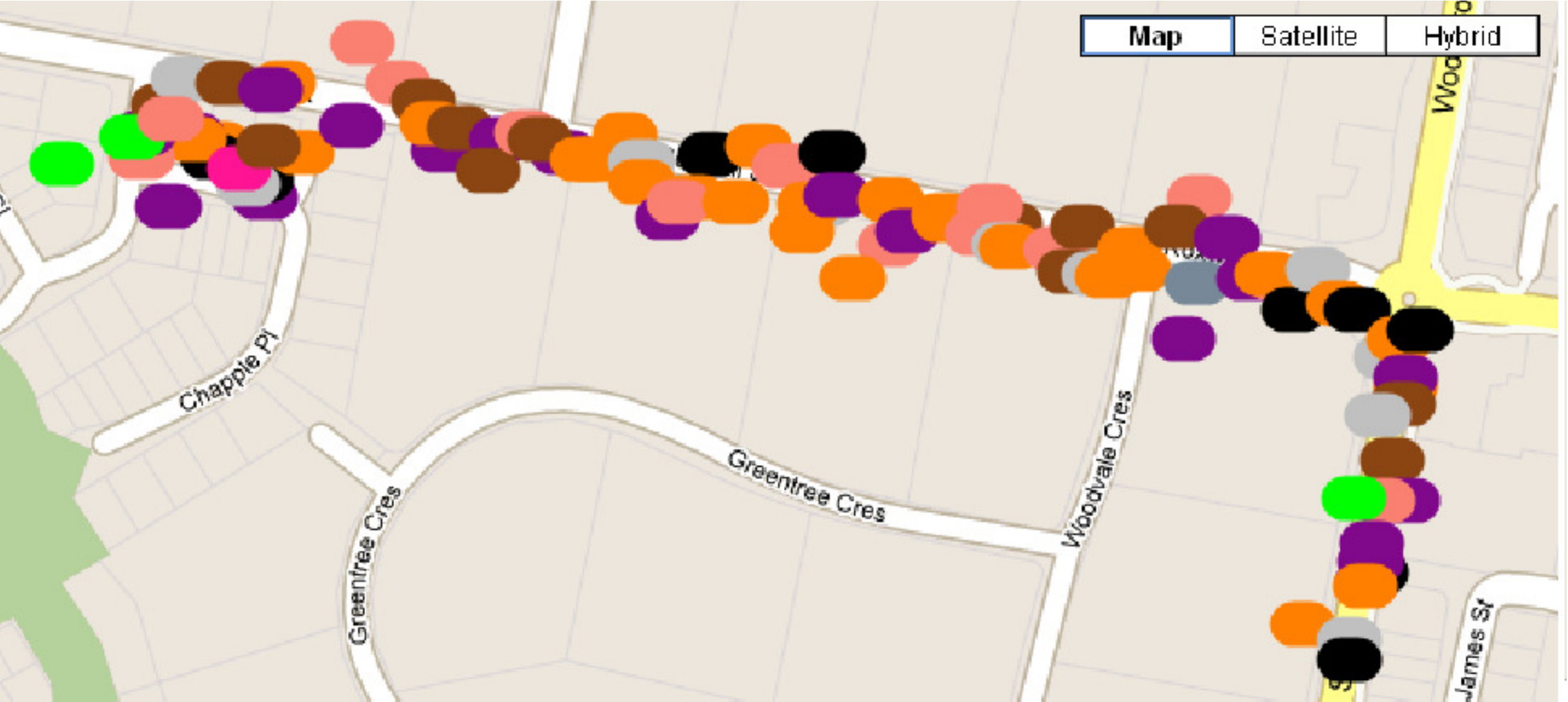}
\label{fig:3offpeak}
}
\subfigure[]{
\includegraphics[height=.2\columnwidth,width=.45\linewidth]{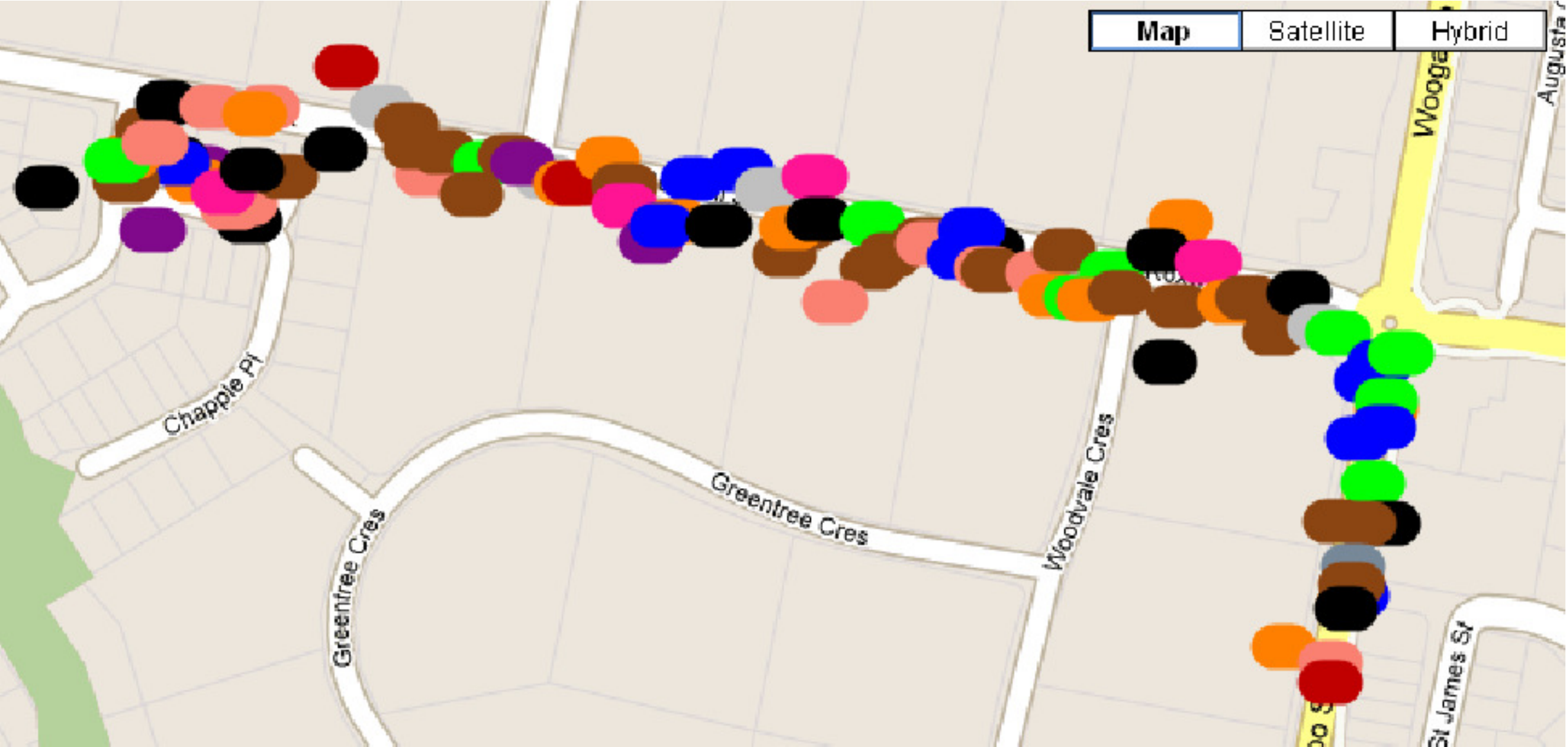}
\label{fig:5offpeak}
}
\subfigure[]{
\includegraphics[height=.2\columnwidth,width=.45\linewidth]{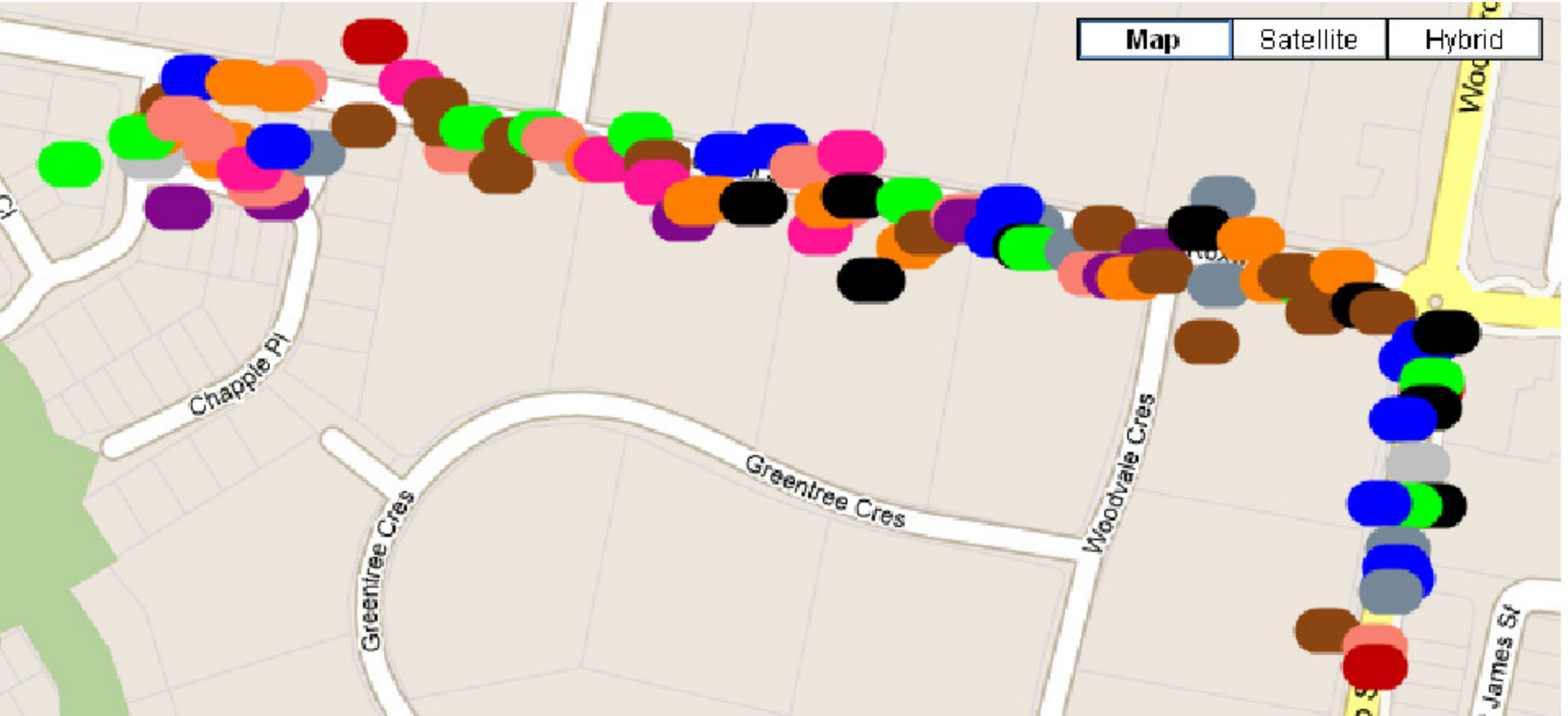}
\label{fig:off7peak}
}
\caption{Noise map reconstruction during peak traffic hour (7:30-8:30) at Roxell Street. (a) Ground Truth (b) Reconstruction using 90\% missing samples, (c) Reconstruction using 50\% missing samples, and (d) Reconstruction (very close to the ground truth) using 70\% missing samples.}
\label{fig:peak}
\end{figure*}

\subsection{Performance Evaluation}
To evaluate the performance of Ear-Phone as an end-to-end system, we conducted several outdoor experiments. Our primary goal 
is to investigate the impact of data (\laeqt{1s} samples) availability on the reconstruction
performance ( Based on our simulation results in Section~\ref{sec:simulation}, $\ell_1$-norm offers the best reconstruction accuracy compared to linear interpolation, nearest neighbor interpolation and Gaussian process interpolation. We therefore reconstruct using $\ell_1$-norm in the outdoor experiments.). In the experiments, we reconstructed the 
noise map along two different road intersections in Brisbane, Australia. 
The first intersection includes the intersection of Woogaroo Street and Roxell Street and Chapple place, located in a residential suburb named Forest Lake. The second intersection is in a semi commercial suburb, Oxley. This intersection includes a major road named Blunder Road and few other suburban roads. In Blunder Road, road traffic does not vary too much in peak and off-peak, however in other two intersections peak hour traffic is significantly higher than off-peak traffic. 

We reconstructed an hourly noise map separately during peak (
7:30-8:30 for Roxell Street, 10:00-11:00 for Blunder road) 
and off-peak (
20:00-21:00 for Roxell Street and 15:00-16:00 for Blunder Road) hours. To collect noise samples, five subjects walked along these segments several times within the one 
hour period with Ear-Phone running on the Nokia N97. 
We first constructed the ground truth by using all the noise samples. The ground truth along Roxell Street intersection during peak and off-peak are shown in Fig~\ref{fig:1peak} and \ref{fig:1offpeak}, respectively. 
We randomly chose various percentages of the total data (collected from all users) and investigated the corresponding reconstruction performance. Due to similarity of reconstruction, we only show the reconstruction results at Roxell Street; reconstruction result at Blunder Road can be found in the Appendix B. In Fig.~\ref{fig:noise_map_recon_error}, we summarize the reconstruction error for various percentage of data for all three of these intersections. We observe that during peak hour, when the percentage of missing data is 30\%, the reconstruction error can be brought down to approximately 3dB.

\section{Simulation}
\label{sec:simulation}
Real experiments certainly provide valuable information. However, real experiments are not repeatable. Furthermore, conducting 
a real experiment on a large scale is expensive and time consuming. We therefore conducted simulation experiments where factors such as the number and mobility patterns of volunteers can be varied easily. In this section, we will first describe how we perform measurement campaigns to collect noise profiles. These noise profiles will be fed into the simulation as ground truth. Next, we will describe the simulation method and performance evaluation.
\subsection{Simulation Design}
\subsubsection{Noise Profile}
As in Section~\ref{sec:implementation}, we limit our consideration to noise measurements along a road, which can be modelled as a scalar field over a 2-dimensional lattice of cells with one spatial and one temporal dimension. 
We assume that each cell has a spatial width of $\Omega$ meters and a temporal width of $T$ seconds. We use the ordered pair $(i,j)$ to refer to the cell bounded by the spatial interval $[(i-1)\Omega,i\Omega]$ and temporal interval $[(j-1)T,jT]$. Assuming that $i \in N_s = \{ 1,2,...,n_s \}$ and $j \in N_t = \{ 1,2,...,n_t \}$, the reference lattice covers a length of $n_s \Omega$ meters and a duration of $n_t T$ seconds. We assume that the equivalent noise 
level \emph{\laeqt{T}} measured over each cell is almost constant. Now let $d(i,j)$ denote the equivalent noise level \emph{\laeqt{T}} measured in cell $(i,j)$, then a {\sl noise profile} $S$ is defined as the set of all \emph{\laeqt{T}} measured
over the defined lattice, i.e. $S=\{d(i,j)\}_{(i,j) \in N_s \times N_t}$.

\subsubsection{Measurement Campaign}

Our first task is to conduct a number of measurement campaigns to obtain {\sl reference noise profiles} which we can feed into the simulation as ground truth. We conducted four experiments to collect \emph{\laeqt{1s}} under a variety of noise conditions and settings. The experimental conditions and parameters used are summarized in Table~\ref{tab:sound_level}\footnote{The experiments were conducted during saturdays, however, due to the proximity of the experimental areas to the lake, where every saturdays amusement events like boat rowing, jumping castle etc. take place, there was high density traffic during the afternoon and evening. }. During each of these experiments, we measured \emph{\laeqt{1s}} along Forest Lake Boulevard, which is a major artery road in Brisbane. This road has two-way traffic with $2$ lanes in each direction. The traffic flow was reasonably high as indicated by the mean noise level in Table~\ref{tab:sound_level}. We used 6 \mobiles (Nokia N97) to capture the reference noise profile and placed them in 6 equidistant locations along the road with the microphone pointed towards the road. Different spatial separations are used in the experiments, see Table~\ref{tab:sound_level}. The clocks on the phones were synchronized to ensure that all phones start and stop sampling at the same time. The \mobiles measured \emph{\laeqt{1s}} during the experiment and stored the data in a text file which was downloaded to a computer at the end of the experiment. From each experiment, we created a reference noise profile, where $|N_s|$ = 6 and $|N_t|$ is the experimental 
duration in seconds. We deliberately conducted one experiment (see Table~\ref{tab:sound_level}) with a side road (Woogaroo st) between the mobiles to create a reference profile with high noise variation. A side road divides the traffic flow, therefore noise levels on either side of the road typically have high difference.

\subsubsection{Simulation Method}
Let $d_i \in [0,n_s \Omega]$ denote the position of the agent (simulated volunteers) at time $i T$ seconds. 
The location of this agent at time $(i+1)T$ is 
given by $d_{i+1}=d_i+V_i T$ where $V_i$ is the average speed (in ${\rm ms^{-1}}$) of the agent in the time 
interval $[iT,(i+1)T]$. The value of $V_i$ is assumed to be uniformly distributed in $[0,1.31]$ since pedestrian speed range from 1.25 to 1.31 ms$^{-1}$ ~\cite{walk_speed_1}. The sign of $V_i$ determines the direction of movement. We used the probabilistic random walk model~\cite{mobility} to generate the mobility pattern. People typically do not move in purely random manner (moving back and forth), instead they keep on moving in a given direction for a while before they change their direction. Probabilistic random walk model implements this behavior using a three state Markov Chain model. The state transition parameters of our Markov Chain model are reported in Fig.~\ref{fig:prm}. 
\begin{figure}[h]
\centering
\includegraphics[width=0.8\linewidth]{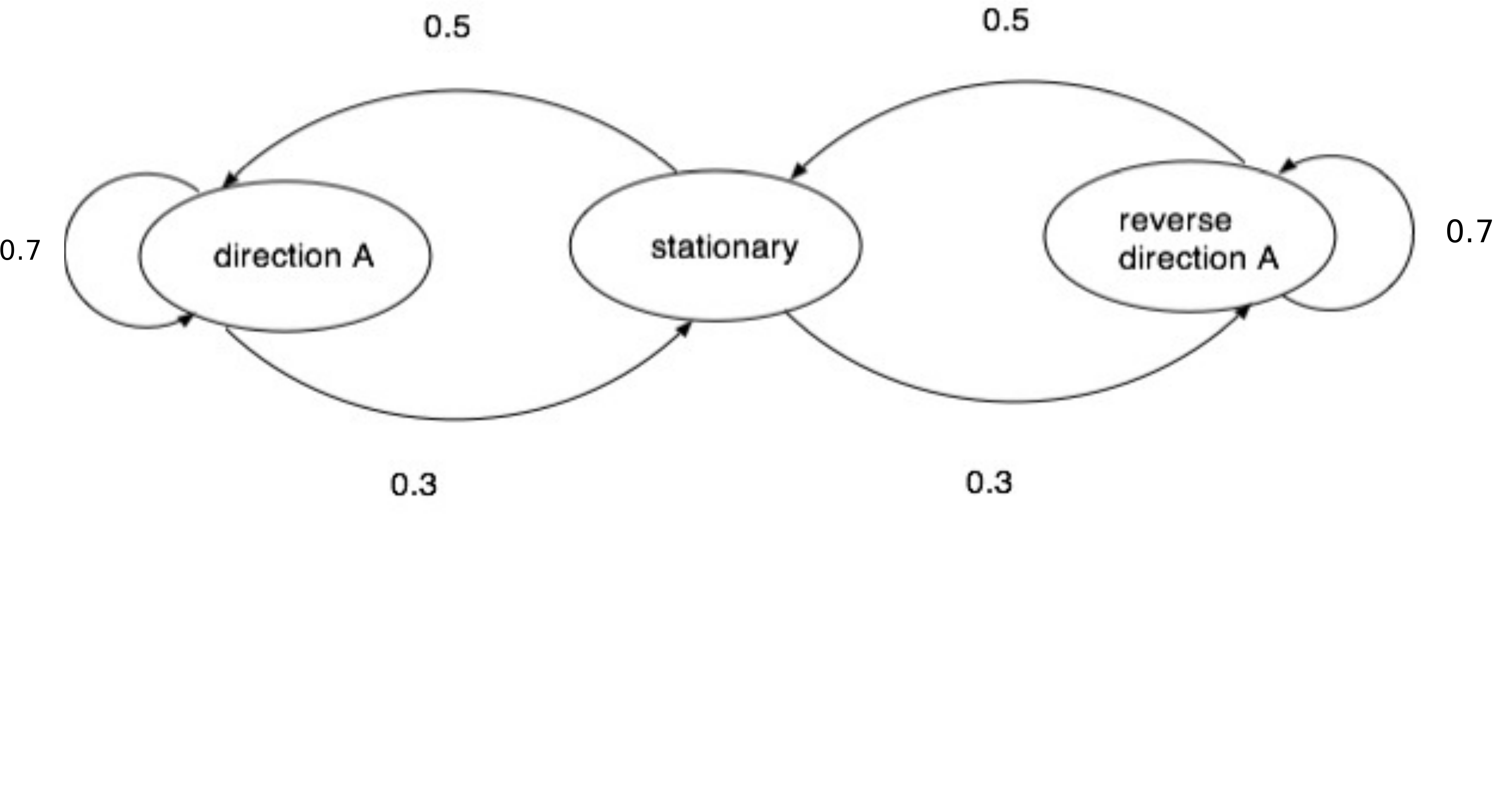}
\caption{Probablistic random walk model parameters.}
\label{fig:prm}
\end{figure}
The agent is in cell $(\lceil \frac{d_i}{\Omega} \rceil, i) \in N_s \times N_t$ at time $i T$, where $\lceil u \rceil$ denotes the smallest integer that is greater than or equal to $u$. Consider $W \subset N_s \times N_t $ denote all the cells visited by this particular agent. To simulate urban sensing, we assume that an agent does not take samples at all visited cells (Due to privacy concerns, volunteers may not contribute samples near their home or office. The microphone may be in use for conversation). Let $\tilde{W} \subset W$ denote the set of all cells whose data is contributed by this agent. In the simulation we vary $\tilde{W}$ to understand the relationship between reconstruction accuracy and percentage of missing data.

%
%

\begin{table}[t]
\centering
\caption{Experimental settings for collecting the reference noise profiles.}
\resizebox{!}{1 cm}{
\begin{tabular}{|l|c|c|c|c|c|c|}
\hline
Exp& Date and time & Mean, Standard & Spatial & Duration & Continuous road & \% of DCT coefficients \\
No.& & Deviation of &separation  &(min)&segment without & to approximate the profile within\\
& & sound level (dBA) &(meters) & &side roads &1 dBA RMS error  \\
\hline
1 & 17/11/12 11:00 am  & 63.05,3.15 & 10 & 60 & yes & 22.83\\ \hline
2 & 17/11/12 2:30 pm & 67.11,4.43 & 10 & 60 & yes & 30.15\\ \hline
3 & 24/11/12 4:14 pm & 69.43,6.01 & 50 & 60 & yes & 37.33\\ \hline
4 & 24/11/12 6:24 pm & 73.22,6.79 & 50 & 60 & no & 43.91\\ \hline
\end{tabular}
}
\label{tab:sound_level}
\end{table}

\subsection{Reconstruction Performance}
\label{subsec:performance}
We now proceed to study the trade-off between the reconstruction accuracy and the percentage of missing data for different reconstruction algorithms.  We use the four different noise profiles as reference and evaluate the 
reconstruction performance under varied mobility patterns and number of agents.

In 
Figs.~\ref{fig:ex_1} to~\ref{fig:ex_4} we plot the reconstruction accuracy versus sampling requirements. We observe that among all the reconstruction algorithms, $\ell_1$-norm has better reconstruction accuracy for all 4 reference profiles. Precisely, except for profile 4, Ear-Phone can reconstruct the profiles to within 3 dB (recall that a 3 dB difference is not perceptible by human beings) error with 40\% missing samples. However, for linear, nearest neighbor and Gaussian process interpolation methods, the percentage of missing samples must be less than $5\%$ in order to achieve the same accuracy. 
Profile $4$ can accept the smallest missing samples, which. This increase in sampling requirements for profile 4 can be explained in terms of the compressibility of the noise profile. 
One way to determine the compressibility of a noise profile is to study the percentage of transform coefficients needed 
to approximate a profile to a given level of accuracy. The last column of Table \ref{tab:sound_level} shows that profile $1$ is the most compressible, since it required the smallest percentage of coefficients to represent the signal within 1 dBA RMS error. On the other hand, profile 4 is the least compressible, since it required the largest percentage of coefficients to represent the signal within 1 dBA RMS error. Note that profile 4 spans over two road segments with varying traffic, therefore, contains relatively high variance in noise levels.

\section{Discussion}
\label{sec:discussion}
\subsection{Need for Incentives}
In this paper we propose participatory sensing to generate the noise map in an urban environment. We propose $\ell_1$-minimization to reconstruct the noise map from the \emph{incomplete} and \emph{missing} samples contributed by mess, which is a prevalent problem in participatory sensing. We demonstrate that $\ell_1$ minimization offers a good compressibility- accuracy trade-off in reconstructing the noise map, however, the sampling requirements may still be high for crowd sourced application. We therefore recommend incentive mechanism as a complement to $\ell_1$ reconstruction. Note that incentive mechanisms are likely to be necessary for many crowdsourcing applications in general in order  to obtain the requisite data.

Designing an incentive mechanism is out of the scope of our paper. ``Platform-Centric Model'' proposed by Yang et al. in~\cite{Yang:2012} offers one potential solution, which reward people for holding the phone in hand wherein the holding time can be determined from the output of our context classifier. 

%
\begin{figure}
\centering
\subfigure[]{
\includegraphics[width=0.45\linewidth]{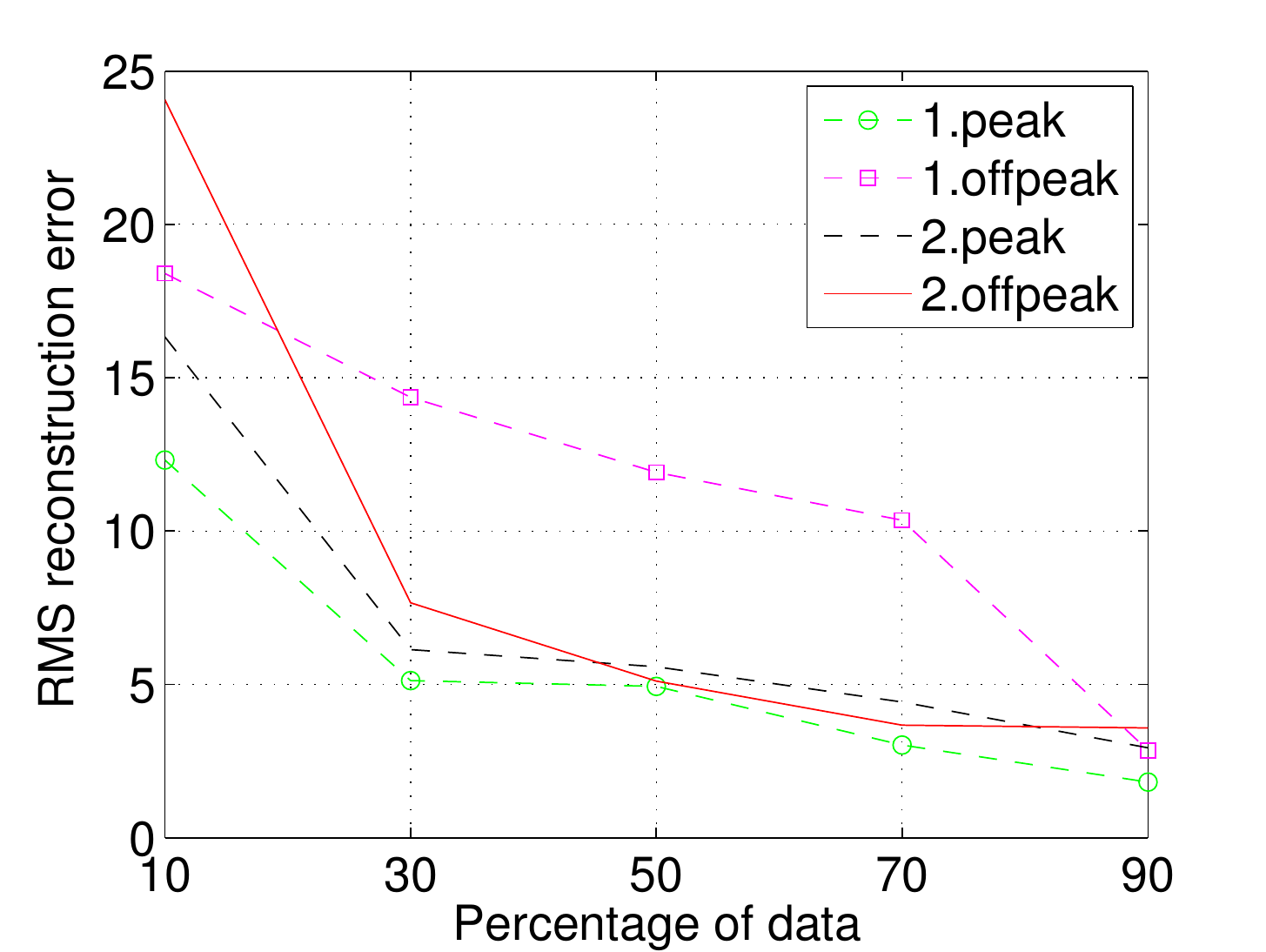}
\label{fig:noise_map_recon_error}
}
\subfigure[]{
\includegraphics[width=0.45\linewidth]{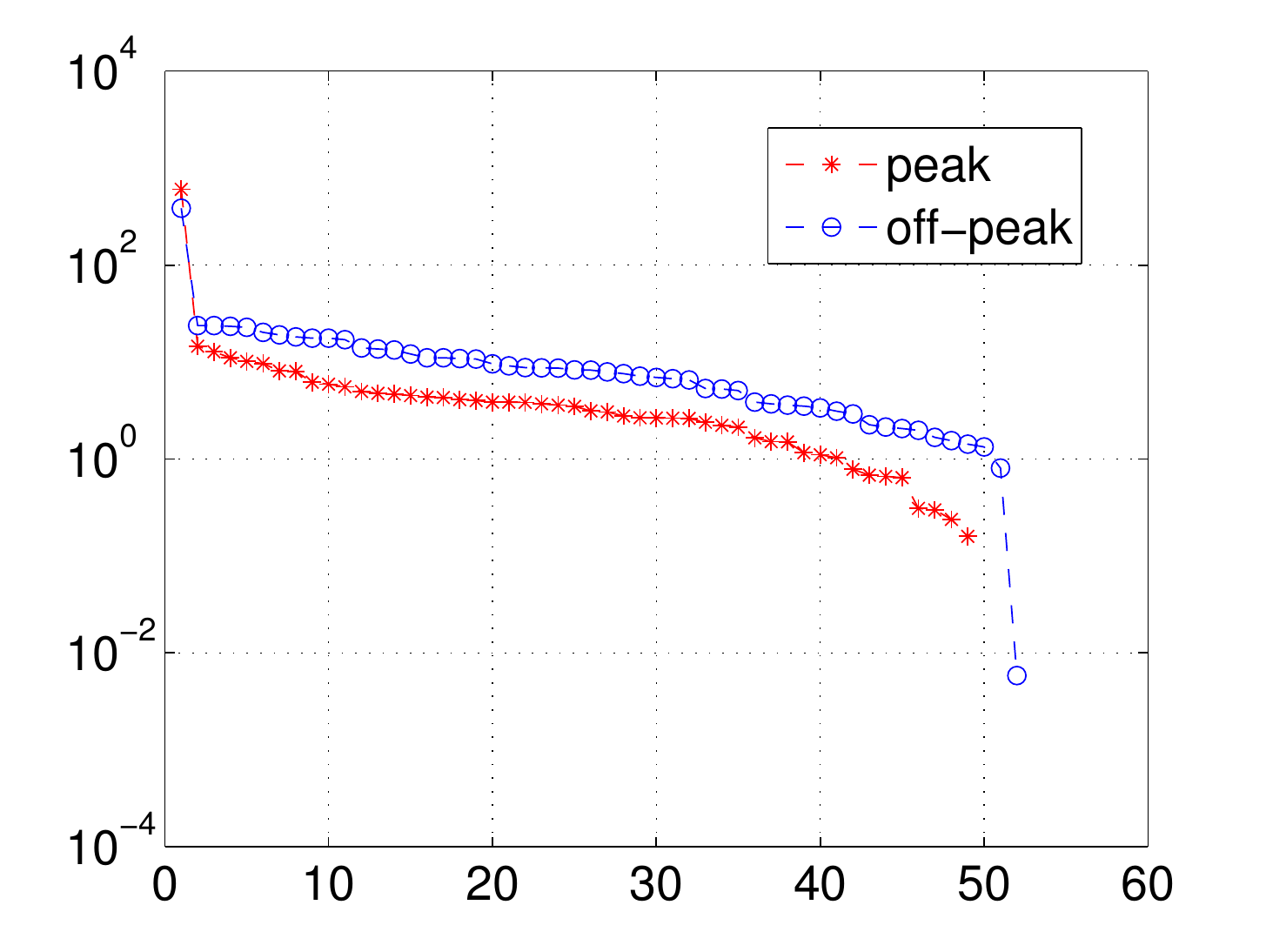}
\label{fig:compressibility_relation}
}
\caption{(a) 
1's are the error curves for the Roxell Street intersection and 2's are the error curves for the Blunder Road intersection (b) Compressibility plot for Roxell Street intersection. To avoid cluttering the image, compressibility of noise level at only Roxell Street is shown. To obtain this plot, we first compute the DCT coefficients of the true spatial-temporal noise map (ground truth) and then sort the absolute value of the coefficients (vertical axis) in descending order. }
\label{fig:various}
\end{figure}
\subsection{Noise Level Dynamics in Peak and Off-Peak}
Based on our datasets we observed one interesting property of noise map reconstruction: reconstruction during peak hour requires fewer samples compared to reconstruction during off-peak. To investigate the reason we compared the compressibility of sound profile during peak and off-peak (see Fig~\ref{fig:compressibility_relation}) and found that the ground truth during peak hour is more compressible compared to ground truth during off-peak. In order to determine why the noise profile during peak is more compressible compared to that of off-peak hour, we further computed the variance of noise level during peak and off-peak hour and found that during off-peak the variance is higher than that of peak hour. Intuitively, during off-peak hour due to intermittent vehicles passing, there is a higher level of variance in sound level, whereas, during the peak hour due to near continuous vehicle passing, the sound level is less varying. 
\label{sec:discussions}

%

\begin{figure}[ht]
\centering
\subfigure[]{
\includegraphics[width=0.45\linewidth]{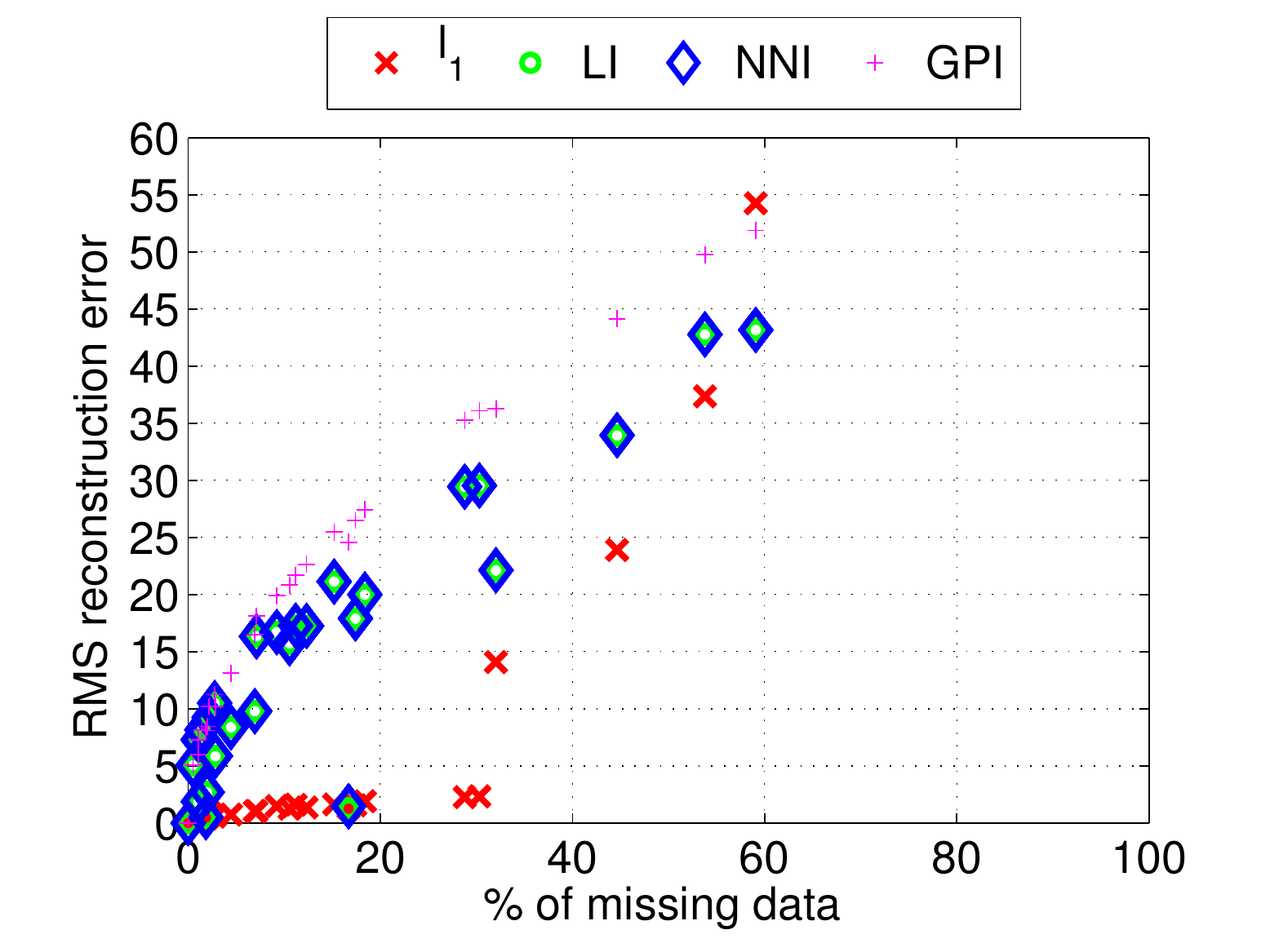}
\label{fig:ex_1}
}
\subfigure[]{
\includegraphics[width=.45\linewidth]{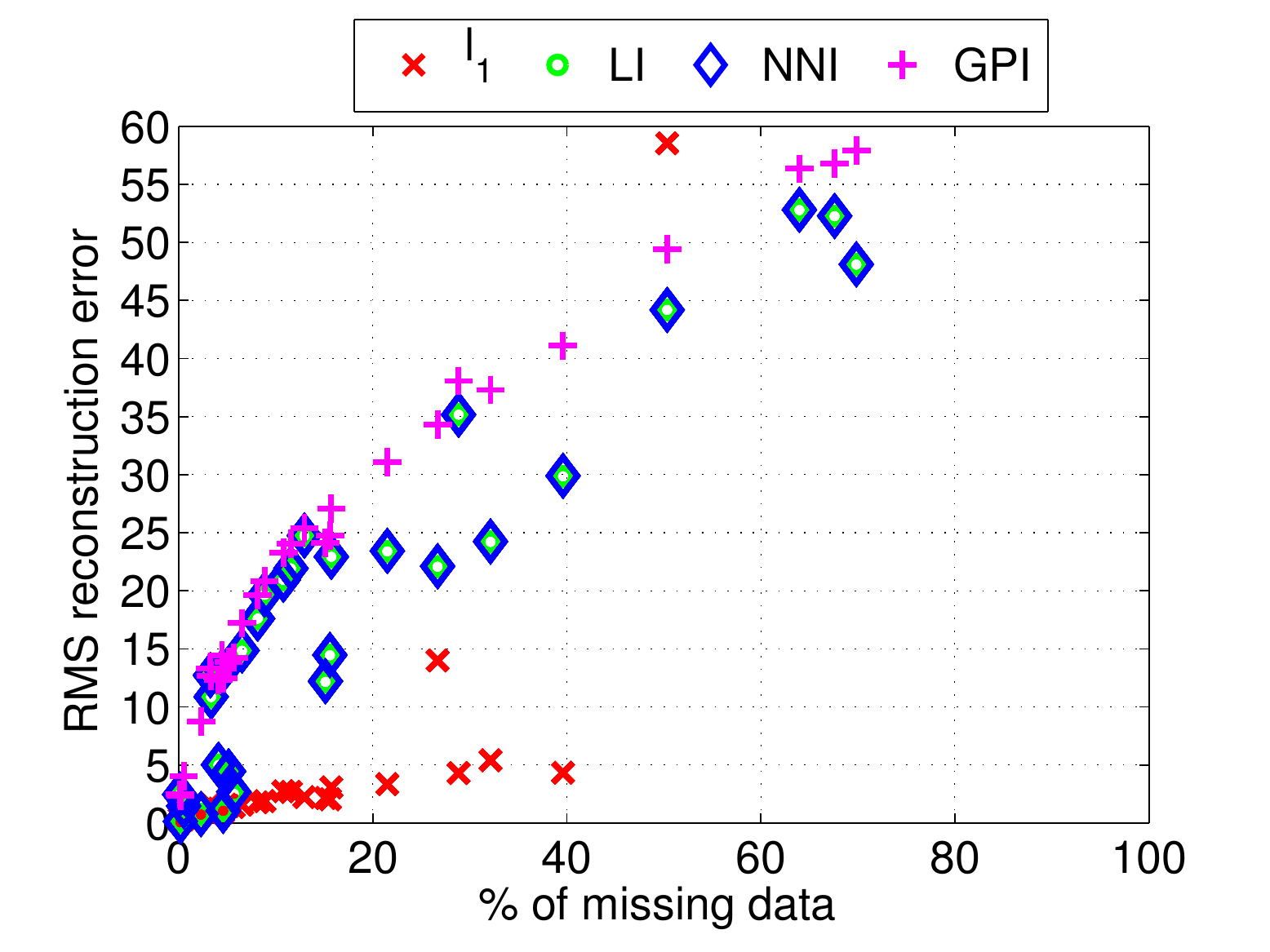}
\label{fig:ex_2}
}
\subfigure[]{
\includegraphics[width=0.45\linewidth]{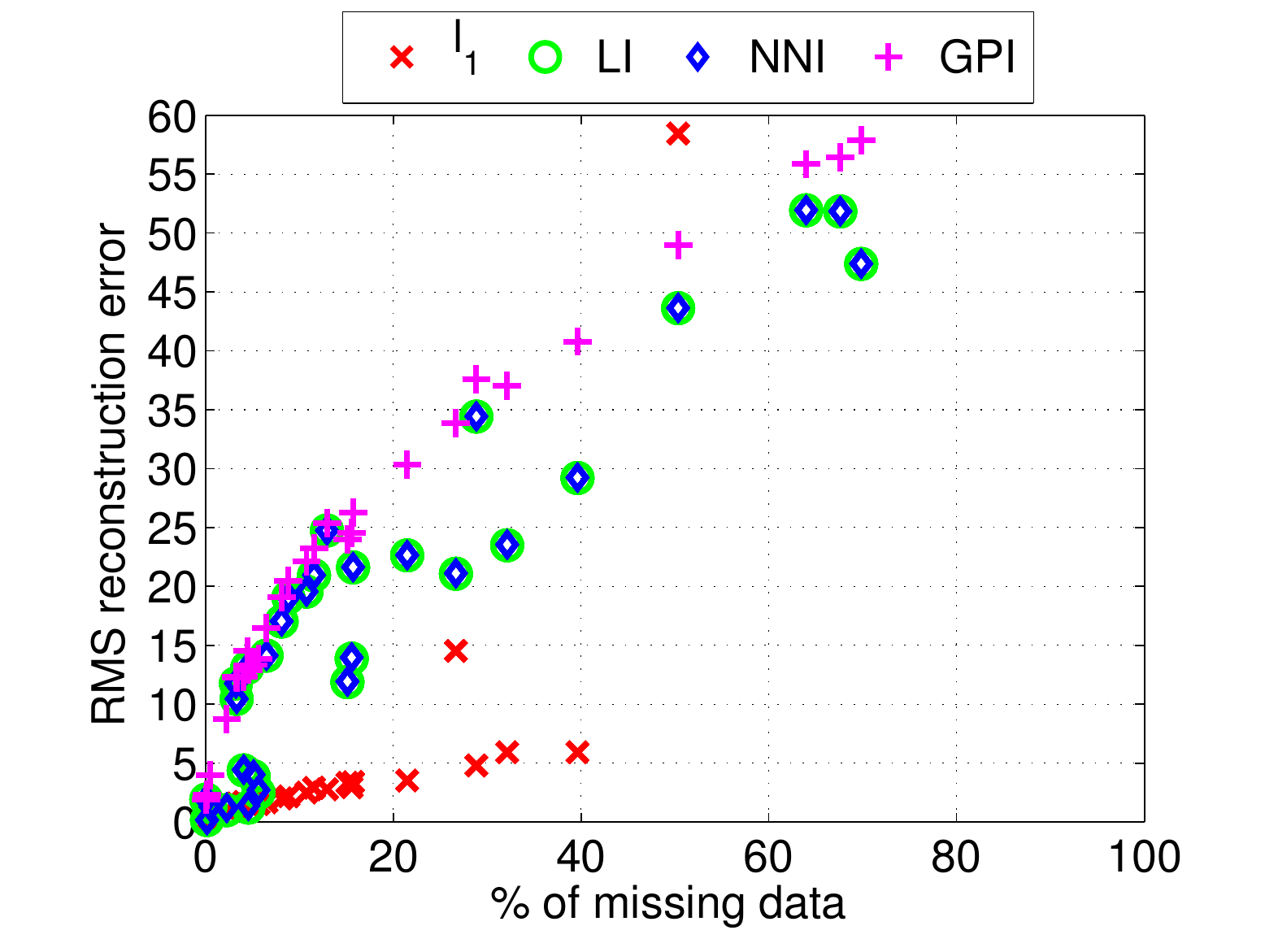}
\label{fig:ex_3}
}
\subfigure[]{
\includegraphics[width=0.45\linewidth]{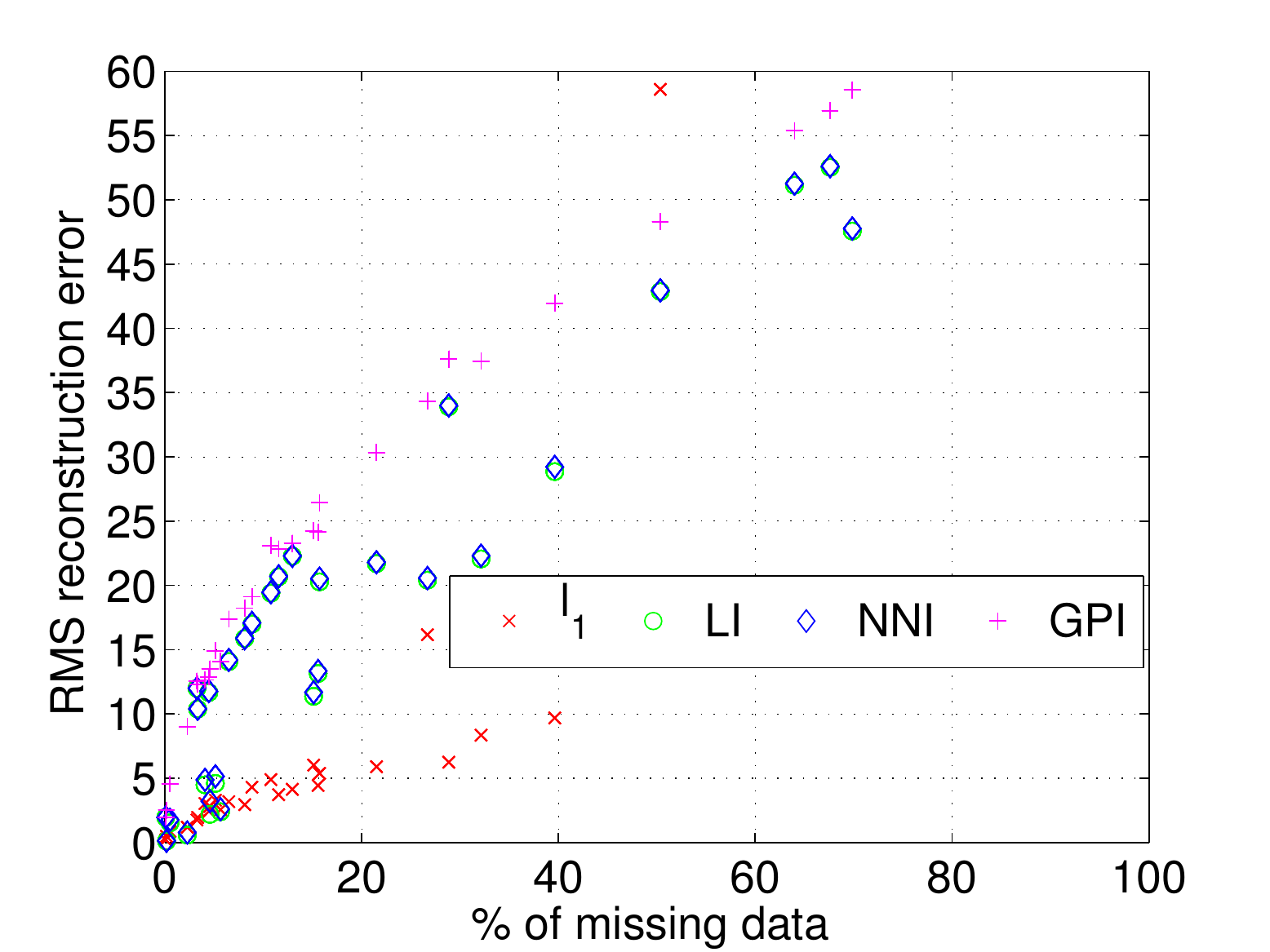}
\label{fig:ex_4}
}
\caption{Comparison of various reconstruction algorithms. Along the $x$-axis is the percentage of missing data points and its impact on reconstruction accuracy expressed in RMS error is given in the $y$-axis. The reconstruction methods are $\ell_1$ $=$ $\ell_1$-minimization; LI $=$ linear interpolation; NNI $=$ nearest neighbor interpolation; and GPI $=$ Gaussian process interpolation. Note that LI (represented by green circles) and NNI (represented by blue diamonds) give almost identical results, so the blue diamonds obscure the view of the green circles. }
\label{fig:performance}
\end{figure}

\section{Conclusions}
\label{sec:conclude}
In this paper, we present the design, implementation and evaluation of Ear-Phone, an end-to-end noise pollution mapping 
system based on participatory urban sensing. Ear-Phone comprises signal processing software to measure noise pollution, calibration software to  perform in-situ calibration, and context discovery software (to discover user context in terms of phone placement) at the mobile phone, as well as 
signal reconstruction software and query processing software at the central server. We draw the following conclusions from our experiments and evaluations. 

To address the problem of noise map reconstruction from incomplete data samples,  we investigate the feasibility of various interpolation and regularization techniques which include linear interpolation, Gaussian process modelling and nearest neighbor interpolation and $\ell_1$-norm minimization, respectively. We conclude that $\ell_1$-norm offers better accuracy compared to others.
 
Using simulation experiments, we show that Ear-Phone can recover a noise map with high accuracy, allowing nearly 
40\% missing samples.
Furthermore, experiential results from two outdoor noise mapping experiments reveal that Ear-Phone can accurately characterize the noise levels along roads from incomplete samples.

Experiments relating to calculating measurement accuracy in various sensing contexts reveal that only usable noise level data can be acquired while the phone is held in hand. In other contexts e.g. in bag and pocket noise levels get corrupted due to time lag and amplitude shift. The proposed classification algorithm can detect the mobile phone in hand with 84\% accuracy.

%

We also implement an in-situ calibration system which can be easily operated by the general public. 

In summary, our work demonstrates that it is feasible to use mobile phones for environmental sensing applications 
such as noise pollution monitoring, overcoming the real-world challenges of irregular sampling, calibration variations among devices, and irregular placement of the mobile phones. 

\bibliographystyle{elsarticle-num}
\bibliography{reference_ipsn}

\section{Appendix}
\subsection{Basis Selection}
\label{sec:basis_selection}
\begin{figure}[thp]
\centering
\resizebox{!}{4.5 cm}{
\includegraphics[width=.8\columnwidth]{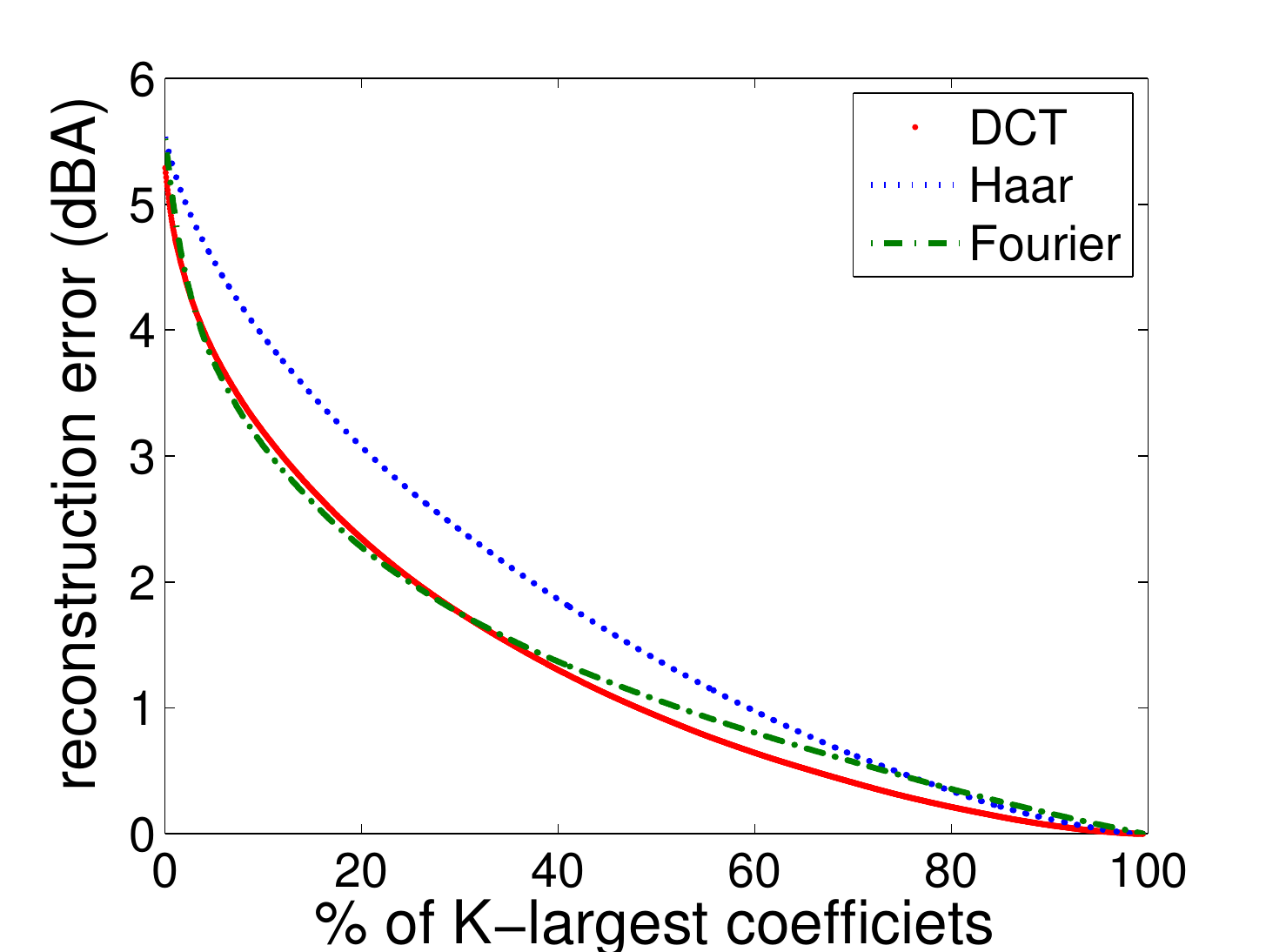}
}
\caption{Compressibility of the noise profile.}
\label{fig:Sparsity_Map}
\end{figure}
In order to study the compressibility of noise profile, we compute their representations in a number of transform bases, which include DCT, Fourier and different wavelets such as Haar, Daubechies, Symlets, Coiflets, and Splines etc. For each basis, we compute the root mean square (RMS) error between the original profile and its approximation by retaining only the largest $k$ ($k = 1,2,... $) coefficients in that basis. Fig.~\ref{fig:Sparsity_Map} is a representative plot that shows the compressibility of noise profile in DCT, Haar and Fourier basis (The results in Figure \ref{fig:Sparsity_Map} is obtained from reference profile $4$ mentioned in Section \ref{sec:simulation}. We have carried out similar study using the other collected noise profiles, and they give similar results.). We observe that for same number of coefficients, the representation in DCT gives a lower error compared to other bases. In the last column of Table~\ref{tab:sound_level}, we have summarized the percentage of DCT coefficients required to approximate the profiles collected in all experiments within 1 dBA RMS error.
\begin{figure}[]
\centering
\subfigure[]{
\includegraphics[height=0.3\columnwidth,width=0.3\linewidth]{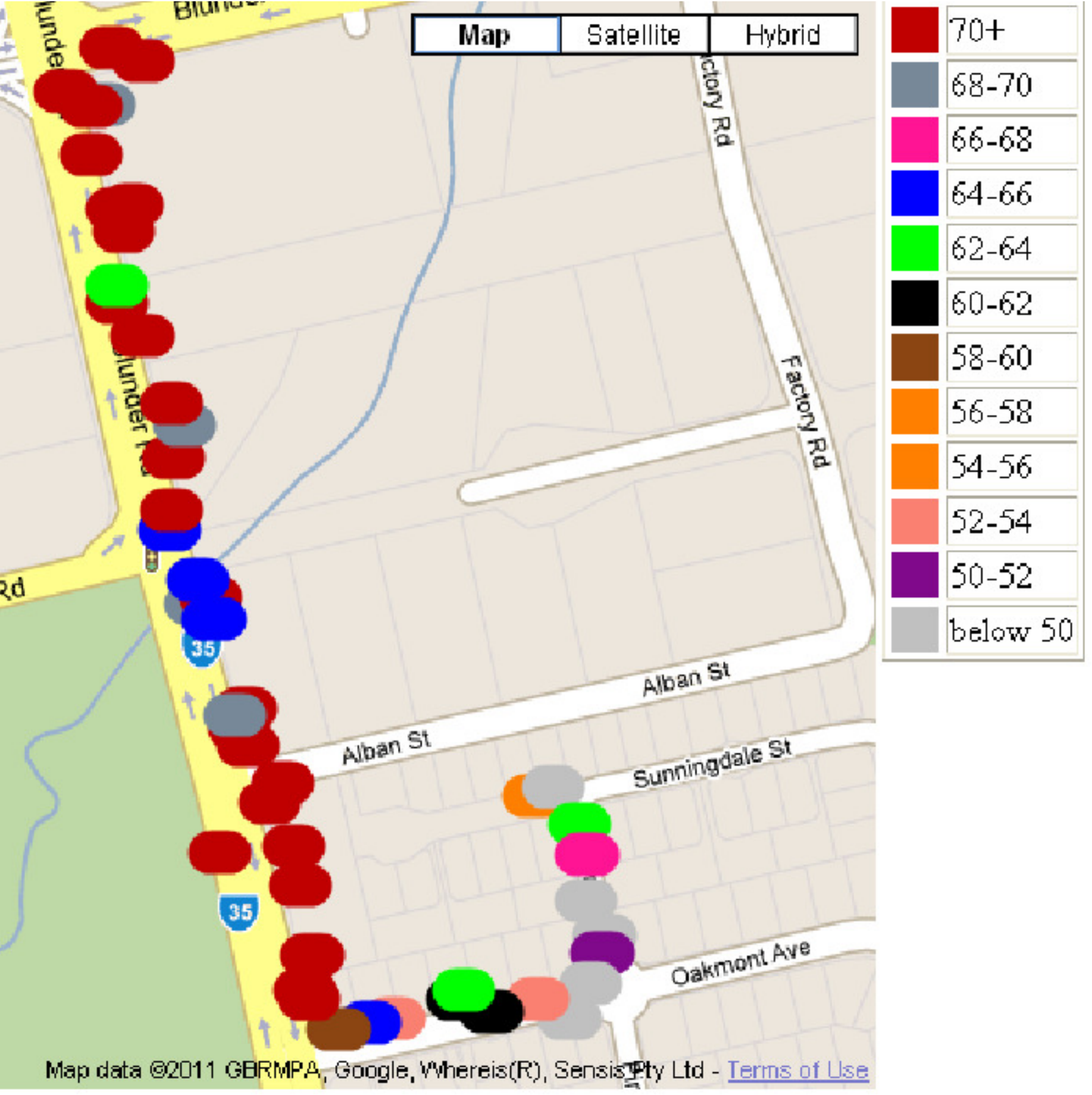}
\label{fig:gt_day_3_peak}
}
\subfigure[]{
\includegraphics[height=0.3\columnwidth, width=0.3\linewidth]{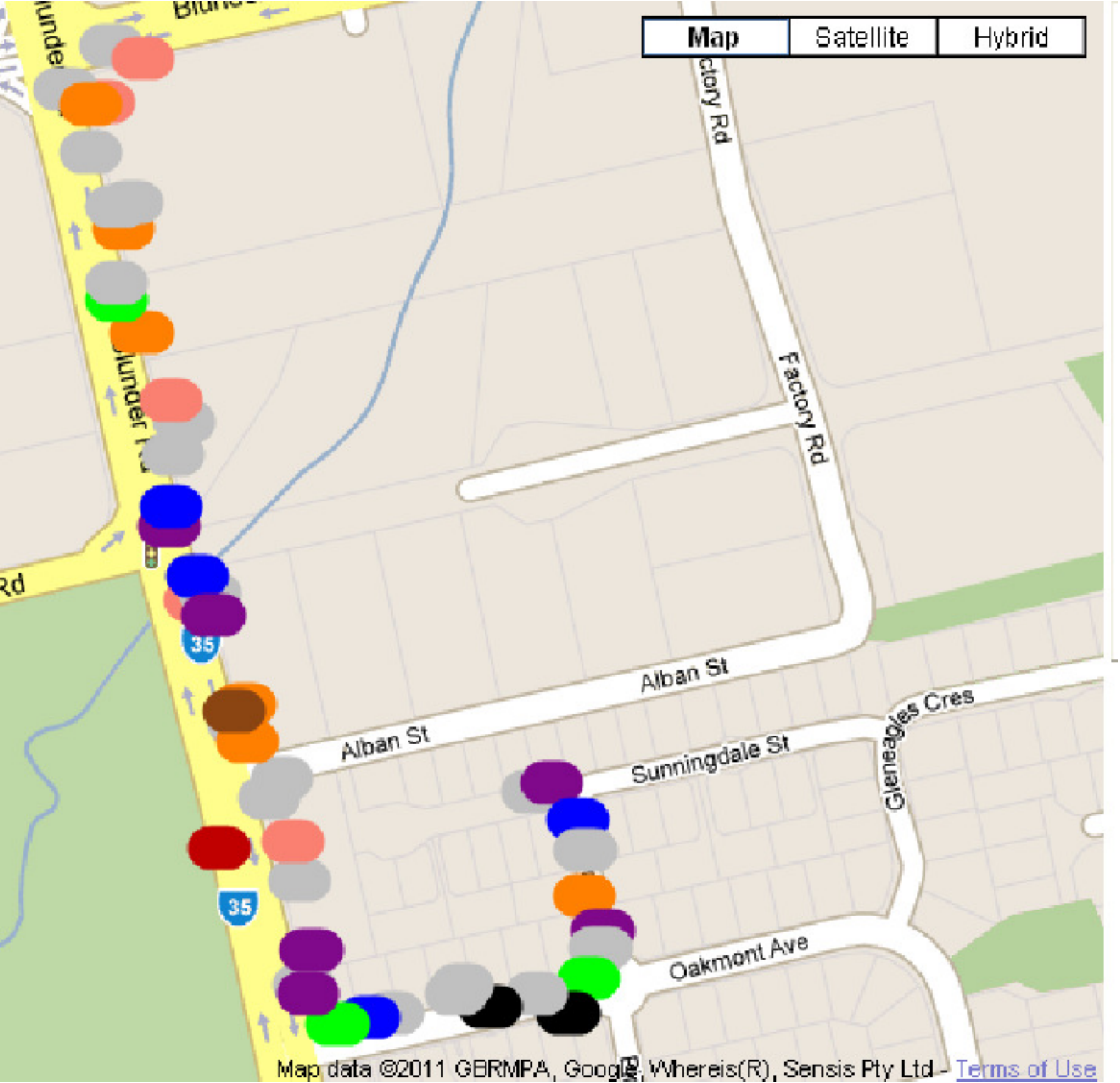}
\label{fig:3peak}
}
\subfigure[]{
\includegraphics[height=0.3\columnwidth, width=0.3\linewidth]{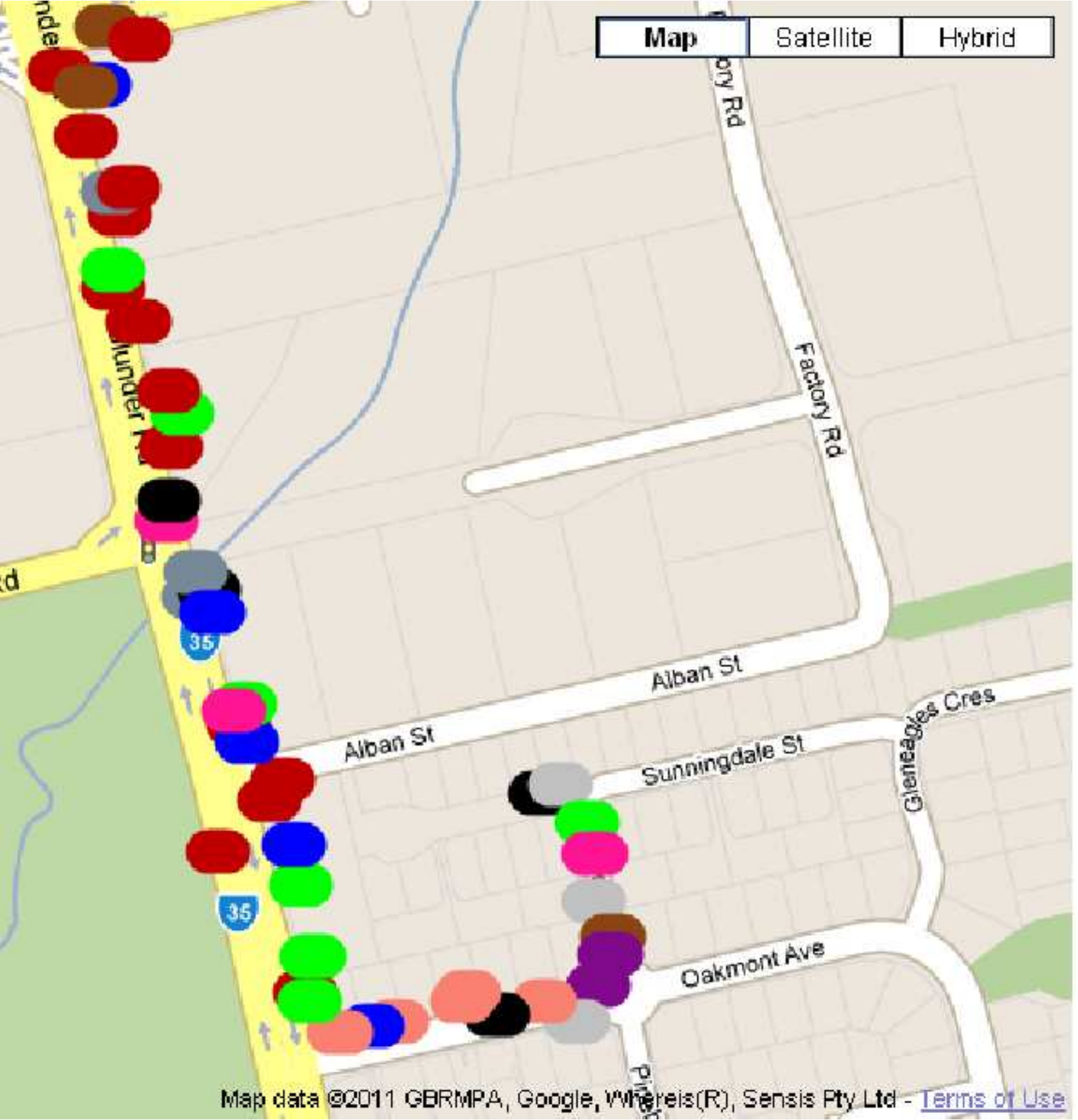}
\label{fig:5peak}
}
\subfigure[]{
\includegraphics[height=0.3\columnwidth, width=0.3\linewidth]{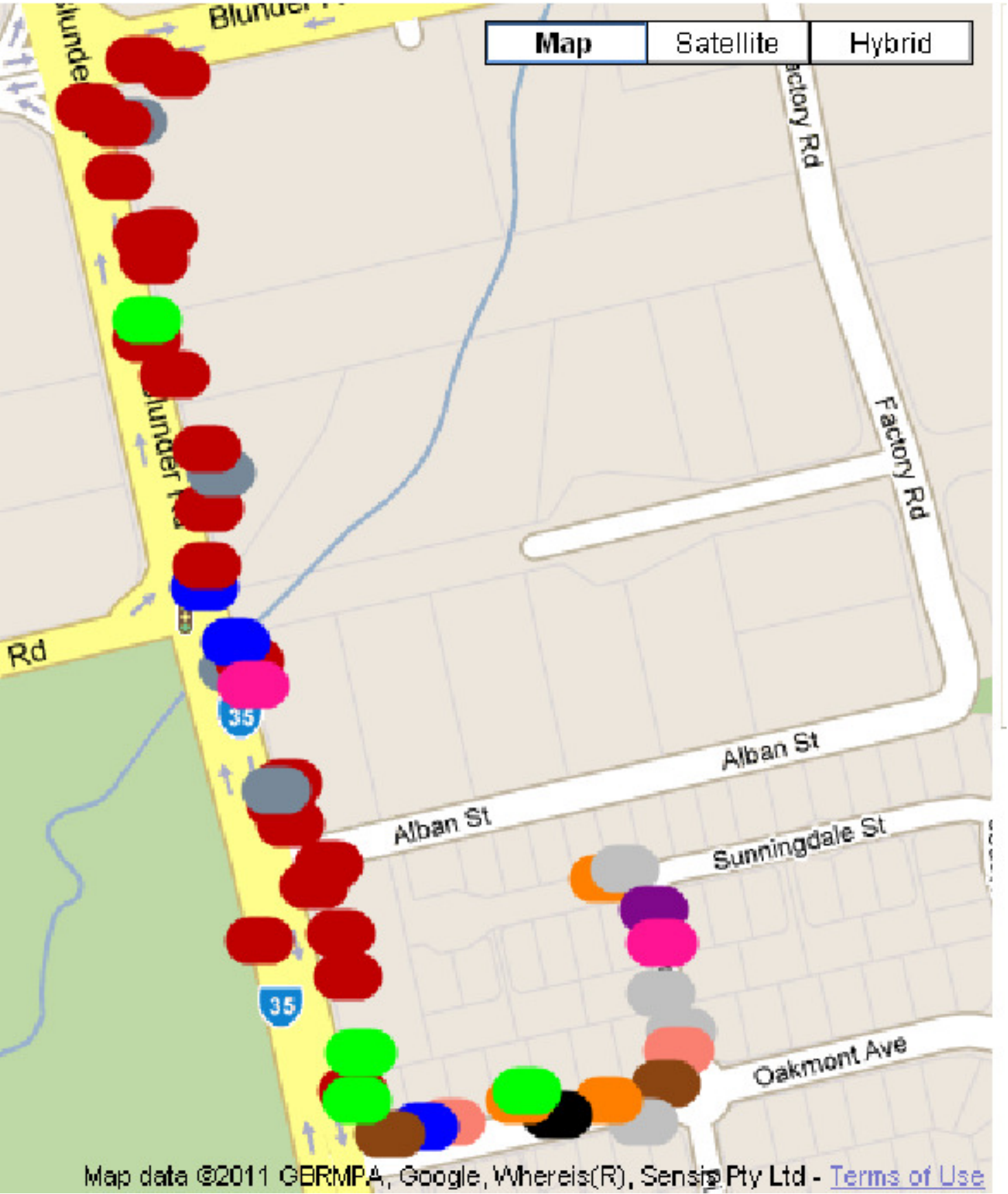}
\label{fig:7peak}
}
\caption{Noise map reconstruction during peak traffic hour (10:00 -11:00 ) at Blunder Road. (a) Ground Truth (b) Reconstruction using 90\% missing samples, (c) Reconstruction using 50\% missing samples, and (d) Reconstruction (very close to the ground truth) using 30\% missing samples. }
\label{fig:2peak}
\end{figure}

\subsection{Reconstruction performance}
Reconstruction performnace at Blunder Road during peak and off-peak hours are shown in Fig~\ref{fig:2peak} and Fig~\ref{fig:2offpeak}, respectively.
\begin{figure}[tbp]
\centering
\subfigure[]{
\includegraphics[height=.3\columnwidth,width=.3\linewidth]{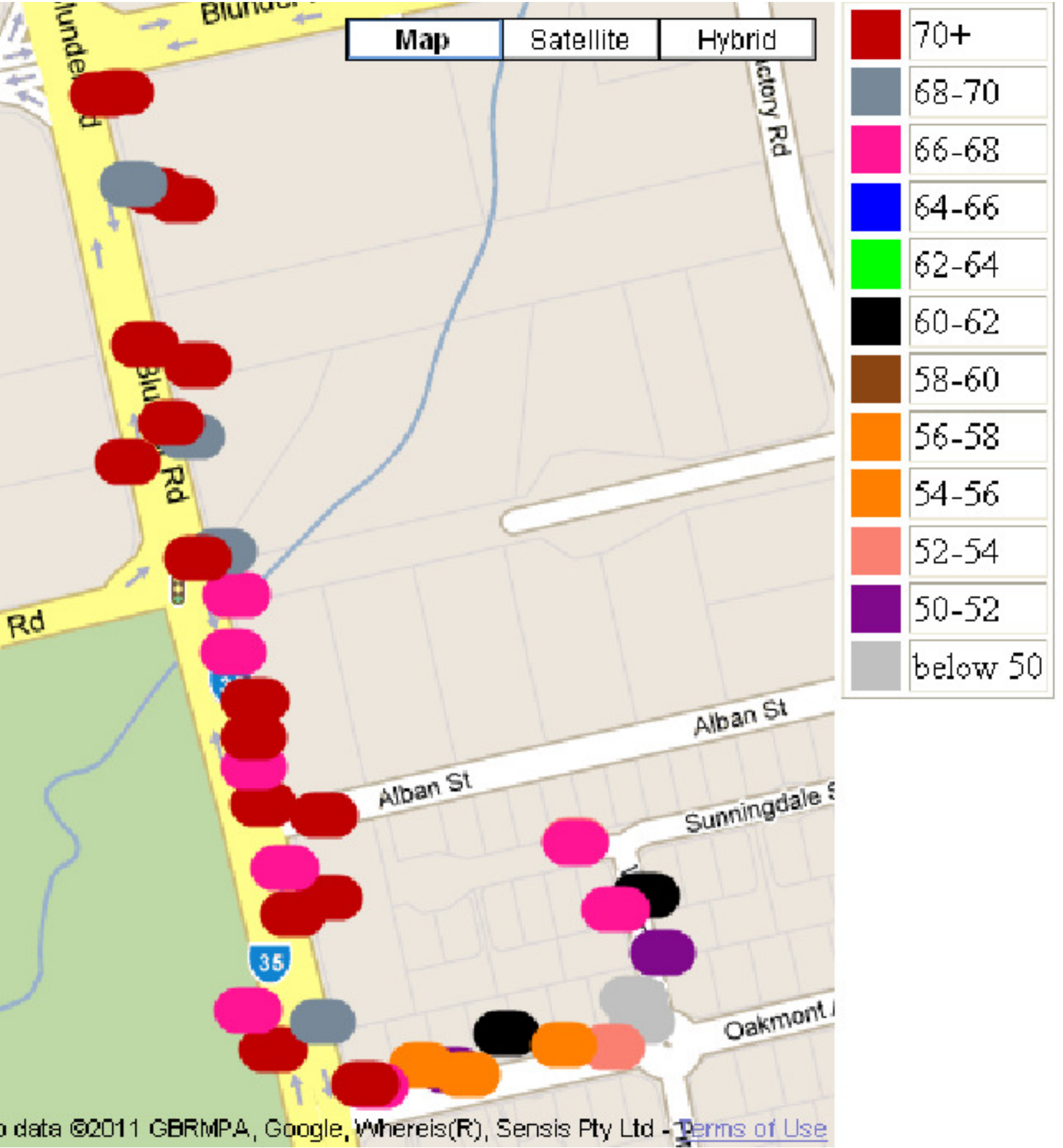}
\label{fig:gt_day_3_offpeak}
}
\subfigure[]{
\includegraphics[height=.3\columnwidth,width=.3\linewidth]{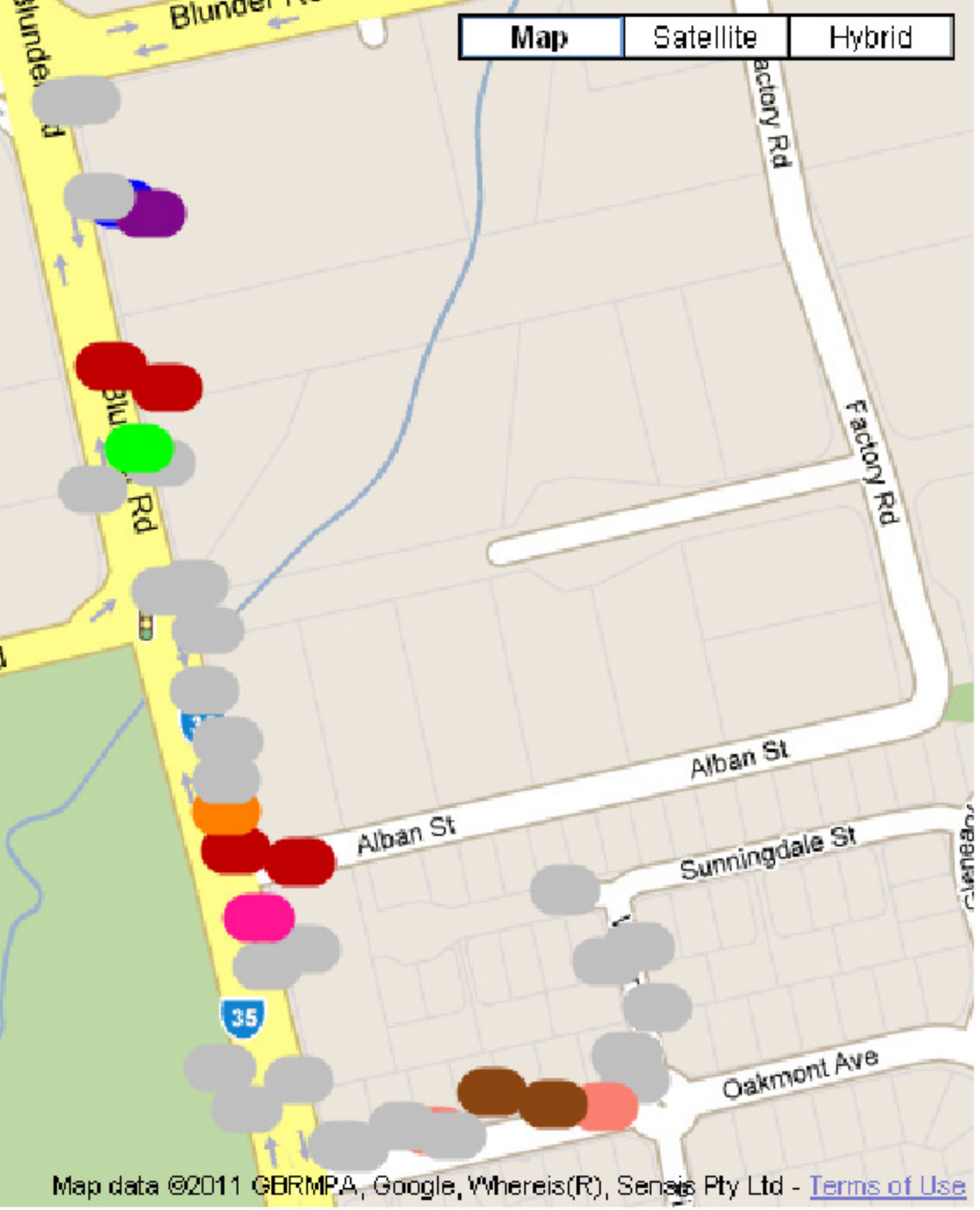}
\label{fig:3peak}
}
\subfigure[]{
\includegraphics[height=.3\columnwidth,width=.3\linewidth]{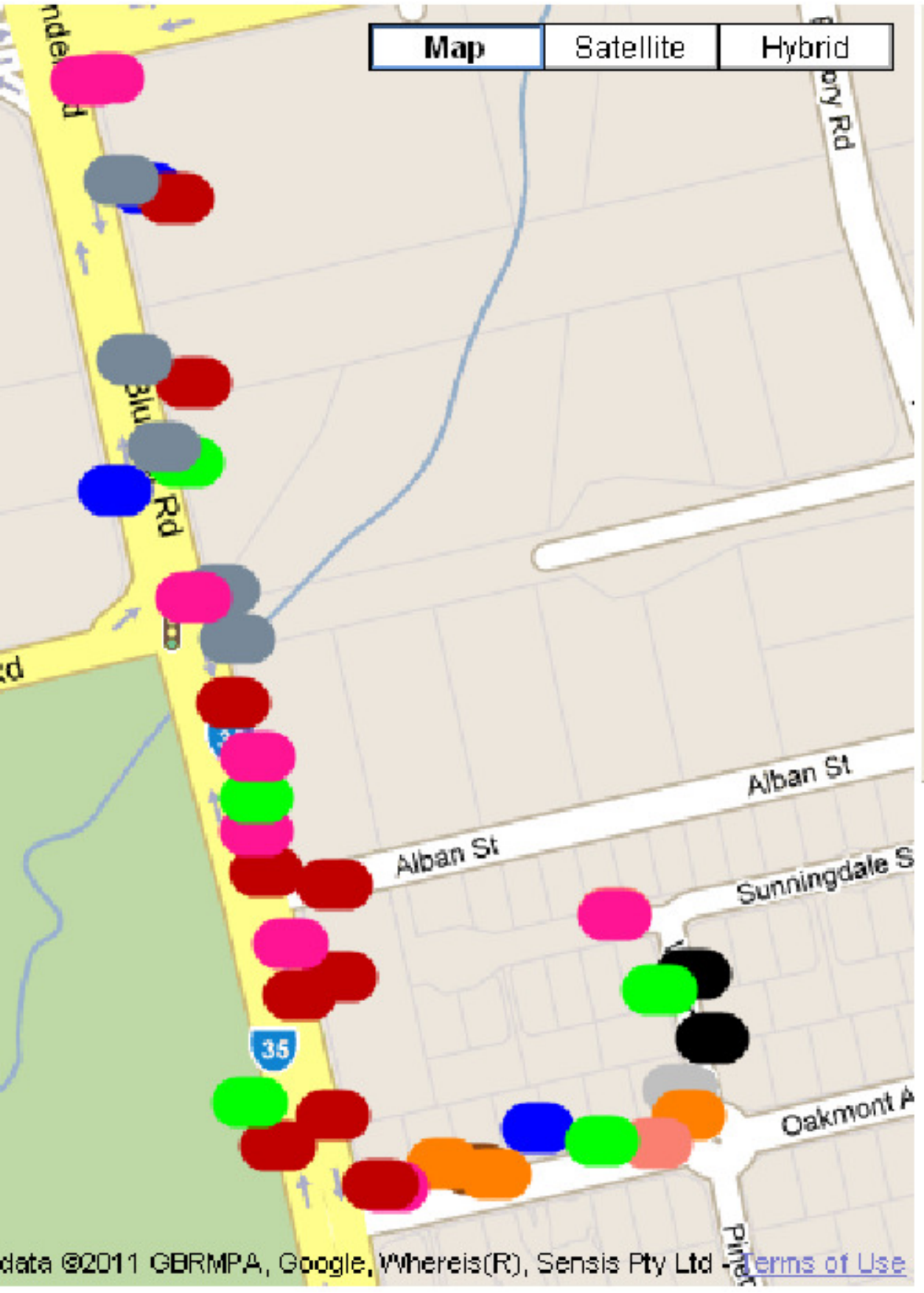}
\label{fig:5peak}
}
\subfigure[]{
\includegraphics[height=.3\columnwidth,width=.3\linewidth]{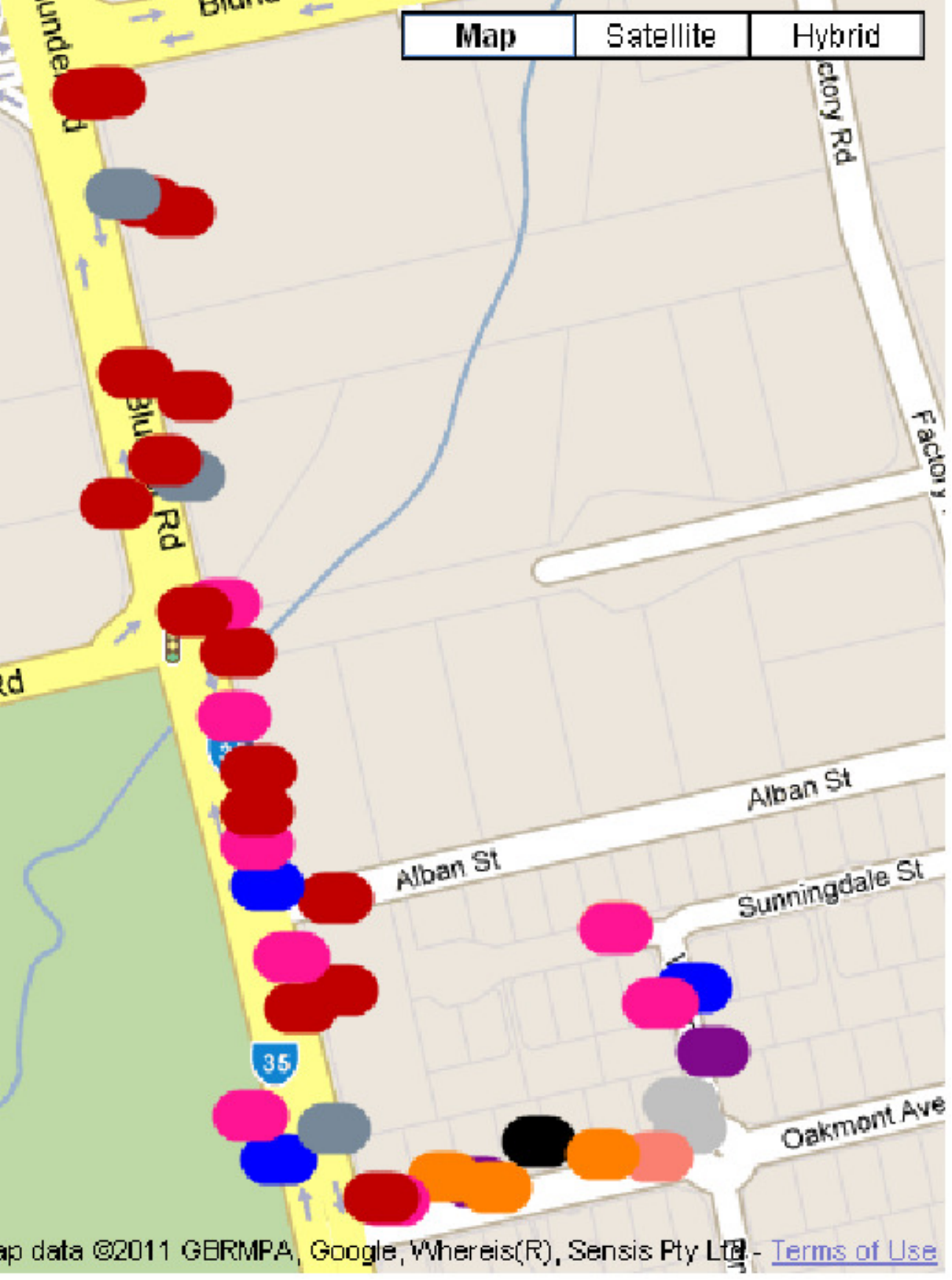}
\label{fig:7peak}
}
\caption{Noise map reconstruction during off-peak traffic hour (15:00 -16:00 ) at Blunder Road. (a) Ground Truth (b) Reconstruction using 90\% missing samples, (c) Reconstruction using 50\% missing samples, and (d) Reconstruction (very close to the ground truth) using 30\% missing samples. }
\label{fig:2offpeak}
\end{figure}

\begin{table}[t]
\small
\centering
\caption{Coefficient of the digital filter that approximates A-weighting.}
\setlength{\tabcolsep}{1pt}
\begin{tabular}{|c|r|r|r|r|r|r|r|r|r|r|r|} \hline
$\ell$     & 0 & 1 & 2 & 3 & 4 & 5 & 6 & 7 & 8 & 9 & 10 \\ \hline
$b_\ell$ & 0.9299 &  -2.1889  &  0.7541  &  1.3229  & -0.7728  &  0.1025 &  -0.2398  & -0.0098  &  0.1154 &  -0.0103  & -0.0033 \\ \hline
$a_\ell$ &  & 2.1856  & -0.7403 &  -1.0831 &   0.6863  & -0.2274 & 0.2507 &  -0.0058  & -0.0821  &  0.0153  &  0.0004 \\ \hline
\end{tabular}
\label{tab:filter_coeff}
\end{table}

\end{document}